\documentclass{article}
\usepackage[margin=1in]{geometry} 
\usepackage{comment}
\usepackage{amsmath}
\usepackage{xcolor}
\usepackage{graphicx}
\usepackage{subcaption}
\usepackage{float}
\usepackage{multirow}
\usepackage{boldline}
\usepackage{booktabs}
\usepackage{bm}
\usepackage{txfonts}
\usepackage{natbib}
\usepackage{hyperref}
\usepackage{authblk}

\usepackage[utf8]{inputenc}
\newcommand{\acosmos}{A$^{3}$COSMOS}
\newcommand{\arcseccustom}{$^{\prime\prime}$}
\newcommand{\z}{\bf z}
\newcommand{\y}{\bf y}
\newcommand{\xzero}{{\bf x}_0}
\newcommand{\noise}{\boldsymbol{\epsilon}}
\newcommand{\x}{\mathbf{x}}
\newcommand{\DDPM}{g_{\boldsymbol \theta}}

\newcommand{\xrecon}{\hat{\bf x}_0}
\newcommand{\nbruns}{n}
\newcommand{\image}{{\bf s}}
\newcommand{\aggregatedetect}{\textit{aggregate}-\textit{detect}}
\newcommand{\detectaggregate}{\textit{detect}-\textit{aggregate}}
\newcommand{\detectaggregatetable}{etect-aggregate}
\newcommand{\outputswidth}{0.10\textwidth}
\newcommand{\mystyle}[1]{\ifmmode\mathit{#1}\else\textit{#1}\fi}

\begin{document}

\title{Radio-astronomical Image Reconstruction with Conditional Denoising Diffusion Model}

\author[1]{M. Drozdova}
\author[1]{V. Kinakh}
\author[2]{O. Bait}
\author[1]{O. Taran}
\author[1]{E. Lastufka}
\author[2]{M. Dessauges-Zavadsky}
\author[1]{T. Holotyak}
\author[2]{D. Schaerer}
\author[1]{S. Voloshynovskiy\thanks{Corresponding authors: svolos@unige.ch, mariia.drozdova@unige.ch}}

\affil[1]{Department of Computer Science, University of Geneva, 7 route de Drize, 1227 Carouge, Switzerland}
\affil[2]{Geneva Observatory, University of Geneva, 51 Chemin Pegasi, 1290 Versoix, Switzerland}

\maketitle

\begin{abstract}
    
{Reconstructing sky models from dirty radio images for accurate source extraction, e.g. source localisation and flux estimation, is a complex yet critical task. It has important applications in galaxy evolution studies at high redshift, particularly in deep extragalactic fields using for example the Atacama Large Millimetre Array (ALMA).
With the development of large-scale projects like the Square Kilometre Array (SKA), we anticipate the need for more advanced source extraction methods. Existing techniques, such as CLEAN and PyBDSF, currently struggle to effectively extract faint sources, highlighting the necessity for the development of more precise and robust methods.
}
{The success of source extraction task critically depends on the quality and accuracy of image reconstruction. Since the imaging process represents an information-lossy operator, the reconstruction is characterized by uncertainty. The current study proposes the application of stochastic neural networks for the direct reconstruction of sky models from dirty images. This approach allows to localize radio sources and to determine their fluxes with the corresponding uncertainties, providing a potential advancement in the field of radio source characterization.}
{We utilize a dataset of 10164 images simulated with the CASA tool simalma based on the ALMA Cycle 5.3 antenna configuration. We apply conditional Denoising Diffusion Probabilistic Models (DDPMs) to directly reconstruct sky models from these dirty images, followed by processing through Photutils to extract the coordinates and fluxes of the sources. To test the robustness of the proposed model which was trained on a fixed water vapour value, we examine its performance under varying levels of water vapor.}
{ The proposed approach demonstrates state-of-the-art in source localisation, achieving over 90\% completeness at a signal-to-noise ratio (SNR) as low as 2.  Additionally, the described method offers an inherent measure of prediction reliability thanks to the stochastic nature of the chosen model. In terms of flux estimation, the proposed model excels past PyBDSF performance, accurately extracting fluxes for 96\% of the sources in the test set, a notable improvement over CLEAN+ PyBDSF's 57\%.}
{Conditional DDPMs is a powerful tool for image-to-image translation, yielding accurate and robust characterisation of radio sources, and outperforming existing methodologies. While this study underscores its significant potential for applications in radio astronomy, we also acknowledge certain limitations that accompany its usage, suggesting directions for further refinement and research. \textbf{The Python implementation of the proposed framework
is publicly available at \url{https://github.com/MariiaDrozdova/diffusion-for-sources-characterisation}} }

\end{abstract}


%

\section{Introduction}
\begin{figure*}[ht]
  \centering
  \includegraphics[width=\linewidth]{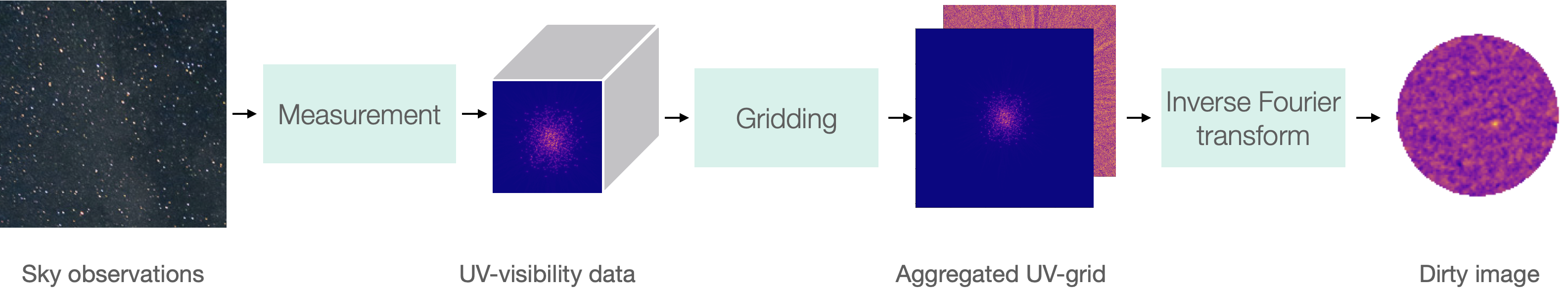}
  \caption{Simplified pipeline for real and simulated observations. Each pair of antennas collects sky observations that yield UV-visibility data. These data are then aggregated via a gridding process into an aggregated UV grid. Finally, an inverse Fourier transform is applied along with the effect of a primary beam to generate the dirty image.}
  \label{fig:observation_model}
\end{figure*}

 The advent of the Square Kilometer Array (SKA) will revolutionize radio astronomy at centimeter (cm) and millimeter (mm) wavelengths. Already, the various SKA precursors, such as the Australian Square Kilometre Array Pathfinder \citep[ASKAP;][]{ASKAP}, the Murchison Widefield Array \citep[MWA;][]{MWA}, the LOw-Frequency ARray \citep[LOFAR;][]{LOFAR}, and the MeerKAT radio telescope \citep[][]{MeerKAT}, have started to produce interesting results. The Atacama Large Millimeter Array (ALMA), owing to its excellent sensitivity, led to several interesting studies. Some of the studies relevant to our current work are the  deep extragalactic continuum surveys, which focus on the gas and dust in high-redshift galaxies; for example, the Reionization Era Bright Emission Line Survey \citep[REBELS;][]{REBELS}, the ALMA SPECtroscopic Survey in the Hubble Ultra-Deep Field \citep[ASPECS; ][]{ASPECS}, the ALMA Large Program to Investigate [CII] at Early times \citep[ALPINE][]{ALPINE1, ALPINE2, ALPINE3}, and the GOODS-ALMA survey at 1.1 mm \citep{GOODS-ALMA} and the Automated Mining of the ALMA Archive in the COSMOS Field (\acosmos) data set\footnote{\url{https://sites.google.com/view/a3cosmos/home?authuser=0}} \citep{liu19a3cosmos}.
 
These modern radio interferometers lie at the cutting edge of instrumentation in terms of the receivers, correlators, low-noise amplifiers, high-speed data links, and other technologies. They are also expected to produce data on exabyte scales \citep{Scaife20}, which will require novel and efficient data reduction and analysis techniques. However, in order to convert the observed visibilities to science-ready images, the processing of the resulting data still relies on the traditional CLEAN algorithm \citep{hogbom1974aperture} for imaging and deconvolution.

In radio astronomy, images are reconstructed using visibilities measured at discrete locations depending on the antenna configuration and integration time (aperture synthesis imaging). If the field of view of the antenna is relatively small, these visibilities can be converted into ``dirty images'' by taking an inverse Fourier transform. The dirty images are further improved to remove the various artifacts coming from the side lobes using a deconvolution process. This entire process of constructing the deconvolved images from the observed visibilities is referred to as image reconstruction. The most popular algorithm for such an imaging and deconvolution process is the CLEAN algorithm \citep{Hogbom74, Clark80}, which essentially relies on identifying intensity peaks modeled as point sources in the dirty image. These peaks are iteratively removed (or CLEANED) after convolving them with the dirty beam. In each iteration, peaks of  progressively lower intensity are CLEANED until a user-defined threshold is achieved. Several variants of the CLEAN algorithm have been introduced to mitigate its various limitations; for example, the multifrequency synthesis algorithm \citep[mfs;][]{Conway90} and multiterm mfs \citep[mtmfs;][]{Rau11} for wide-band imaging. These are widely used by the community with radio interferometers, such as ALMA, Very Large Array (VLA), MeerKAT, and so on. However, all CLEAN-based algorithms do not address the fundamental issue of image reconstruction in radio astronomy. As the visibilities are only measured at discrete locations, imaging in radio astronomy is an ill-posed problem, such that the same visiblities can be mapped to multiple sky models. Further, the CLEAN algorithm can benefit from prior knowledge of source positions, which usually requires expert human knowledge.

In addition to the CLEAN method, there exist other methods for reconstruction of sky-model images from dirty images. Notably, Tikhonov regularization \citep{tikhonov1977solutions} and Wiener filtering \citep{wiener_alma99737573406836} represent two classic deterministic methods. These iterative methods operate in an iterative manner to minimize the fidelity loss with some priors on the expected solution:  $\mathcal{L}(\image) = ||\z - H\image||^2_2 + \lambda \Omega(\image)$, where $\z$ is a dirty image, $\lambda$ is a  regularization constant, and $\Omega(\image)$ denotes the regularizer. Considering probabilistic methods with a prior $p(\image)$, one can define $\Omega(\image) \propto - \ln p(\image)$.

Another method used in radio astronomy is based on the maximum entropy method (MEM; see e.g., \citealt{maximum_entropy, Cornwell85, Narayan86}), whereby the distribution
with the greatest entropy is chosen among all the distributions that fit the data. Further techniques include bias methods, the total variation prior \citep[]{variation_methods}, and Gaussian mixture methods \citep[]{bishop2007}. These methods are often handcrafted and deterministic, complementing the CLEAN method.

Various deep learning models have recently been proposed to solve the problem of image reconstruction in radio astronomy. \citet{Gheller22} use an autoencoder network to ``denoise'' the dirty images, particularly for diffuse extended emission. \citet{Connor22} used a feed-forward neural network to achieve super-resolution on already CLEANED images. \citet{Schmidt22} used a deep learning(DL)-based approach motivated by super-resolution models, where they use a series of convolutional layers on the sampled amplitude and phases of the visibilities to reconstruct the missing visibilities. Complementing these deep learning models, Bluebild  \citep{tolley2023bipp} offers a novel approach, leveraging functional principal component analysis (PCA) to synthesize images by decomposing the sky into eigenimages with  distinct energy levels. 

At the same time, unprecedented features of DL allow visibilities to be  directly translated into estimates of sky models or source locations. This allows the user to skip the stage of dirty-image computation. \cite{Taran_2023} introduced a method for localizing radio sources from reduced UV samples. The approach taken by these authors involves a two-stage pipeline: initially, a fully connected network predicts a binary map where a value of ``1'' signifies the presence of a source and ``0'' indicates the background. Following this, an autoencoder refines the predictions. With the same dataset as used in the present study, \cite{Taran_2023} achieved a purity of 91\% and a completeness of 74\%. A detailed comparison of our results and theirs can be found in Section \ref{sec: results}.

In the present work, we propose an alternative DL-based solution: we use a denoising diffusion probabilistic model (DDPM; \citealt{sohl2015deep, ho2020denoising}) to reconstruct the sky model from dirty images and then localize the sources. DDPMs are a family of generative models that use a diffusion process to represent data distribution.  In its basic unconditional formulation, this type of model learns to map artificially generated noise with the desired target distribution.

DDPMs have shown remarkable performance on a wide range of tasks: unconditional image generation in \cite{ho2020denoising}, conditional image generation in \cite{dhariwal2021diffusion}, and text-to-image generation in \cite{ramesh2022hierarchical}. DDPMs are also applicable to other image-to-image translation tasks, such as super-resolution, colorization, and inpainting (\cite{saharia2022palette}). \citet{Wang23} demonstrate for the first time that DDPM can be used for image reconstruction for radio astronomical images trained on data generated for the Event Horizon Telescope \citep{EHT19}. In summary, the stochastic nature of DDPMs is beneficial in addressing ill-posed problems where multiple solutions exist for the same observation.

The present paper is organized as follows. In Section \ref{sec: radio_data}, we provide a brief overview of the simulated radio data used in the DDPM training. Section \ref{sec: proposed method} details the proposed pipeline, encompassing preprocessing steps, the DDPM model, post-processing techniques, and the range of metrics employed to evaluate performance. Then, in Section \ref{sec: results}, we present our findings in terms of the performance of the DDPM approach in reconstruction, localization, and flux estimation. Additionally, we analyze the performance in relation to S/N and the varying noise amplitudes caused by water vapor. In Sections \ref{sec: discussion} and \ref{sec: conclusions}, we consider the limitations of the approach and discuss potential directions for future research.

\section{Simulated data}
\label{sec: radio_data}

We performed the entire training of the proposed DDPM model on simulated ALMA data for the 12m array. These data are the same as those used in \citet{Taran_2023} for their DL model training. We refer to \citet{Taran_2023} for further details of the simulations. All the simulations are produced using the Common Astronomy Software Applications (CASA) data processing software \textsc{v6.2} \citep{mcmullin2007casa, CASA}. Briefly, we chose a fixed ALMA configuration, cycle 5.3. These simulations are designed to somewhat match the real observations from the \acosmos ~dataset \citep{liu19a3cosmos} with the aim being to apply to them in the future. Therefore, all the simulated pointings are randomly distributed around the COSMOS field (J2000 10h00m28.6s +02d12m21.0s) within a radius of 1 deg. Each pointing corresponds to an independent sky model. In the simulations, a sky model corresponds to an image of 51.2$\times$51.2\arcseccustom ~across. The various sources in this sky model are modeled as a 2D Gaussian overlaid on a zero background. The various source properties, such as their sizes (major and minor axis of the Gaussian), positions on the sky, and fluxes in each sky model are varied randomly. We fixed the ALMA band to Band-6 (230 GHz) split into 240 channels with a channel width of 7.8 MHz. We added the effect of noise from water vapor and receivers using a precipitable water vapor (PWV) parameter value of 1.796. For each sky model, we produced the corresponding visibilities for a total integration time of 20 mins while the phasecenter of the sky model was under transit. We then produced the corresponding dirty images for each pointing using the CASA task \textsc{tclean} by setting the \textsc{niter} parameters equal to zero. This task does the appropriate UV-gridding and inverse Fourier transform of the visibilities, along with the effect of primary beam to produce the corresponding dirty image. Figure \ref{fig:observation_model} shows a block diagram of the simulation pipeline. In total, we have 9164 simulated pointings (sky models) and the corresponding dirty images. This comprises a total of 27632 sources, plus 1000 simulations without sources. 

\begin{figure}[ht]
    \centering
    \includegraphics[width=0.3\linewidth]{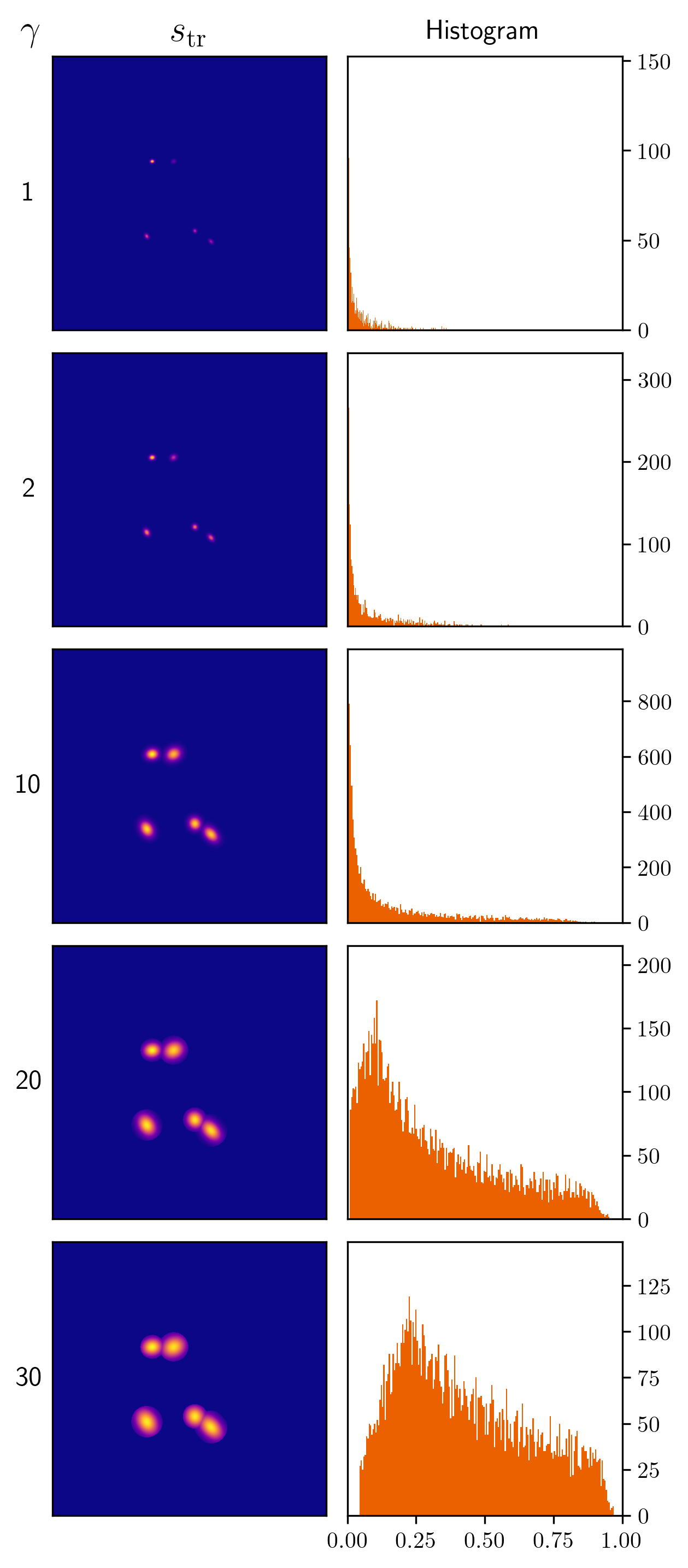}
    \caption{Effect of $\gamma$ value on sky model preprocessing. The preprocessing of the sky model via Equation \ref{eq:normalization} is significantly affected by the $\gamma$ value. The proposed method modifies the intensities  of the image pixels to improve the network's capacity for accurate reconstruction. By employing a nonlinear root operation, nonzero values are pushed closer to one, effectively making the histogram more uniform. However, this modification also impacts the relative distances. Therefore, an optimal $\gamma$ value is required to ensure a compromise between these two tendencies. The histogram y-axis limits are set so that the height of the  third bin comprises 80\% of the image.}
    \label{fig:images_normalization}
\end{figure}

In this study, these dirty images do the conditioning for the DDPM model, which predicts the corresponding sky model. In order to study the effect of varying noise, we also add simulations with the same configuration, but change the PWV values from 0.472 to 5.186. We note that these additional simulations are not used for training the model, but simply for testing the effect of varying noise, as in the analysis of \citet{Taran_2023}. 

\section{Proposed method}
\label{sec: proposed method}
\subsection{Preprocessing}

In this work, we apply machine learning based on DDPM to dirty images in order to predict sky models. Both dirty images and sky models require preprocessing. In machine learning, natural images are usually mapped to the $[-1, 1]$ range by a simple scaling. As we work with measurement data, which have their own properties and statistics, we need to adjust the data-preprocessing step.

The sky models are non-negative and are typically very sparse and of very small magnitude, of order $10^{-5}$ Jy. We would like to not only change the range of values for sky models but make the distribution of the values flatter. A sky model is what the model predicts, and so the loss functions will be applied directly to it.  In this study, we use mean square error (MSE) due to the model choice. MSE loss works better with dense images (the ones whose values have flatter histograms). We therefore need a transform that makes the image histograms flatter and is also invertible, because after getting predictions, we need to return to the original ranges.

 We take inspiration from the preprocessing techniques used in particle physics \citet{Finke_2021} to face the sparsity and low signal magnitude. These use blur and root functions. As the sky models already contain sources modeled as Gaussians, we retain only the root function. We use
the following scaling in the experiments:
\begin{equation}
\image_{\text{tr}} = \left(\frac{\image}{c}\right)^{\frac{1}{\gamma}},
\label{eq:normalization}
\end{equation}
where $\image$ is a sky model, $c$ is a constant, and $\gamma$ is a variable parameter for which we explore a range of values, spanning from 1 to 30. The value of $c$ is set to $2.96 \times 10^{-5}$, corresponding to the maximum value of image intensity identified empirically within the training set.  This method helps us to enhance the contrast of the image and mitigate the issues arising from low signal strength.
To map the variables to the range of $[-1,1]$, we apply the following formula:

\begin{gather}
\xzero = \frac{\image_{\text{tr}} - 0.5}{0.5},
\label{eq:scaling}
\end{gather}
where $\xzero$ is considered as a target image for the DDPM system. 

The transformation in Equation \ref{eq:normalization} is not linear, and so it does not preserve distances, which means that small relative values in the transformed image correspond to a huge gap between values in the corresponding original image. We illustrate the influence of power $\gamma$ from Equation \ref{eq:normalization} in Figure \ref{fig:images_normalization}. In this visualization, we take one sky model of the generated set and we show how the suggested normalization in Equation \ref{eq:normalization} changes the image as well as its histogram. The choice of $\gamma = 1$ corresponds to the case where only linear transform is applied. As $\gamma$ grows, the sources become wider and brighter, and the histogram becomes flatter, indicating a denser image.

In the experiments, dirty images $\z$ are obtained from aggregated UV grids  $\y$, where

\begin{gather}
 {\z} = \frac{\frac{{\mathcal{F}}^{-1}({\y})}{c_{d}} - 0.5}{0.5},
\label{eq:dirty_scaling}
\end{gather}
where $\mathcal{F}^{-1}$ is a function that maps UV grids to dirty images.  In CASA,  $\mathcal{F}^{-1}$  includes inverse Fourier and artefact correction.  The $0.5$ factor is used to map images approximately to the $[-1,1]$ range, as in Equation \ref{eq:scaling}.

Images after $\mathcal{F}^{-1}$ are already dense and contain values of the order of $c_{d} = 10^{-5}$ Jy. For these images, we apply a scaling operation by dividing each image by a factor of  $c_{d}$. 

These preprocessing steps map the sky model and dirty images to be within standard ranges of typical machine learning models. The described DDPM estimates the sky model denoted $\xrecon$. In order to map variables back to the original scale, we apply the inverse operations:

\begin{gather}
 \hat{\image}_{\text{tr}} = 0.5\xrecon + 0.5, \label{eq:inverse_scaling} \\[0ex]
 \hat{\image} = c  \hat{\image}_{\text{tr}}^{\gamma}, \label{eq:inverse_normalization}
\end{gather}
where $c$ is the same constant as in Equation \ref{eq:normalization}.

\subsection{Image reconstruction based on stochastic models}
In this study, we address image reconstruction in radio astronomy, which is an inherently ill-posed problem. Information is lost due to UV sampling, gridding, and handling of noise coming from the instruments and atmosphere. Given these limitations, the task of accurately estimating the initial sky model from the measured data is naturally ambiguous. Mathematically, it means that many estimates $\hat{\image}$ might correspond to the same measured data $\z$.

To estimate the sky model, we must inject priors into the model. There are two possible approaches to do this: deterministic methods and stochastic methods.
Deterministic methods ---as in CLEAN---  use hand-crafted priors, and traditional machine learning models are based on learnable frameworks; both operate in a straightforward way. They create a one-to-one mapping: they first find optimal parameters based on the training data and then, once these parameters are fixed, they produce a deterministic output for each unique input.
In contrast, stochastic methods allow one-to-many mapping, meaning they can generate multiple probable solutions for a single input. This is achieved by introducing random variables or model-initiation noise into the model during the computation of the output. Therefore, the output depends on both the input data and the realization of the model initiation noise. By varying this model-initiation noise for fixed model parameters, we can get different outputs for the same input data. All such outputs are the most probable to align with the target data manifold and correspond to the given input data (observations). This inherent ambiguity is useful, as by inspecting different reconstructions we can quantify the uncertainty on the predictions.

\begin{figure*}[ht]
  \centering
  \includegraphics[width=1.0\linewidth]{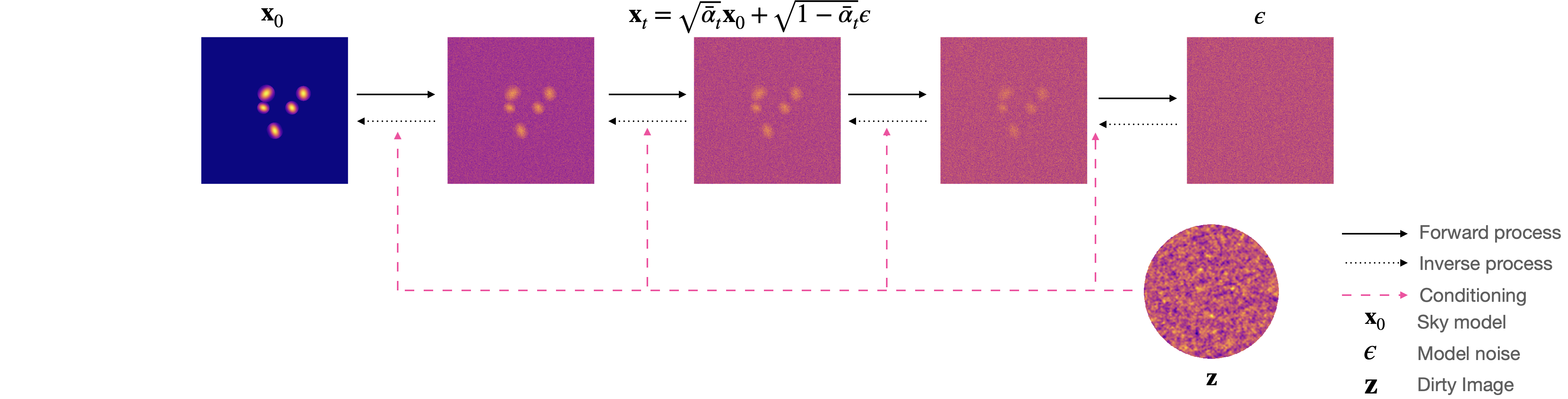}
  \caption{Schematic representation of  diffusion process. The forward process consists in progressively adding noise to the sky model. The inverse process starts with model noise, which is progressively refined with a conditioning on the dirty image.}
  \label{fig:iterative_degrading}
\end{figure*}

\subsection{Denoising diffusion probabilistic models}

\begin{figure*}[ht]
  \centering
  \includegraphics[width=1.0\linewidth]{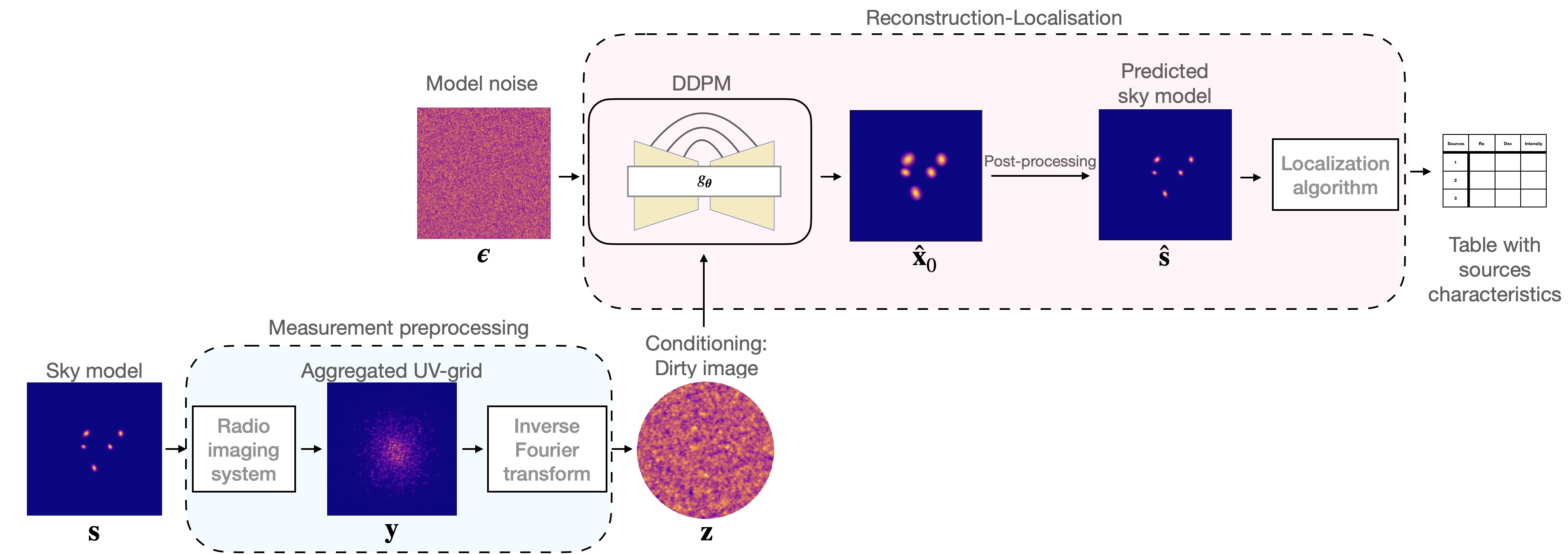}
 \caption{Generalized pipeline of proposed method. The top branch represents inference. The DDPM model acts as a stochastic image-to-image translator, transforming the conditioning $\z$ and the model noise $\noise$ into the target image $\xrecon$ and corresponding $\hat{\image}$. The predicted sky model is then processed by the localization algorithm, outputting the corresponding coordinates and fluxes. The bottom branch represents the simplified measurement process.}
  \label{fig:inference_pipeline}
\end{figure*}

\begin{figure*}[ht]
  \centering
  \includegraphics[width=0.76\linewidth]{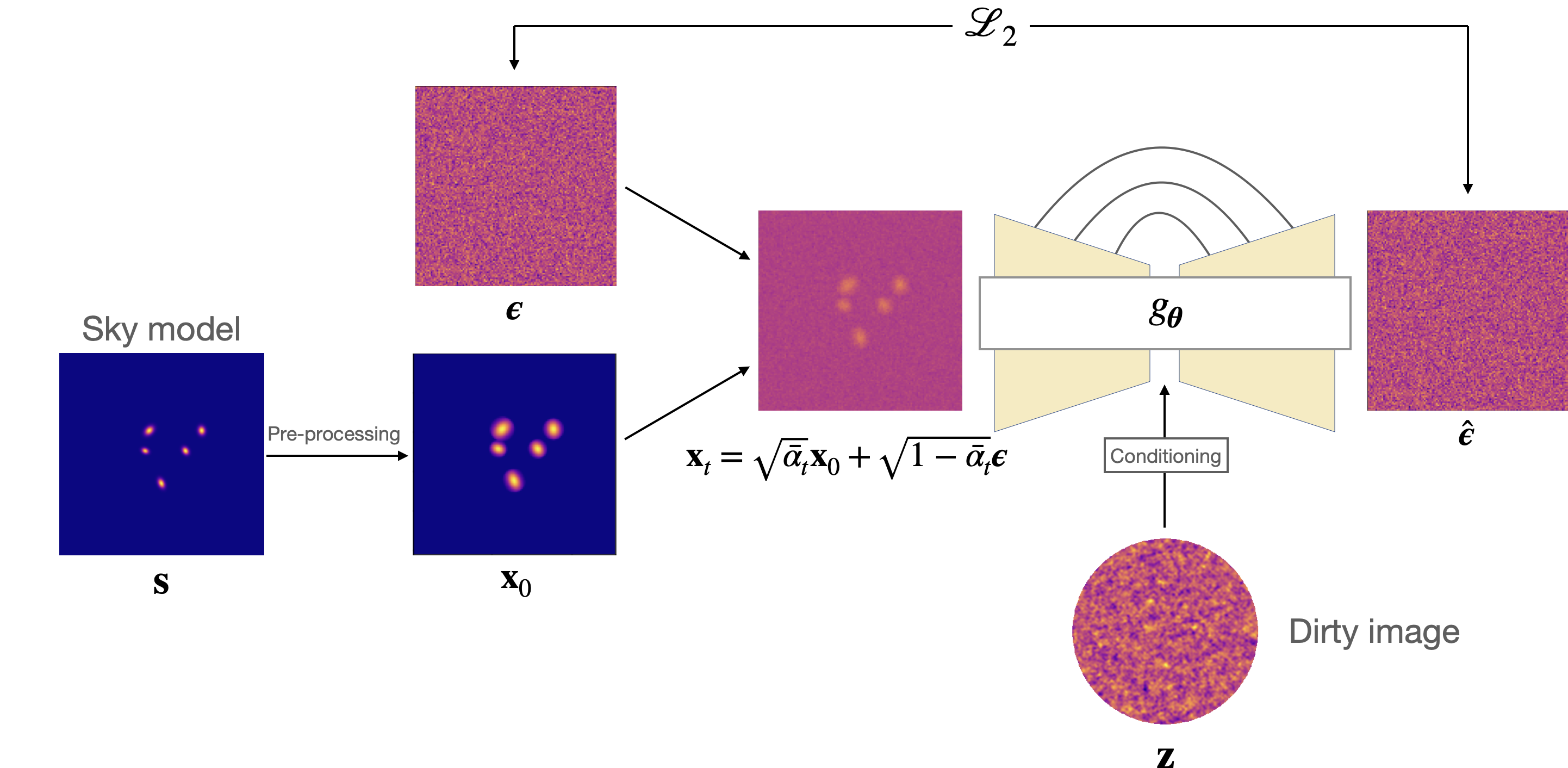}
  \caption{Schematic representation of  training pipeline for time step $t$. The preprocessed sky model $\xzero$ is corrupted by the DDPM model noise $\noise$. The U-Net, given the corrupted target image $\x_{t}$ and conditioned with the dirty image $\z,$ tries to estimate the model noise $\hat{\noise}$ to denoise the $\x_{t}$. The model is trained to minimize the distance between ${\noise}$ and $\hat{\noise}$.}
  \label{fig:training_pipeline}
\end{figure*}

We use the image-to-image DDPM Palette \citet{saharia2022palette}. This diffusion operates in an iterative manner, as visualized in Figure \ref{fig:iterative_degrading}. There are two directions: forward and inverse. During the forward direction stage, we add model noise $\noise$ to the data. This noise, an internal state parameter of the model, is distinct from the astronomical noise arising from antennas, the atmosphere, and so on. In DDPMs, this model noise is generally sampled from a zero-mean Gaussian distribution. During the inverse direction stage, the aim is to remove the added noise using a trained denoiser. This denoiser is a crucial component of the denoising diffusion process and, in this case, is represented by a U-Net \citep[]{unet} { with self-attention layers \citep[]{vaswani2017attention}}. The parameters of U-Net are trained to predict the model noise added to the target image. The iterative process of applying this U-Net multiple times for progressive denoising is the function of the DDPM.

At the forward process stage for each iteration $t \in \{1, T\},$ the added model noise has its own variance, which we denote ${\alpha}_1$ to ${\alpha}_T$, respectively, where each ${\alpha}_t \forall t$ is between 0 and 1, and each subsequent value is greater than the previous one  ${\alpha}_T > {\alpha}_{T-1} > ... > {\alpha}_1 $. Variances are chosen in such a way that, after $T$ steps, we have a meaningless image corresponding to the noise, essentially a sample from the Gaussian distribution. In this case, the reverse process can be initiated from the noise.

Mathematically, the forward process can be written as

\begin{equation}
        \x_t = \alpha_{t} \x_{t-1}+\sqrt{1-{\alpha}_t} \noise,
    \label{eq:ddpm_xt_from_xt-1}
\end{equation}
where $\x_t$ is an image produced by DDPM at the time step $t$, and $\noise$ is the added model state noise.

\cite{ho2020denoising} show that  $\x_t$  can be rewritten as
\begin{equation}
        \x_{t} = \sqrt{\bar{\alpha}_t} \x_{0}+\sqrt{1-\bar{\alpha}_t} \noise,
    \label{eq:ddpm_xt_from_x0}
\end{equation}
where $\xzero$ is a target data sample and $\bar{\alpha}_t := \prod_{s=1}^{t} \alpha_{s}$.

At the inverse process stage, we train the model to remove the added internal noise (model noise). After $T$ steps, we assume that the model will encode a target data manifold 
to itself, making it possible to generate data samples in $T$ steps from a randomly initiated model state $\noise$. In our implementation, each step is conditioned on $\z$ by concatenating it to the input. $\z$ represents the dirty image obtained from the aggregated UV grid ${\mathbf y,}$ as shown in Figure \ref{fig:inference_pipeline} (bottom branch). 

The training stage is visualized in Figure \ref{fig:training_pipeline}. {The subset of 5\,082 images is used. During training, we uniformly select time steps $t \in [0,1000]$, which define the noise variance for each instance. The target image (sky model) $\mathbf{x}_0$ is first normalized, and then we generate an intermediate $\mathbf{x}_t$ using Equation \ref{eq:ddpm_xt_from_x0} and the model noise $\noise$. This $\mathbf{x}_t$ is concatenated with the normalized dirty image $\mathbf{z}$, creating a two-channel tensor fed into the U-Net, which serves as our denoiser. The U-Net architecture, which is adapted from \cite{saharia2022palette}, has four stages of downsampling and four stages of upsampling. The time step is encoded using two linear layers with SiLU activation functions. The denoiser consists of residual blocks, which includes Group Normalization  (with 32 groups and an epsilon value of $10^{-5}$), and Convolutional Layers with kernel size 3. The model uses self-attention mechanisms. The U-Net accepts a two-channel input and produces a one-channel output. This model contains approximately 95M parameters. The primary training objective is to minimize the $\ell_2$ distance between the actual model noise $\noise$ and the noise estimated by the U-Net $\hat{\noise} = \DDPM(\mathbf{x}_t, \mathbf{z})$, as defined in Equation \ref{eq:ddpm_simple_loss}. }

\begin{equation}
    \begin{aligned}
    {
        \mathcal{L}^{DDPM}(\boldsymbol{\theta}) = \mathbb{E}_{t, {\z}, \xzero, \noise}\left[\left\|\noise-\DDPM\left(\sqrt{\bar{\alpha}_t} \xzero+\sqrt{1-\bar{\alpha}_t} \noise, \z, t\right)\right\|_2^2\right],
    }
    \end{aligned}
    \label{eq:ddpm_simple_loss}
\end{equation}
{where $\xzero$ is a target image (sky model), $\z$ is a conditioning (dirty image), $\noise$ is Gaussian zero-mean unit variance noise $\mathcal{N}({\mathbf 0},{\mathbf I})$, $\DDPM$ is conditional DDPM, and $\bar{\alpha}_t =\prod_{s=1}^{t} \alpha_{s} $ is the noise variance parameter, all at step $t,$ as in Equation \ref{eq:ddpm_xt_from_x0}. We train this  U-Net over 237 epochs with a batch size of 24, and we use the Adam optimizer with a learning rate of $10^{-4}$. We select the best model based on the loss on the validation set consisting of 2541 images. The training process was completed in approximately 16 hours on a NVIDIA RTX 3090 GPU.}

The diffusion model can be seen as a stochastic image-to-image translator whose goal consists in transforming dirty images $\z$ into a sky model $\xrecon$. Therefore, dirty images $\z$ can also be referred to as the input $\z$ to the DDPM alongside the model noise $\noise$.

 At the inference step, we begin with the Gaussian noise (noise of the model $\noise$) and the dirty image $\z$. This stage is not merely a consecutive application of the denoiser to $\noise$ but rather a sophisticated iterative process that progressively refines the estimate of the original sky model, $\xrecon$, starting with  $\noise$ through $\x_t$. Each step of the DDPM involves predicting and then subtracting the added noise from the current image  $\x_t$ each time concatenated with dirty image conditioning $\z$. After removing the predicted noise from $\x_t$ we get an estimate of $\mathbf{x}_0$. Then, a new, smaller-variance Gaussian noise is introduced to get $\x_{t-1}$, which is then estimated and removed in the subsequent step. Due to computational limitations, this process is repeated for a total of 250 steps instead of 1000; it gradually reduces the noise and converges toward an accurate reconstruction of $\mathbf{x}_0$.
Subsequently, with Equations \ref{eq:inverse_scaling} and \ref{eq:inverse_normalization}, $\xrecon$ is mapped back to the original range of values  $\hat{\image}$. Finally, we apply the extraction algorithm to get the properties of the sources from the predicted sky model. The process is shown schematically in Figure \ref{fig:inference_pipeline} (top branch).

\subsection{Source localization}

\begin{figure}[ht]
\centering
\begin{subfigure}{0.12\textwidth}
\centering
\includegraphics[trim = 65bp 110bp 125bp 80bp, clip, width=\textwidth]{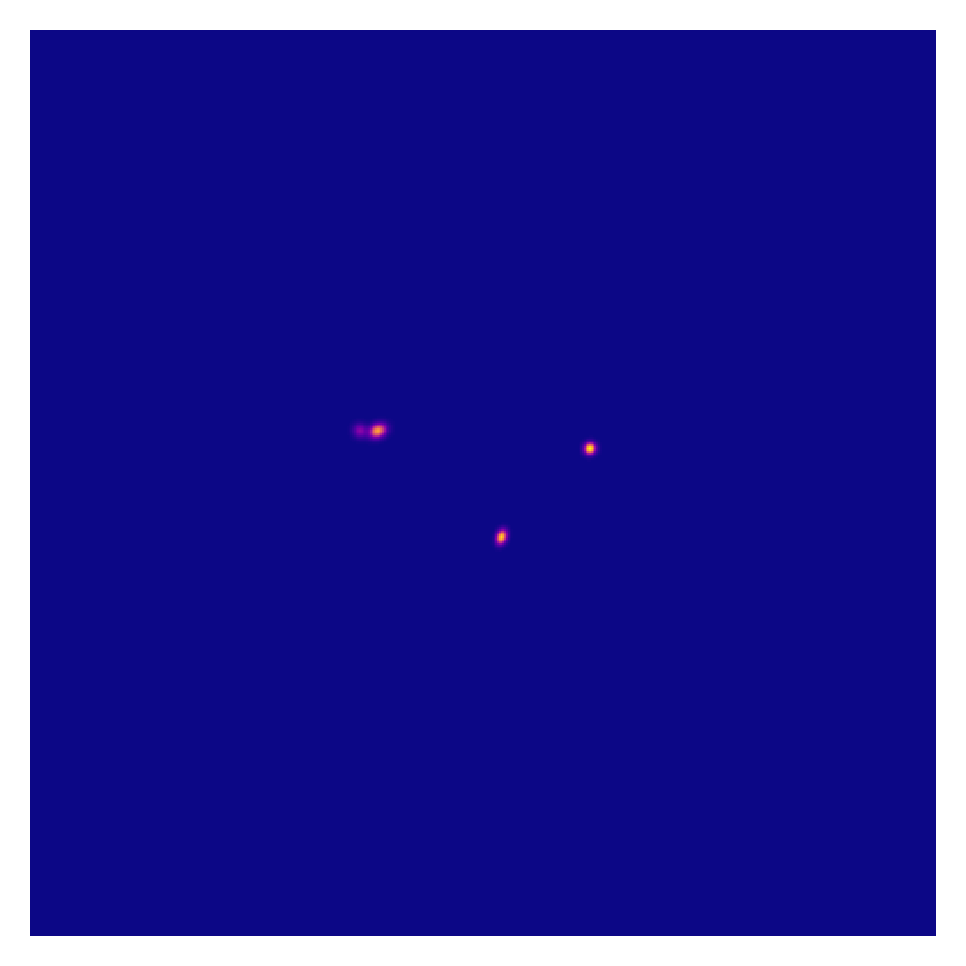}
\caption*{(a)}
\end{subfigure}%
\hspace{1em} 
\begin{subfigure}{0.12\textwidth}
\centering
\includegraphics[trim = 65bp 110bp 125bp 80bp, clip, width=\textwidth]{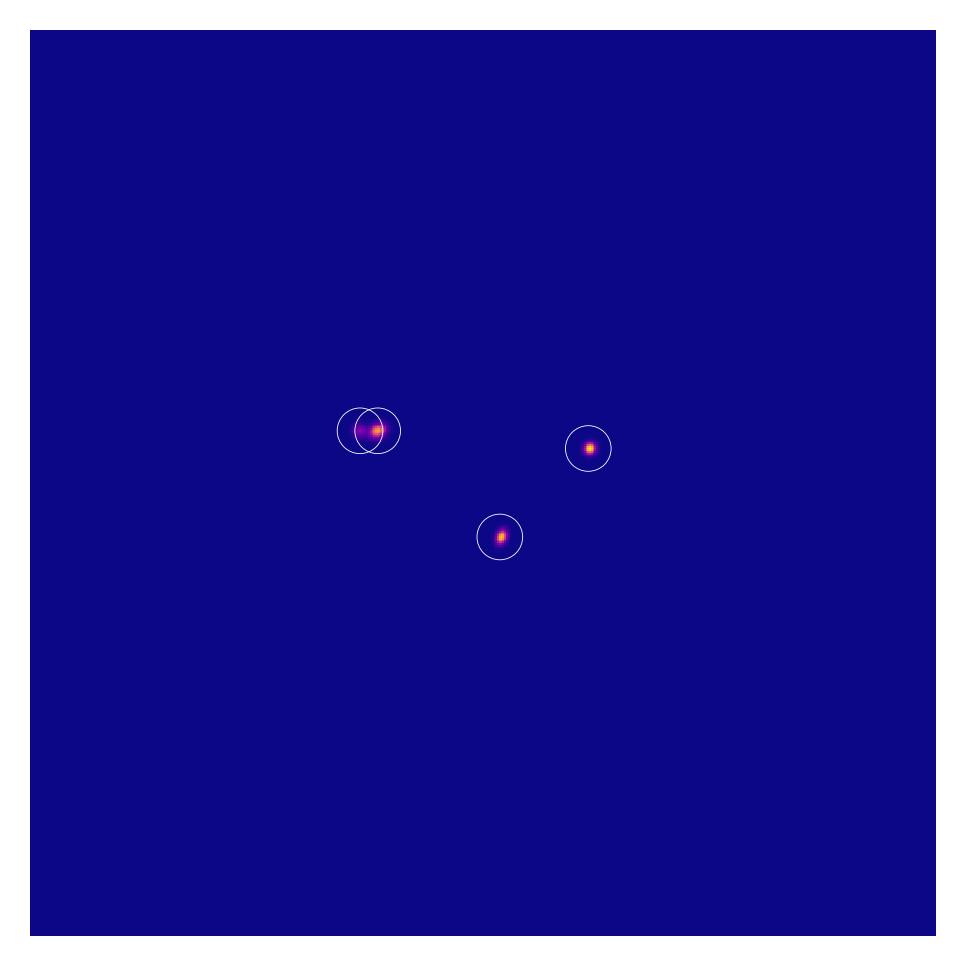}
\caption*{(b)}
\end{subfigure}%
\hspace{1em} 
\begin{subfigure}{0.12\textwidth}
\centering
\includegraphics[trim = 65bp 110bp 125bp 80bp, clip, width=\textwidth]{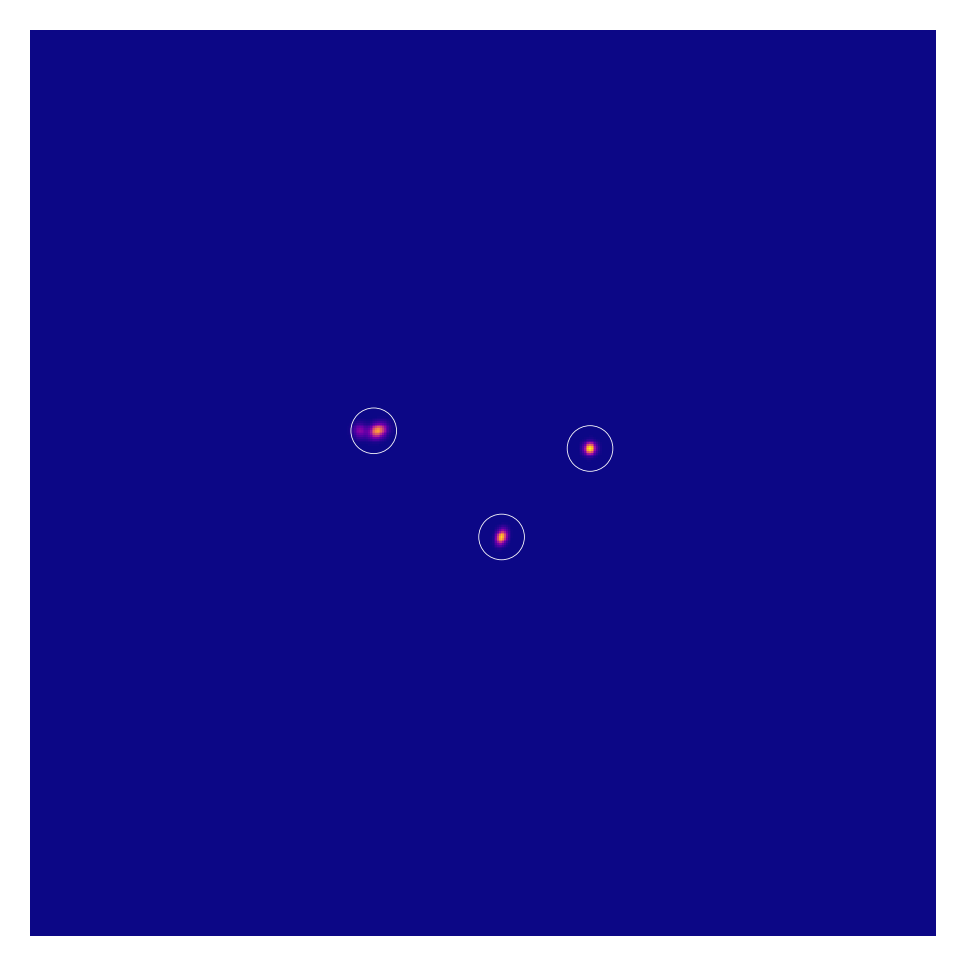}
\caption*{(c)}
\end{subfigure}

\vspace{1em} 

\begin{subfigure}{0.12\textwidth}
\centering
\includegraphics[trim = 125bp 117bp 65bp 73bp, clip, width=\textwidth]{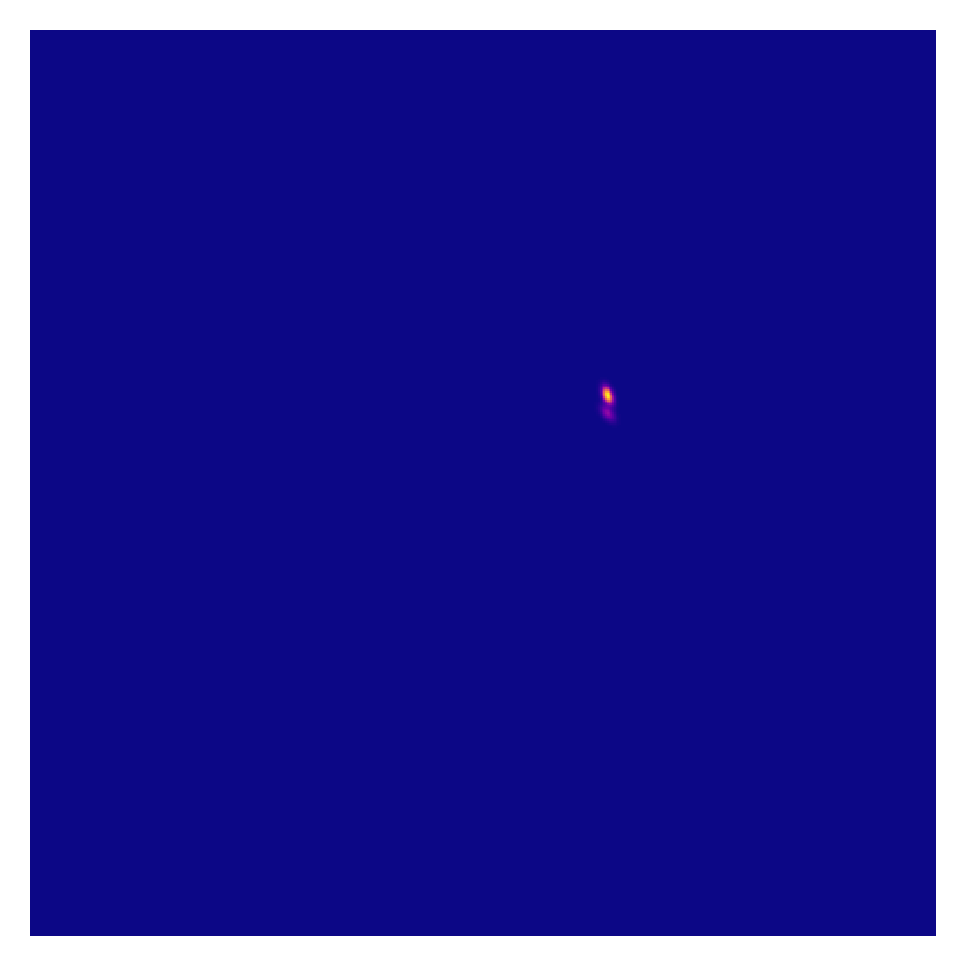}
\caption*{(d)}
\end{subfigure}%
\hspace{1em} 
\begin{subfigure}{0.12\textwidth}
\centering
\includegraphics[trim = 125bp 117bp 65bp 73bp, clip, width=\textwidth]{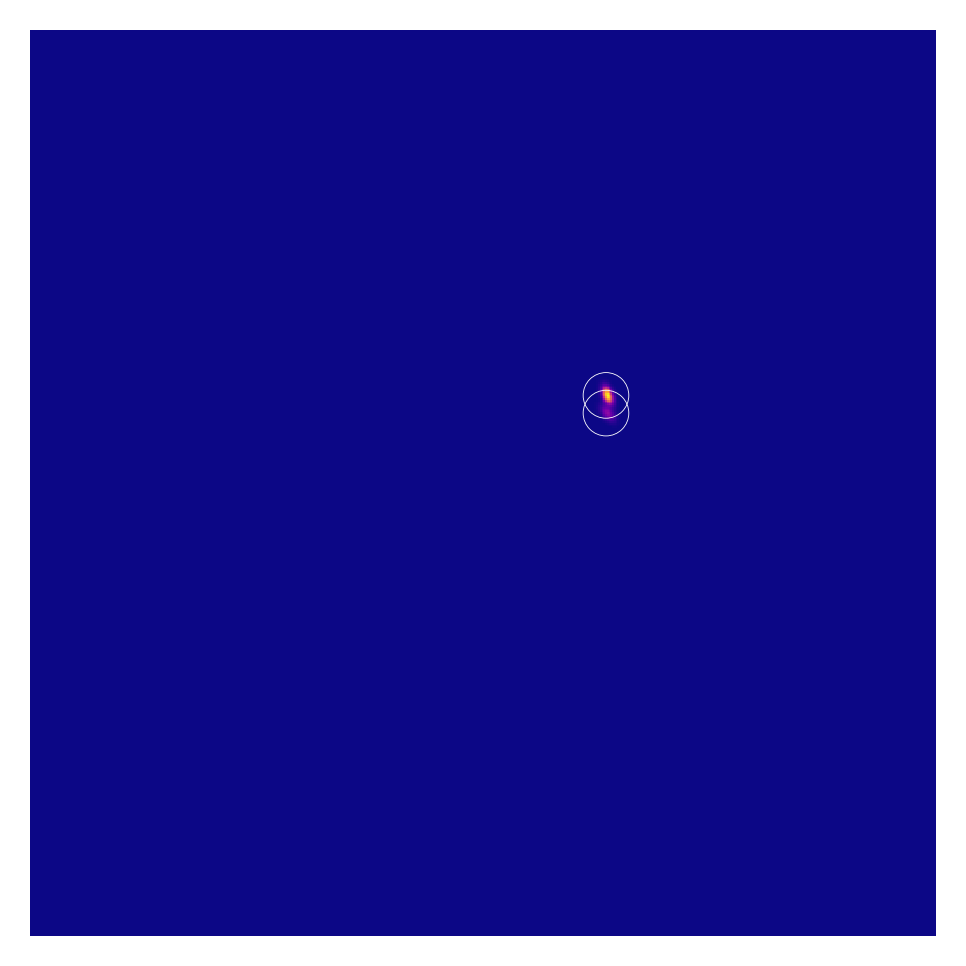}
\caption*{(e)}
\end{subfigure}%
\hspace{1em} 
\begin{subfigure}{0.12\textwidth}
\centering
\includegraphics[trim = 125bp 117bp 65bp 73bp, clip, width=\textwidth]{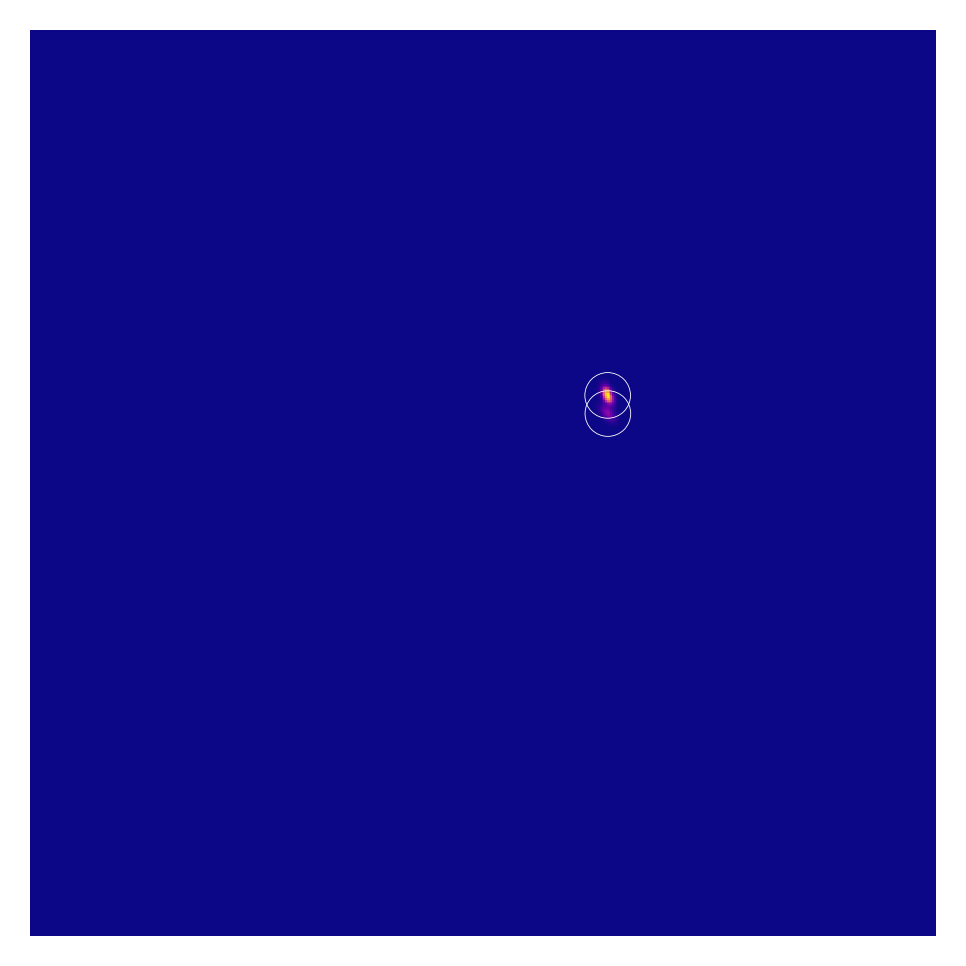}
\caption*{(f)}
\end{subfigure}

\caption{Photutils algorithm behavior on sky models for closely situated sources.Behavior of the Photutils localization algorithm on sky models. Panels (a) and (d) show the sky model; (b) and (e) depict images annotated with the corresponding catalog sources; and (c) and (f) present images annotated with sources detected using the chosen photutils algorithm. As demonstrated in these images, the Photutils algorithm occasionally fails to deblend closely situated sources. This happens for less than 1\% of the sources present in the dataset.
}
\label{fig:failing_photutils}
\end{figure}

The locations, shapes, and fluxes of the sources can be restored from the predicted sky model. While PyBDSF is commonly used for this task, it can only be applied to images containing background noise. As the target sky models do not contain any noise—just point sources—we find this extra step of adding noise to execute PyBDSF destructive. This is especially true considering the relatively straightforward nature of the sky models, which consist solely of Gaussian sources with zero background. These challenges prompted us to explore an alternative method of source localization.

We use an algorithm based on watershedding. We take an implementation of this algorithm from the Photutils library by \cite{photutils} to extract the properties of point sources from predicted sky models. To validate the usage of this algorithm for source localization, we evaluate its performance on all sky models of the simulated dataset. The achieved scores are close to perfect, as shown in Section \ref{sec: results}. The algorithm fails when sources are very close to each other (see Figure \ref{fig:failing_photutils}). Given the small amount of data for which it fails, we use it as a final post-processing step of the pipeline.

The localization algorithm can be deployed directly on a single output from the diffusion model; however, we can also use the stochasticity and run the model with different realizations of model state noise to get multiple sky model predictions. This can potentially enhance the model's performance, but a challenge arises in aggregating these multiple predictions. In this study, we investigate two aggregation techniques to derive a final image from multiple predictions: the mean and the median. These methods are applied at the pixel level, where we assess the value of each pixel across all predictions and compute either its mean or median. This process is repeated for every pixel, resulting in the final output image. 

\subsubsection{Aggregation techniques}

As mentioned above, DDPM can output multiple predictions for one phase of conditioning. We investigate two ways to aggregate the produced predictions via \aggregatedetect  ~and \detectaggregate ~strategies.  The \aggregatedetect ~technique means that first we aggregate the images and then we localize the sources. The \detectaggregate ~method means that we first localize sources from each prediction and then we aggregate the sources into the final table.

For the \aggregatedetect ~strategy, we study the mean and median as an aggregation:

\begin{equation}
\text{{Mean}}(h, w) = \frac{1}{\nbruns} \sum_{k=1}^{\nbruns} {\xrecon}^{k}(h, w),
\end{equation}

\begin{equation}
\text{{Median}}(h, w) = \text{{median}}\left({\xrecon}^{1}(h, w), {\xrecon}^{2}(h, w), \ldots, {\xrecon}^{\nbruns}(h, w)\right),
\end{equation}
where $\nbruns$ is the total number of predicted images for a given $\z$  (20 in this case), ${\xrecon}^{k}$ represents the $k$-th predicted image, and $(h, w)$ represent the pixel coordinates. 

The \detectaggregate ~method consists in aggregating the predicted sources from multiple runs with the same conditioning. In this scenario, the source-extraction algorithm is executed on each image independently. Subsequently, the predictions are merged together; two predictions are considered to concern the same source if their coordinates (right ascension (RA) and Declination (Dec)) are within a radius of $r = 5 \times 10^{-5}$ degrees. In the final aggregated output, each source is listed with its average coordinates and characteristics (such as flux), along with the corresponding standard deviation computed across the multiple predictions. 
  
Additionally, we calculate the number of images in which each source is detected, a value we term the ``reliability'' of the source. For instance, if a source is detected in all 20 images out of 20, it is assigned a reliability of 100\%; but if it is only detected in 10 images, the associated reliability is 50\%. A reliability threshold can be set to determine which sources should be reported. Intuitively, ambiguous sources will not appear in all predictions, because they can also be classified as background noise. This should be particularly true for very faint sources.

We show in Section \ref{sec: results} that as soon as the reliability threshold grows, completeness falls, which means that the model starts to miss an increasing number of sources (while purity grows). 
Reliability is expected to be a proxy for the sensitivity of the proposed algorithm to faint sources.

\subsection{Evaluation metrics}

The task of localization in the described pipeline is divided into two stages: sky model reconstruction from dirty images and source localization. We evaluate the proposed method on both.

\subsubsection{Reconstruction metrics}

To evaluate how well the trained DDPM predicts the sky model from dirty image $\z,$ we use metrics from natural image analysis such as peak S/N, SSIM, and MSE, which are described below.

{The mean squared error (MSE)} quantifies the average squared difference between the predicted and original pixel values:

\begin{equation}
\text{MSE} = \frac{1}{H \times W} \sum_{h=1}^{H} \sum_{w=1}^{W} \biggl(\xrecon(h,w) - \xzero(h,w)\biggr)^2,
\end{equation}
where $H \times W$ represents the image size, $\xzero$ corresponds to the target image, and $\xrecon$ represents the predicted image. A lower MSE indicates better performance.

{The peak signal-to-noise ratio (peak S/N)} measures the quality of the predicted image by comparing it to the target image; it is a calculation of the ratio of the maximum possible pixel value MAX to MSE between the predicted and original images:

\begin{equation}
\text{peak S/N} = 20 \cdot \log_{10}(\text{MAX}) - 10 \cdot \log_{10}(\text{MSE}).
\end{equation}

A higher peak S/N indicates better performance.

{The structural similarity index (SSIM)} introduced by \cite{ssim} assesses the similarity between the predicted $\xrecon$ and target image $\xzero$ based on their structural information:

\begin{equation}
{\mathrm{SSIM}}(\xzero, \xrecon) = \frac{(2 \mu_{\xzero} \mu_{\xrecon} + c_1)(2\sigma_{\xzero\xrecon} + c_2)}{(\mu_{\xzero}^2 + \mu_{\xrecon}^2 + c_1)(\sigma_{\xzero}^2 + \sigma_{\xrecon}^2 + c_2)},
\end{equation}
where $\mu_{\xzero}$ is the global mean value of $\xzero$ of  size $H \times W$; $\mu_{\xrecon}$ is the global mean value of $\xrecon$ of  size $H \times W$; $\sigma_{\xzero}$ is the global variance of $\xzero$;
$\sigma_{\xrecon}$ is the global variance of $\xrecon$; $\sigma_{\xzero\xrecon}$ is the covariance between $\xzero$ and $\xrecon$; and $c_1 = 10^{-4}$ and $c_2 = 9\times10^{-4}$ are two constants chosen to stabilize the division depending on the dynamic range. The SSIM value ranges between 0 and 1, where a higher value indicates better performance. The SSIM implementation used here is used from the skimage package by \cite{skimage}.

\subsubsection{Localization metrics}

To evaluate the performance of the source localization task, we use purity and completeness metrics following the same protocol as in \citet[]{Taran_2023}. These metrics are essentially the same as precision and recall but we define true positives (TPs), false positives (FPs), and false negatives (FNs) differently. The predicted source position is considered a TP if its (RA, Dec) coordinates are within a radius of $r = 5 \times 10^{-5}$ degrees from the true source position, which is known from the simulation. A predicted source is a FP if there is no true source within the same radius $r$. A FN is observed if  there are no predicted sources within the radius around a true source.

Purity, or precision, measures the accuracy of the predicted sources by computing the ratio of the number of detected true sources (TP) to the total number of predicted sources. It can be calculated using the formula:

\begin{equation}
\text{Purity} = \frac{TP}{TP + FP}.
\end{equation}

Completeness, or recall, quantifies the completeness of the predictions by calculating the ratio of the number of detected true sources (TP) to the total number of true sources present in the data:

\begin{equation}
\text{Completeness} = \frac{TP}{TP + FN}.
\end{equation}

In addition to purity and completeness, we also compute the F1-score, which is their harmonic mean:

\begin{equation}
\text{F1-score} = 2 \cdot \frac{\text{Purity} \cdot \text{Completeness}}{\text{Purity} + \text{Completeness}}.
\end{equation}

The F1-score ranges between 0 and 1, where a higher value indicates better performance in terms of both purity and completeness.

\subsubsection{Source characterization metrics}

In the context of source characterization tasks, the goal is to predict the flux values, which represent the integral brightness of the radio source. The accuracy of the flux estimations is measured by calculating the proportion of points for which the flux has been accurately predicted. We refer to a flux prediction as accurate if the predicted value varies from the actual value (established through the simulations) by a quantity equal to or less than the noise amplitude. In the experiments, the noise mean square is set to 50 mJy. Consequently, we introduce a metric known as the "fraction", which is defined as the ratio of the number of points where the flux was accurately predicted to the total number of sources in the simulation.


\section{Results}
\label{sec: results}
\subsection{Image reconstruction}

The first stage of the proposed pipeline is a reconstruction of a sky model from a dirty image.

\subsubsection{Diffusion predictions}
\begin{figure*}[ht]
\centering
\begin{subfigure}{\outputswidth}
\centering
\caption*{Dirty image}
\includegraphics[width=\textwidth]{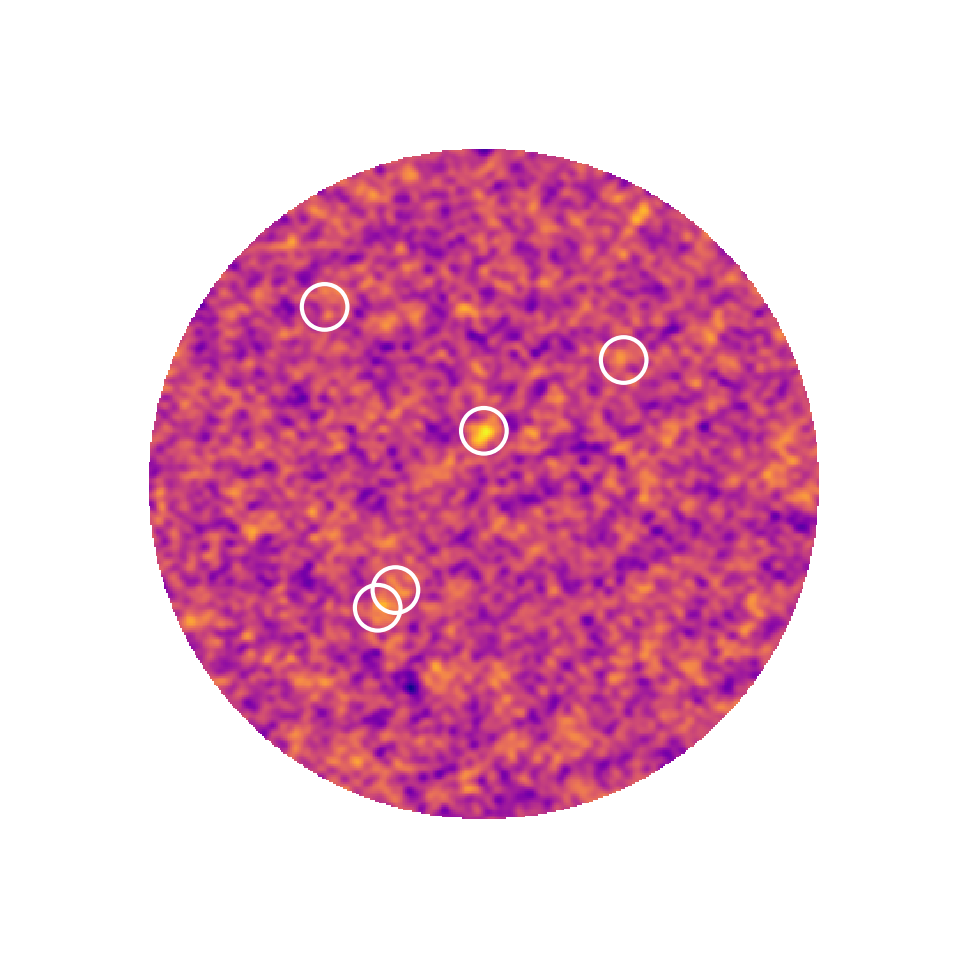}
\end{subfigure}%
\hspace{0.1em} 
\begin{subfigure}{\outputswidth}
\centering
\caption*{Sky model}
\includegraphics[width=\textwidth]{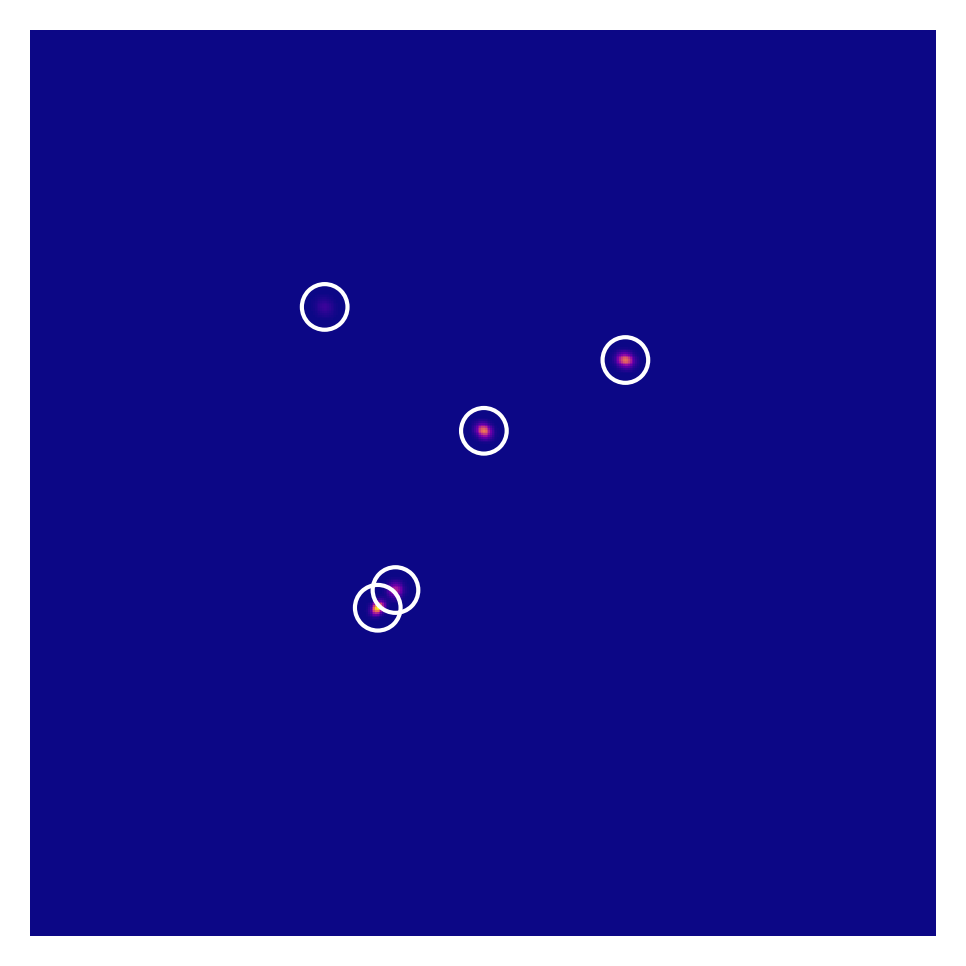}
\end{subfigure}%
\hspace{0.1em}
\begin{subfigure}{\outputswidth}
\centering
\caption*{Median}
\includegraphics[width=\textwidth]{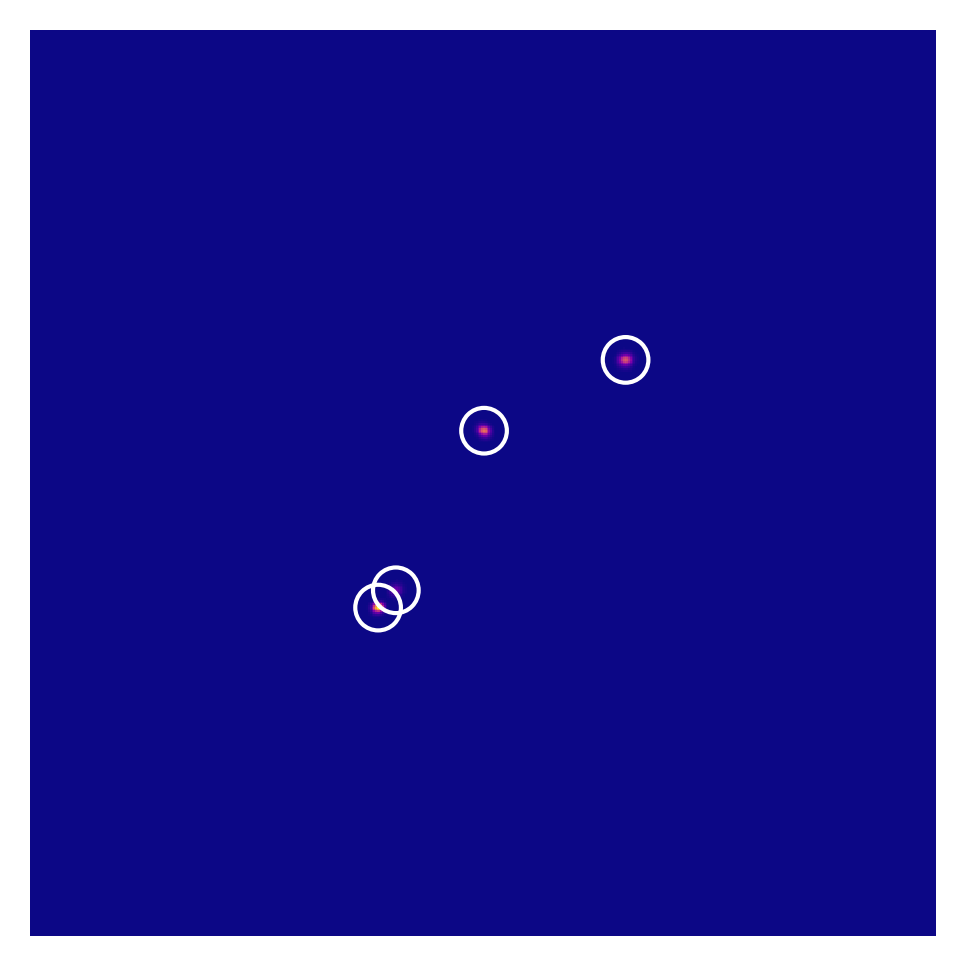}
\end{subfigure}%
\hspace{0.1em} 
\begin{subfigure}{\outputswidth}
\centering
\caption*{Mean}
\includegraphics[width=\textwidth]{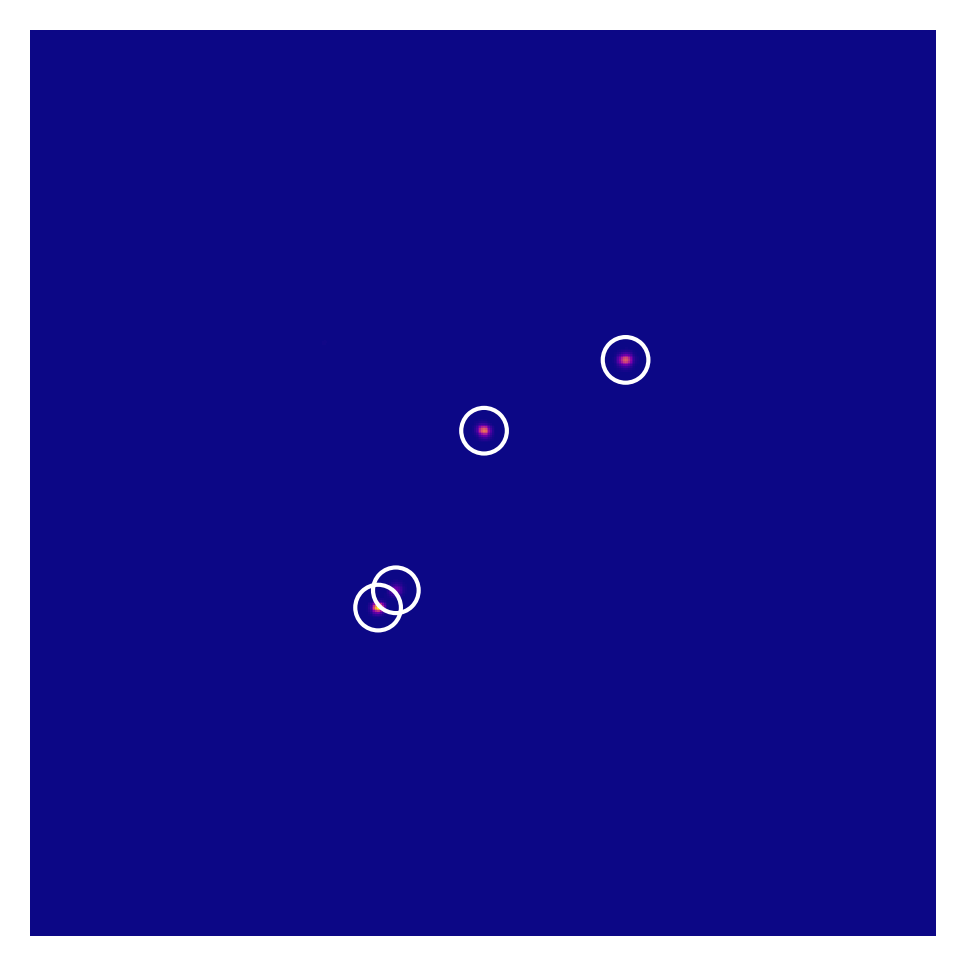}
\end{subfigure}%
\hspace{0.1em} 
\begin{subfigure}{\outputswidth}
\centering
\caption*{Std}
\includegraphics[width=\textwidth]{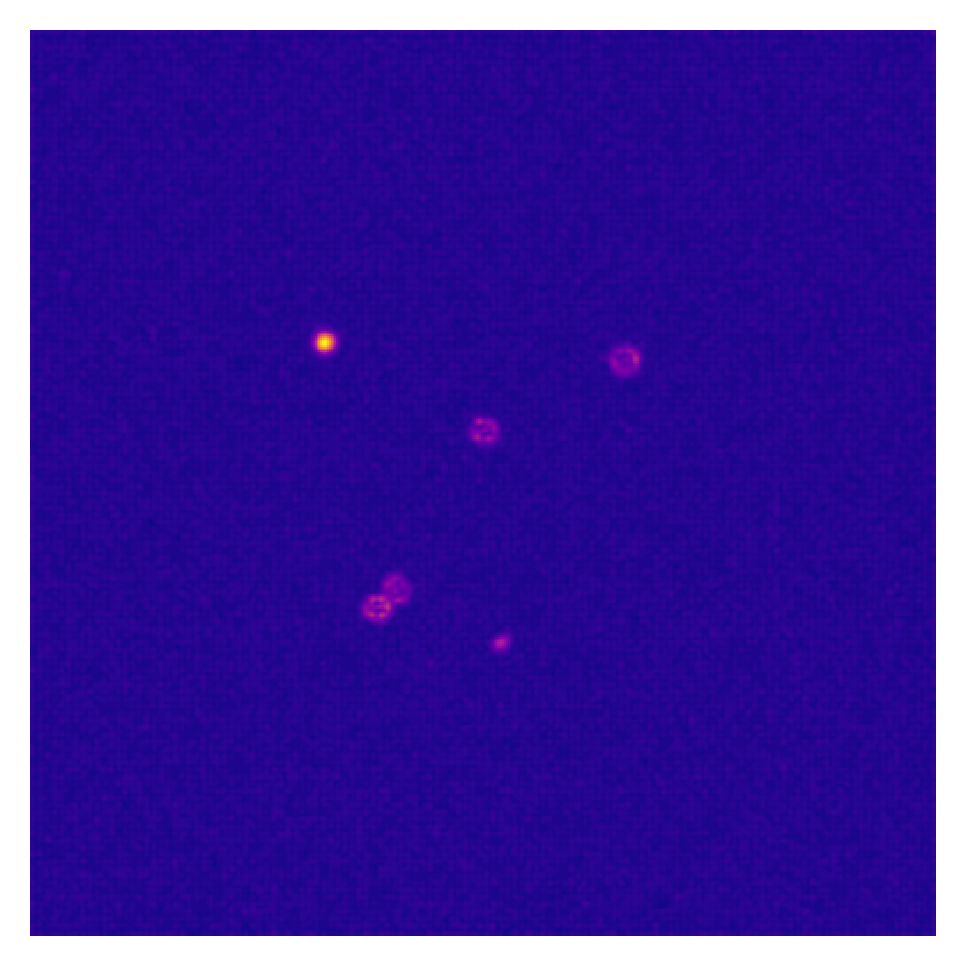}
\end{subfigure}
\par\medskip
\hrule
\par\medskip
\centering Outputs

\par\medskip

\begin{subfigure}{\outputswidth}
\centering
\includegraphics[width=\textwidth]{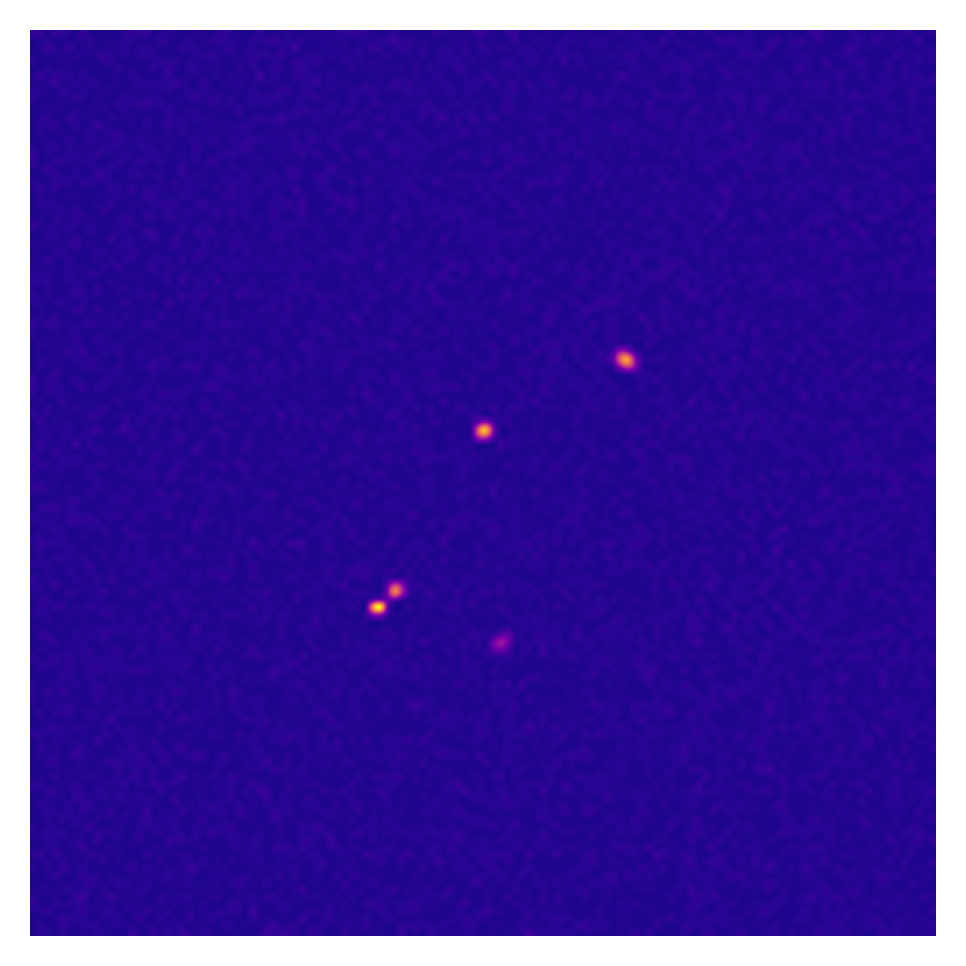}
\end{subfigure}%
\hspace{0.1em}
\begin{subfigure}{\outputswidth}
\centering
\includegraphics[width=\textwidth]{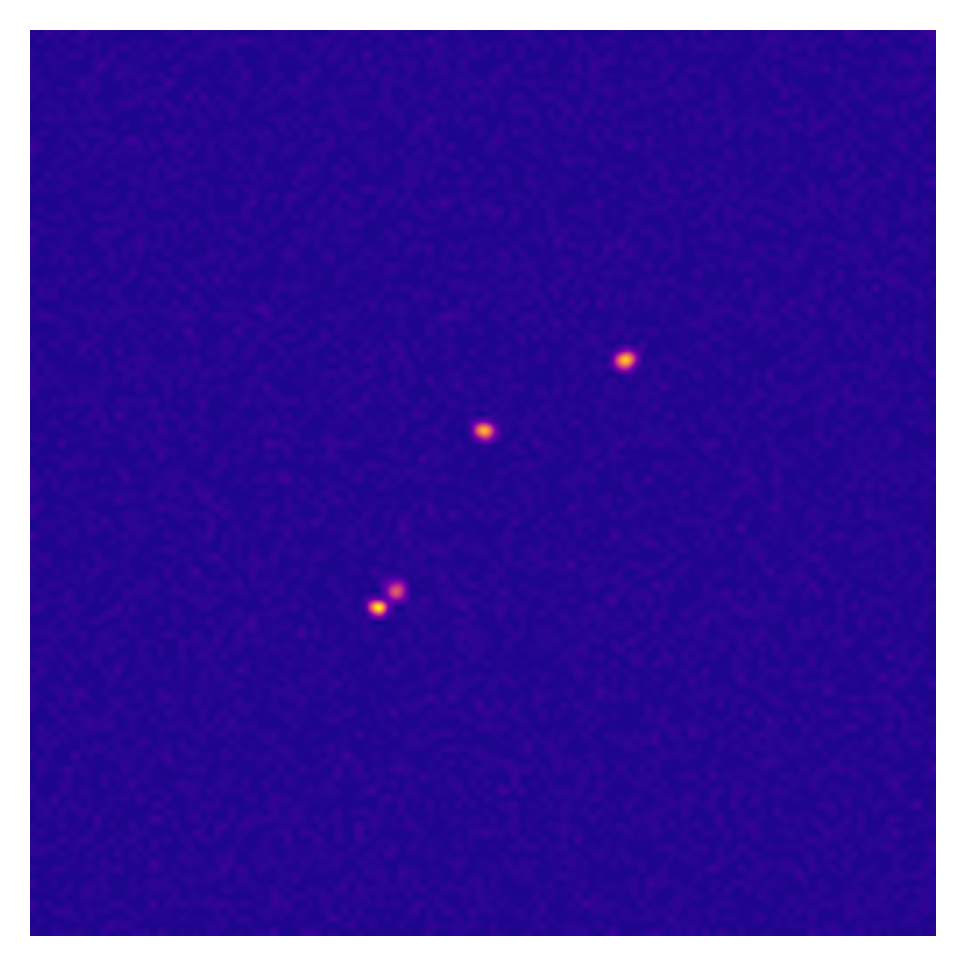}
\end{subfigure}%
\hspace{0.1em} 
\begin{subfigure}{\outputswidth}
\centering
\includegraphics[width=\textwidth]{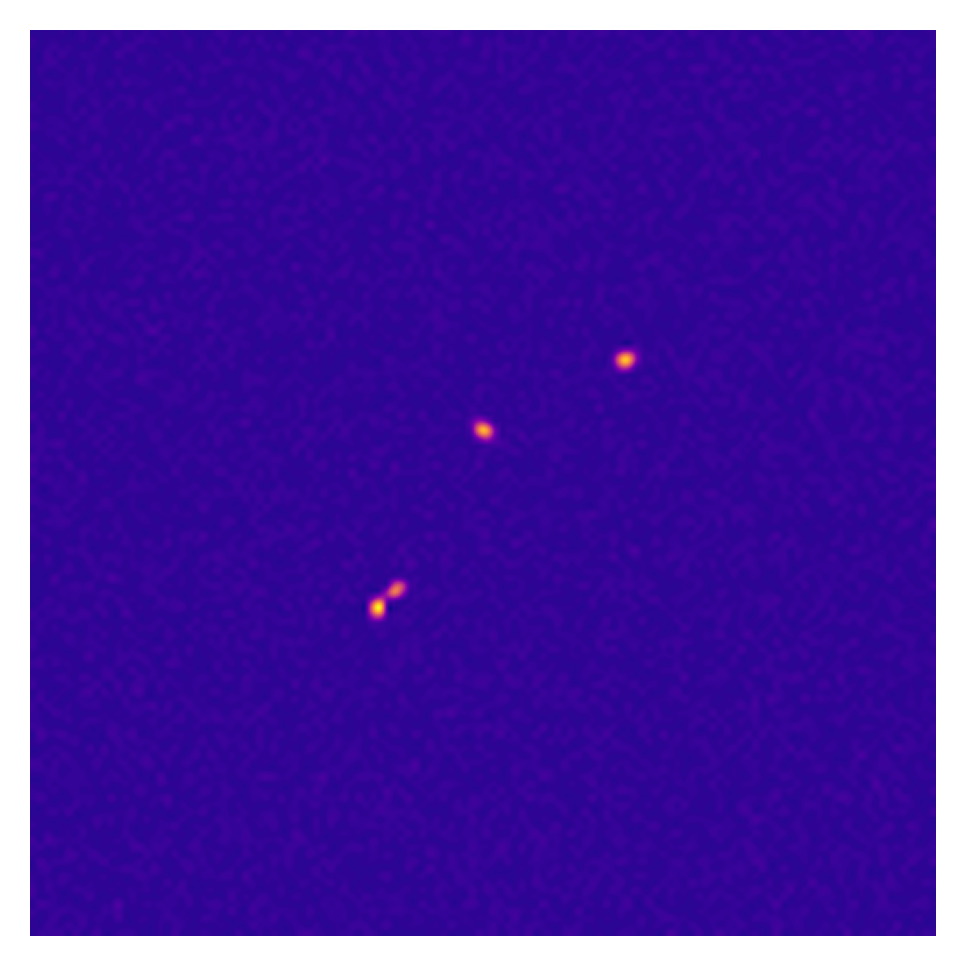}
\end{subfigure}%
\hspace{0.1em} 
\begin{subfigure}{\outputswidth}
\centering
\includegraphics[width=\textwidth]{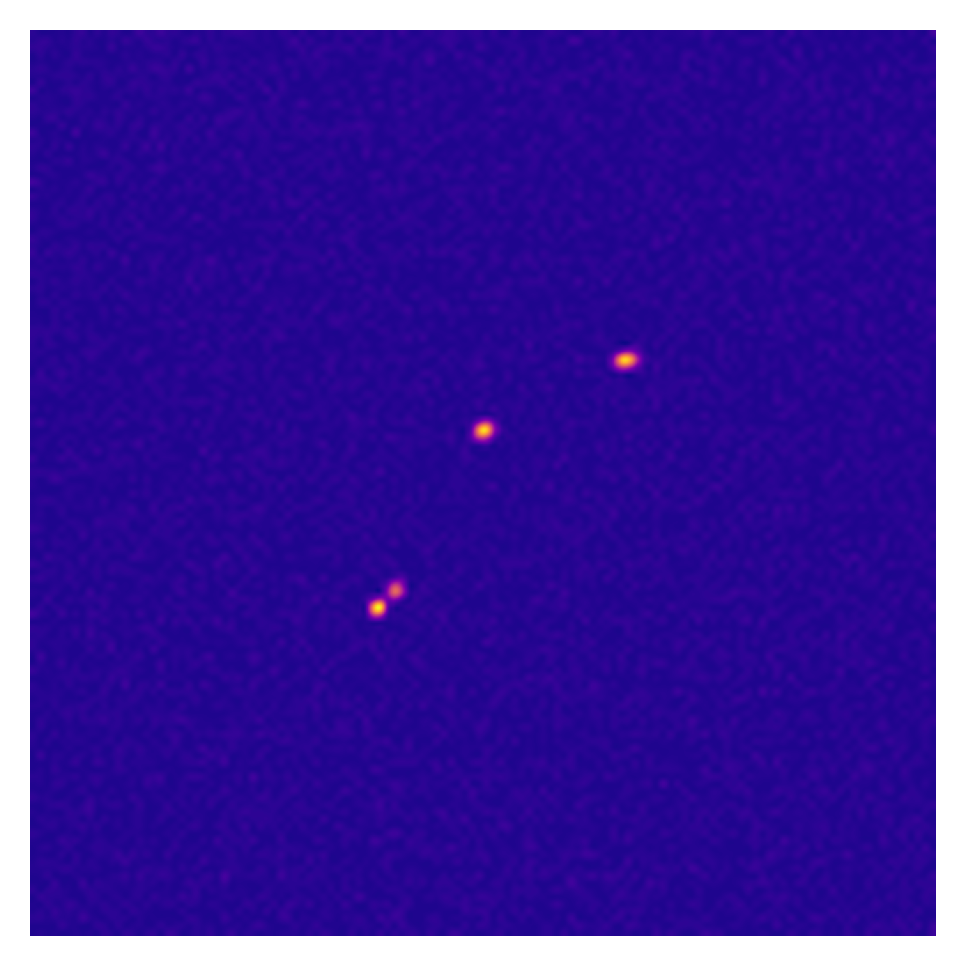}
\end{subfigure}%
\hspace{0.1em} 
\begin{subfigure}{\outputswidth}
\centering
\includegraphics[width=\textwidth]{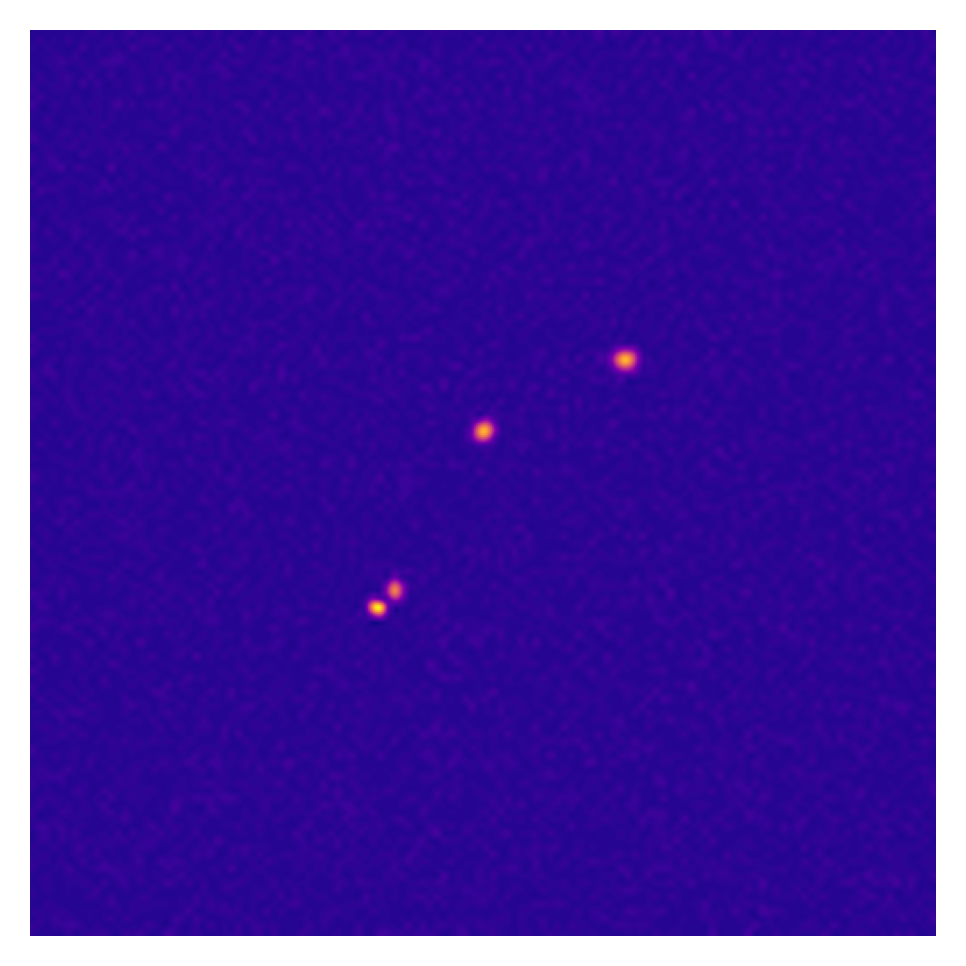}
\end{subfigure}

\vspace{0.1em}

\begin{subfigure}{\outputswidth}
\centering
\includegraphics[width=\textwidth]{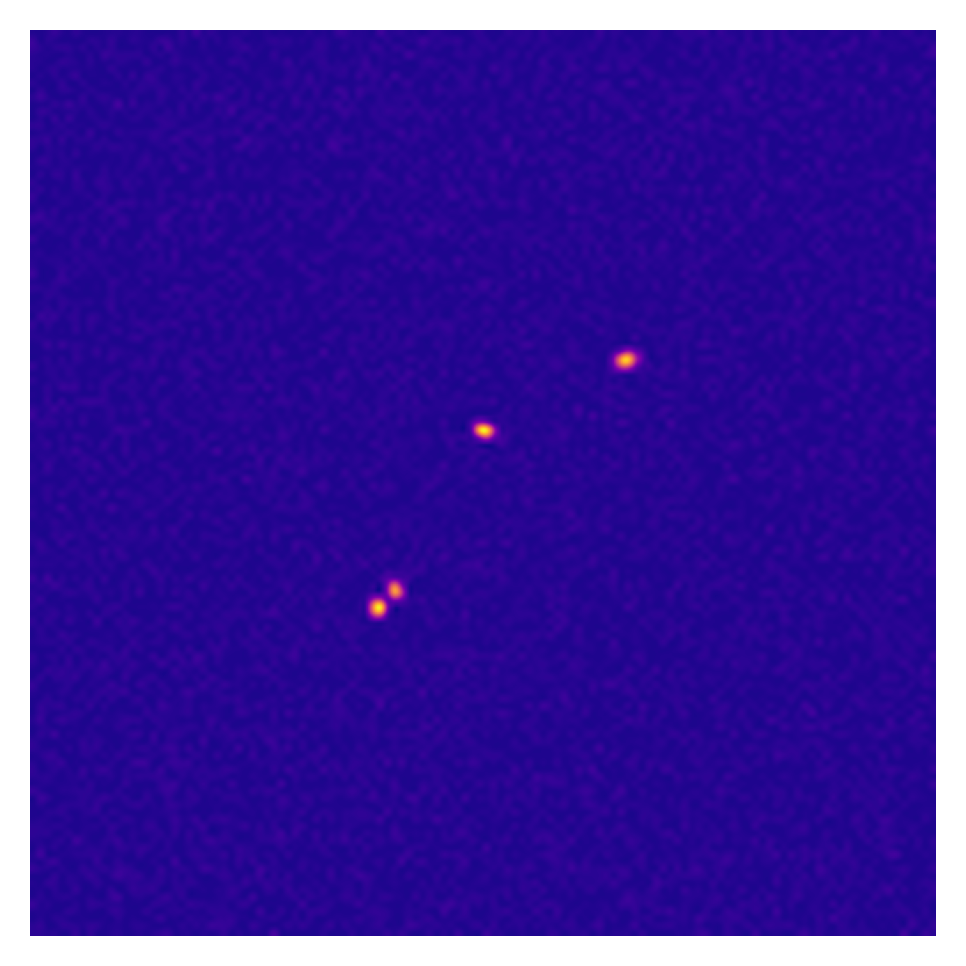}
\end{subfigure}%
\hspace{0.1em} 
\begin{subfigure}{\outputswidth}
\centering
\includegraphics[width=\textwidth]{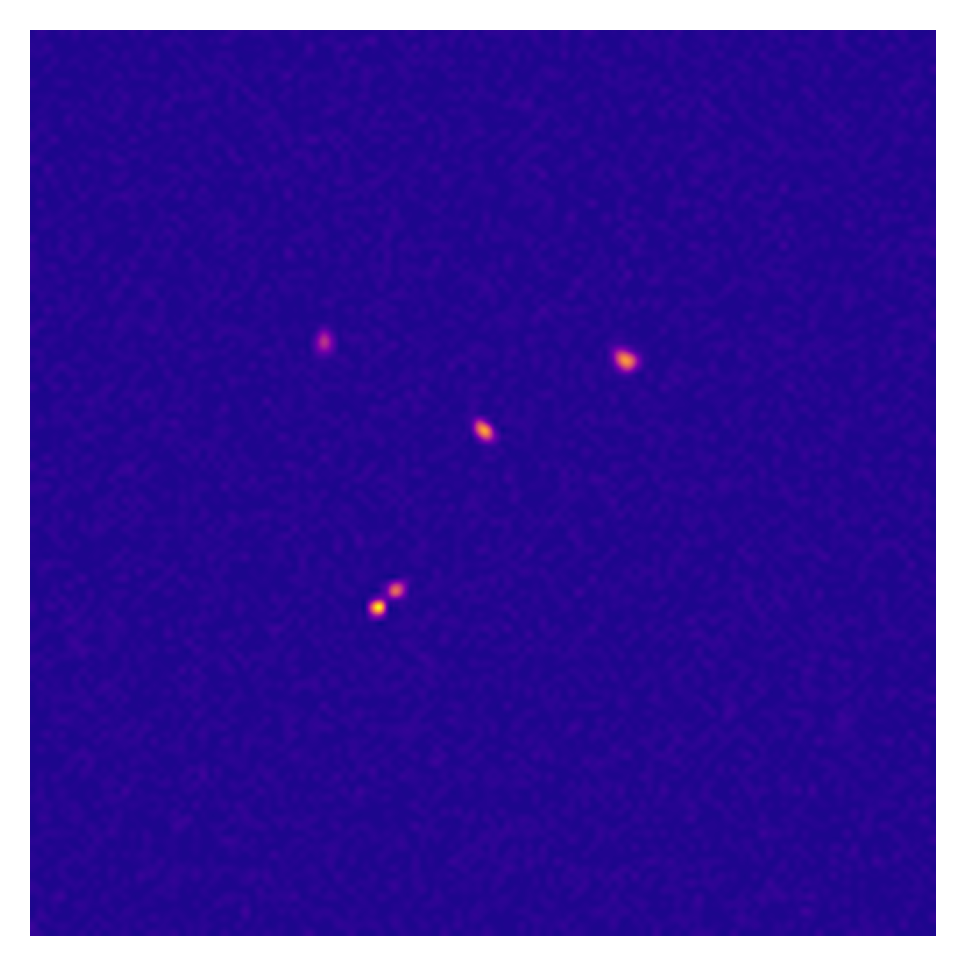}
\end{subfigure}%
\hspace{0.1em} 
\begin{subfigure}{\outputswidth}
\centering
\includegraphics[width=\textwidth]{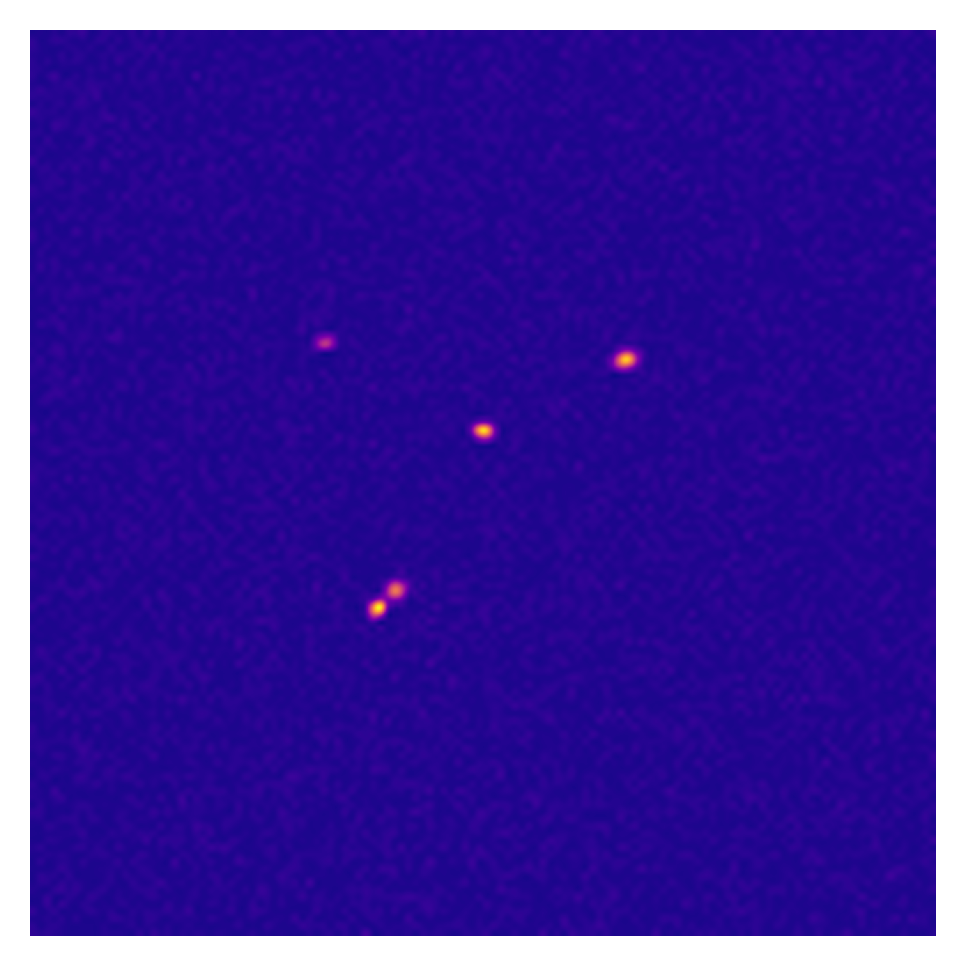}
\end{subfigure}%
\hspace{0.1em} 
\begin{subfigure}{\outputswidth}
\centering
\includegraphics[width=\textwidth]{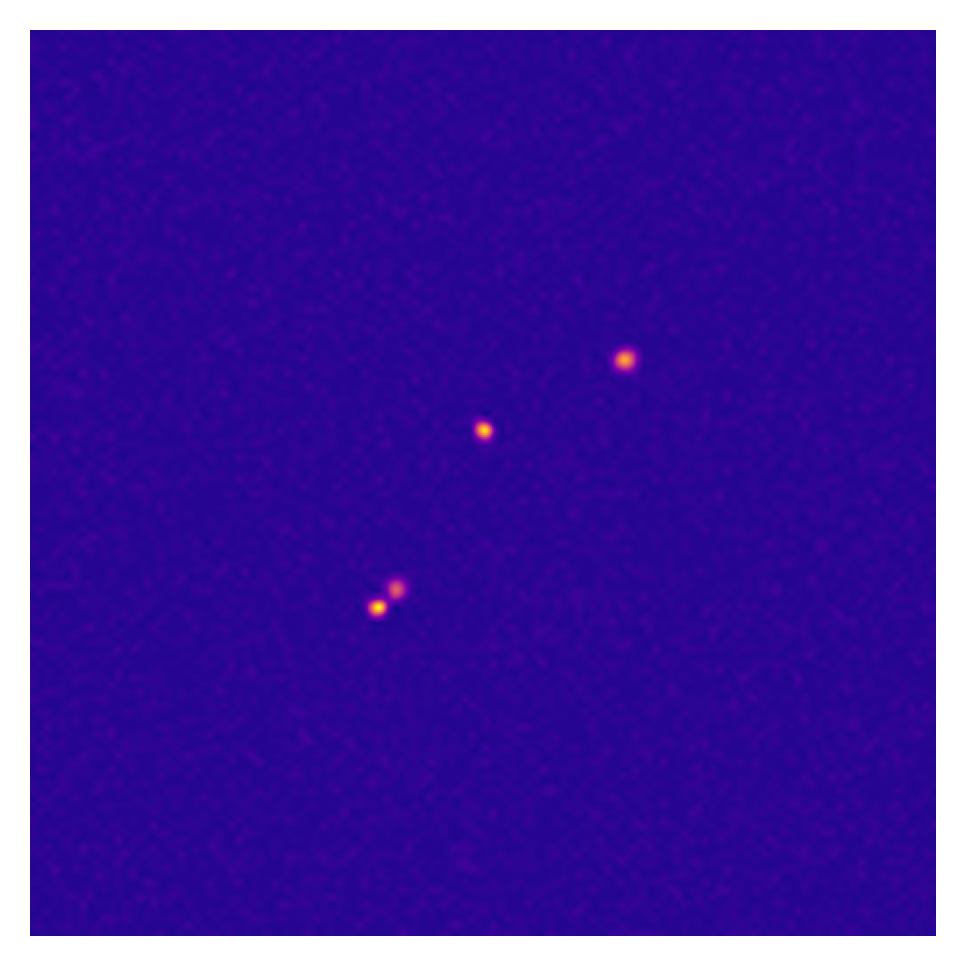}
\end{subfigure}%
\hspace{0.1em} 
\begin{subfigure}{\outputswidth}
\centering
\includegraphics[width=\textwidth]{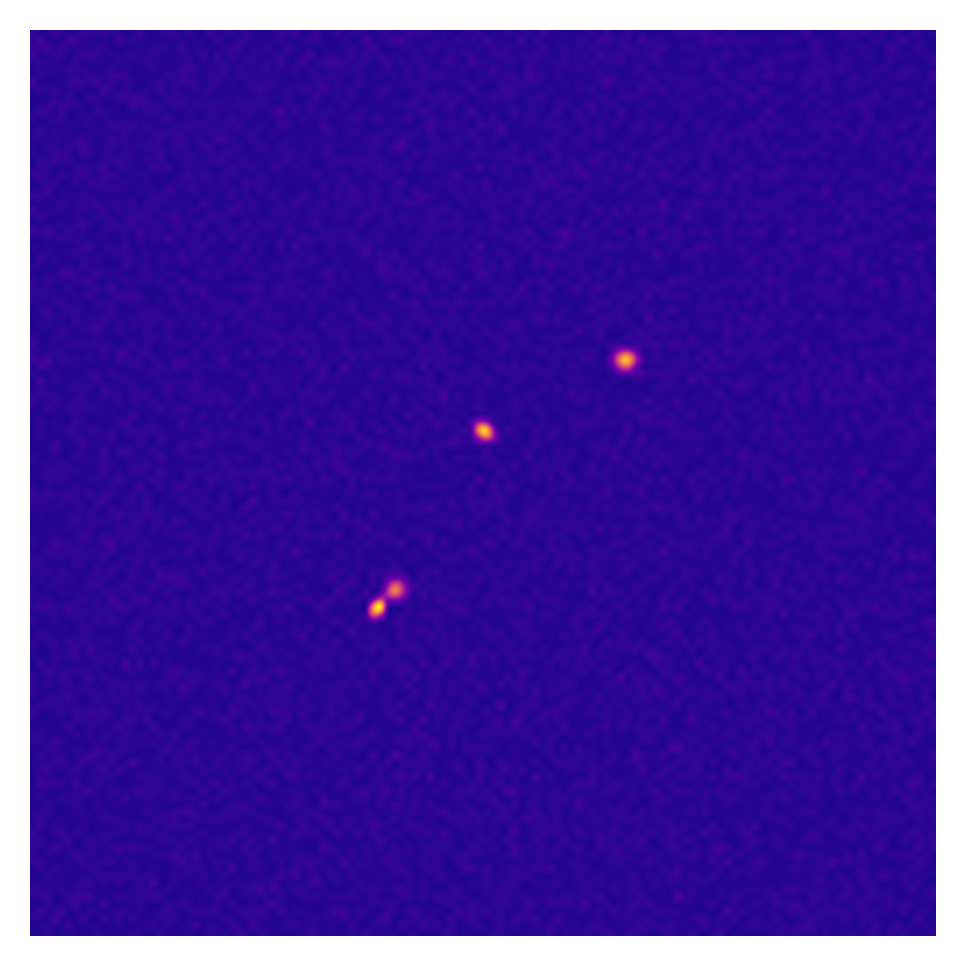}
\end{subfigure}

\vspace{0.1em}

\begin{subfigure}{\outputswidth}
\centering
\includegraphics[width=\textwidth]{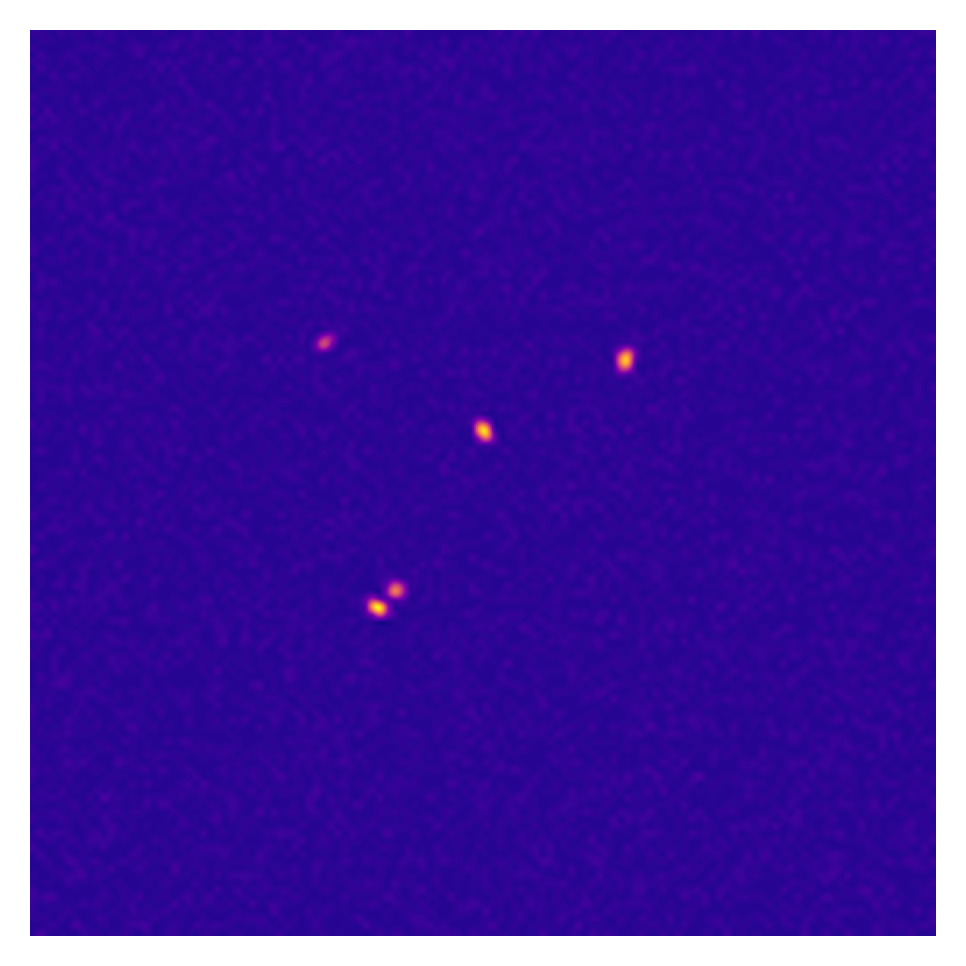}
\end{subfigure}%
\hspace{0.1em} 
\begin{subfigure}{\outputswidth}
\centering
\includegraphics[width=\textwidth]{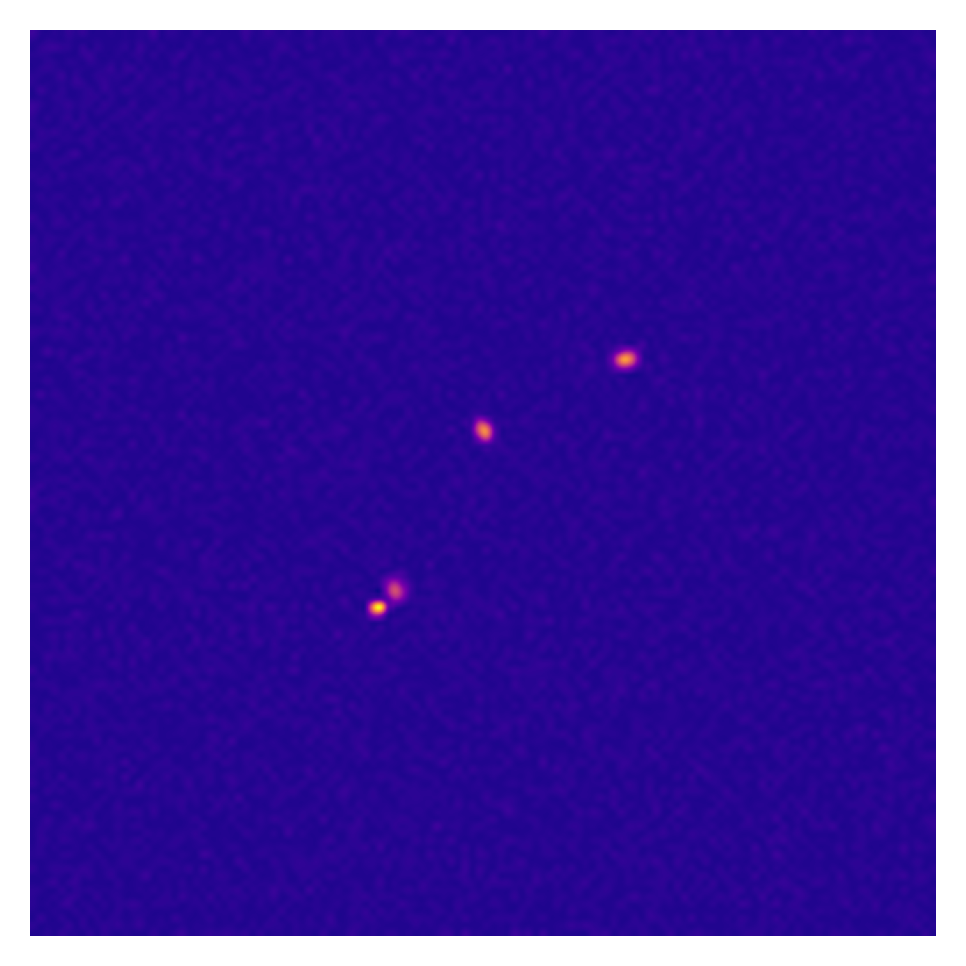}
\end{subfigure}%
\hspace{0.1em} 
\begin{subfigure}{\outputswidth}
\centering
\includegraphics[width=\textwidth]{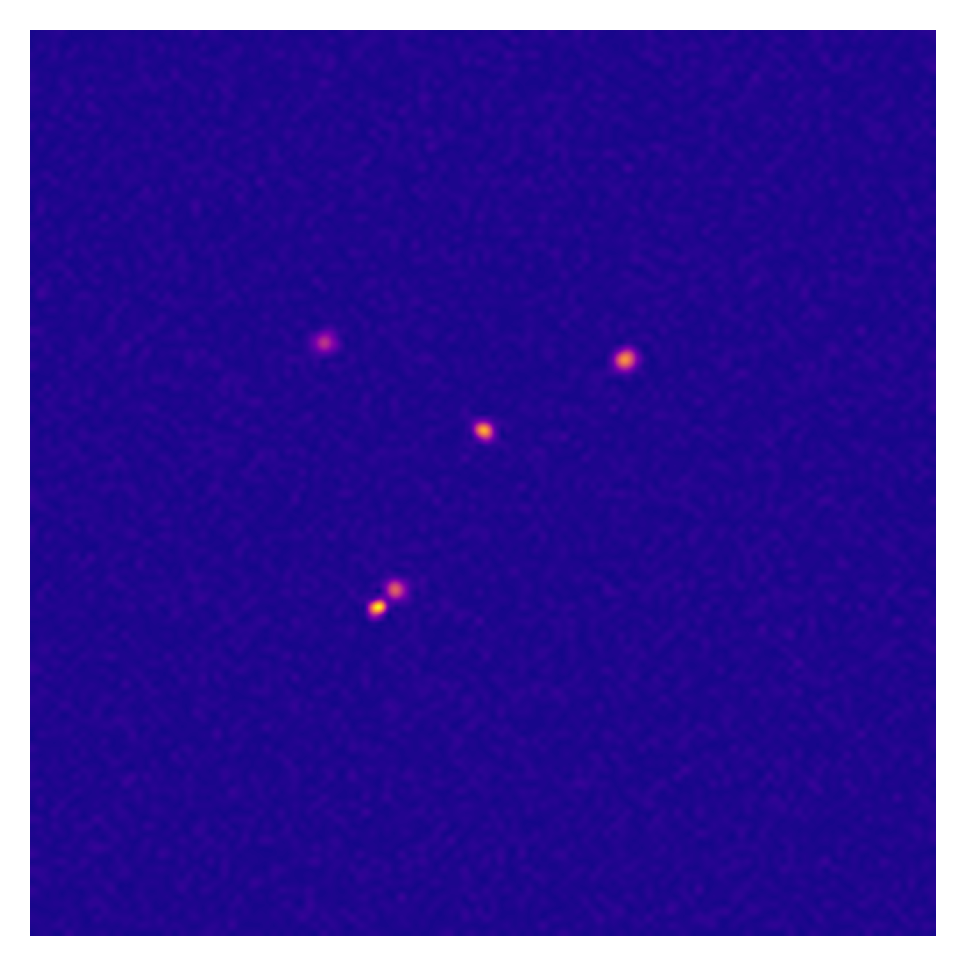}
\end{subfigure}%
\hspace{0.1em}
\begin{subfigure}{\outputswidth}
\centering
\includegraphics[width=\textwidth]{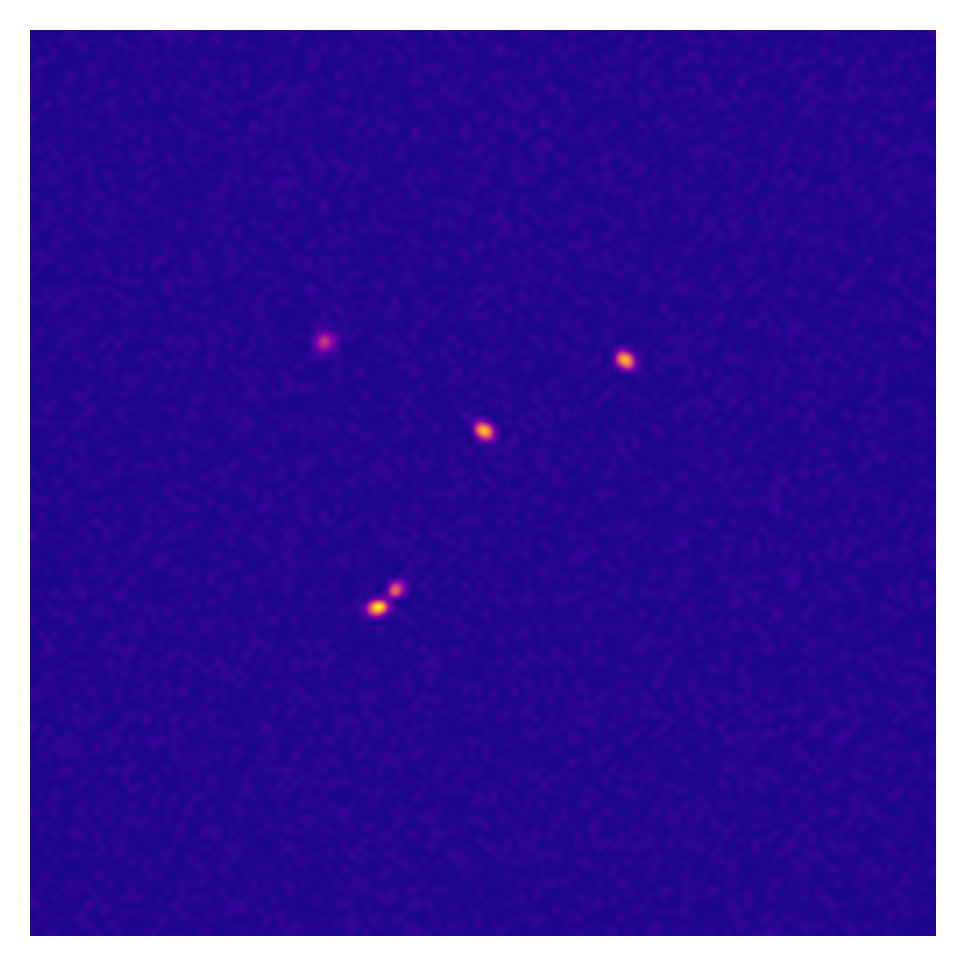}
\end{subfigure}%
\hspace{0.1em} 
\begin{subfigure}{\outputswidth}
\centering
\includegraphics[width=\textwidth]{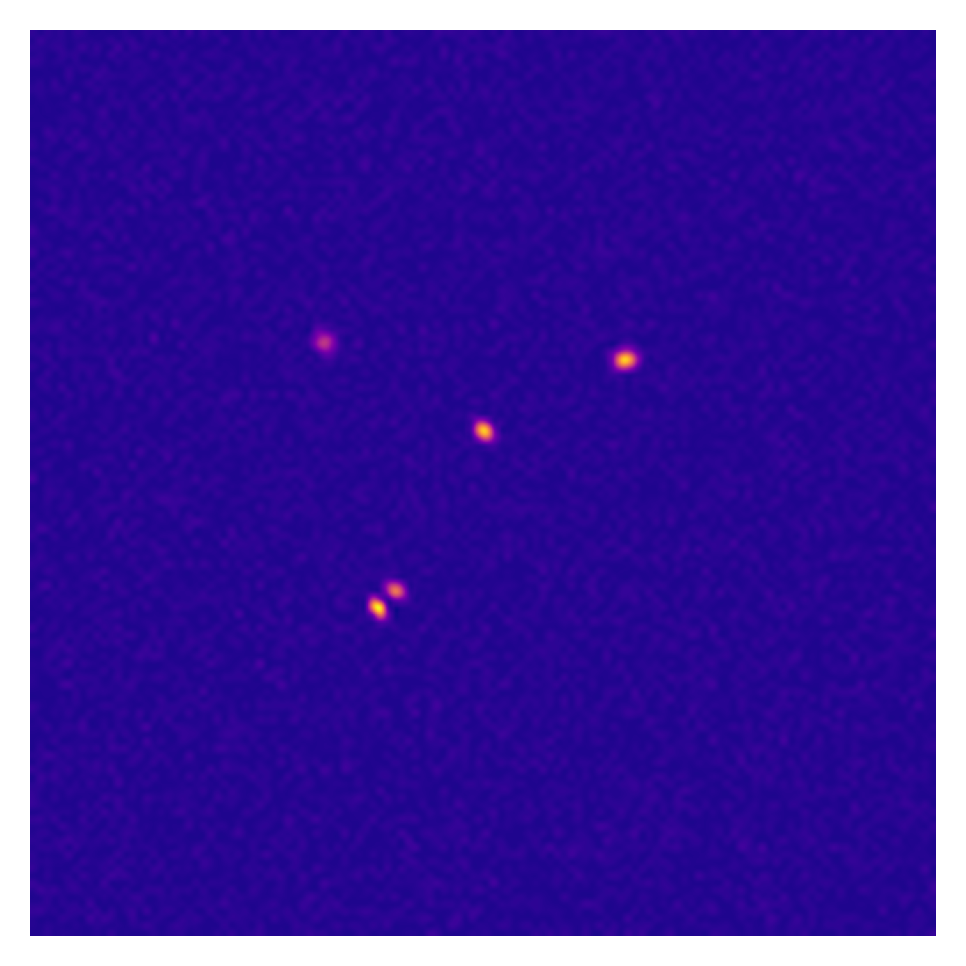}
\end{subfigure}

\vspace{0.1em}

\begin{subfigure}{\outputswidth}
\centering
\includegraphics[width=\textwidth]{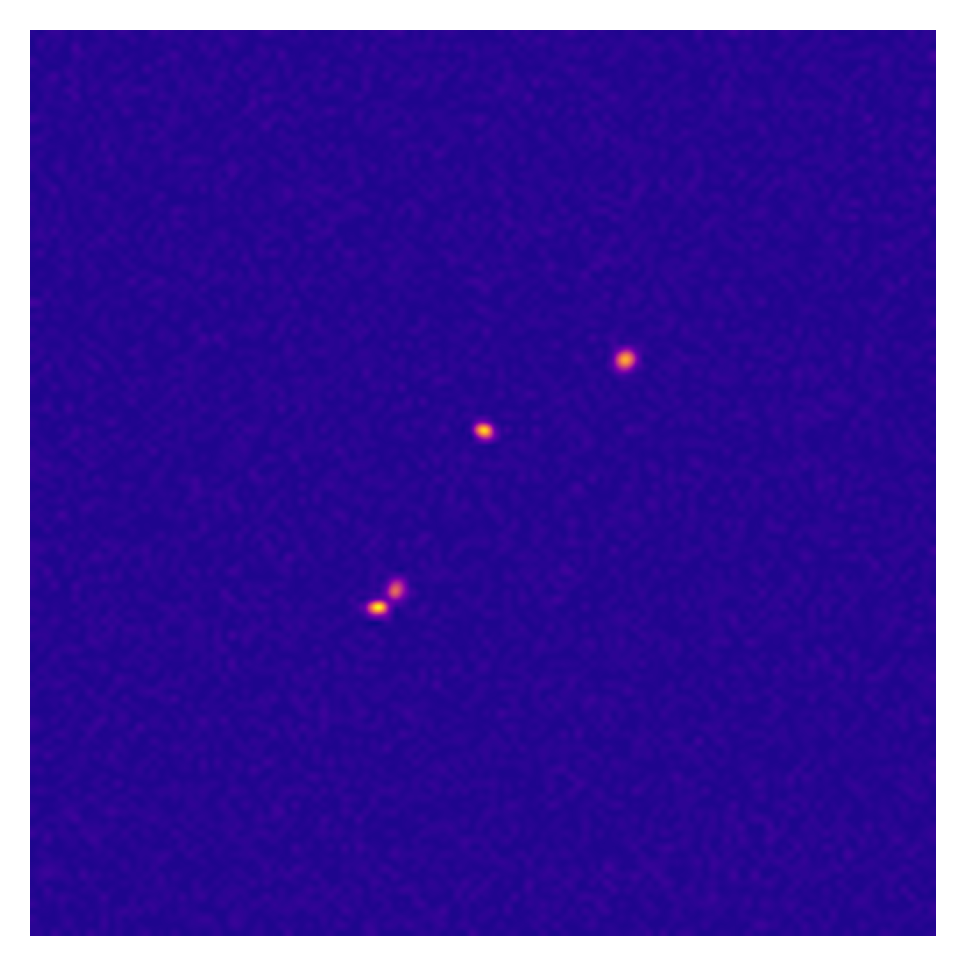}
\end{subfigure}%
\hspace{0.1em} 
\begin{subfigure}{\outputswidth}
\centering
\includegraphics[width=\textwidth]{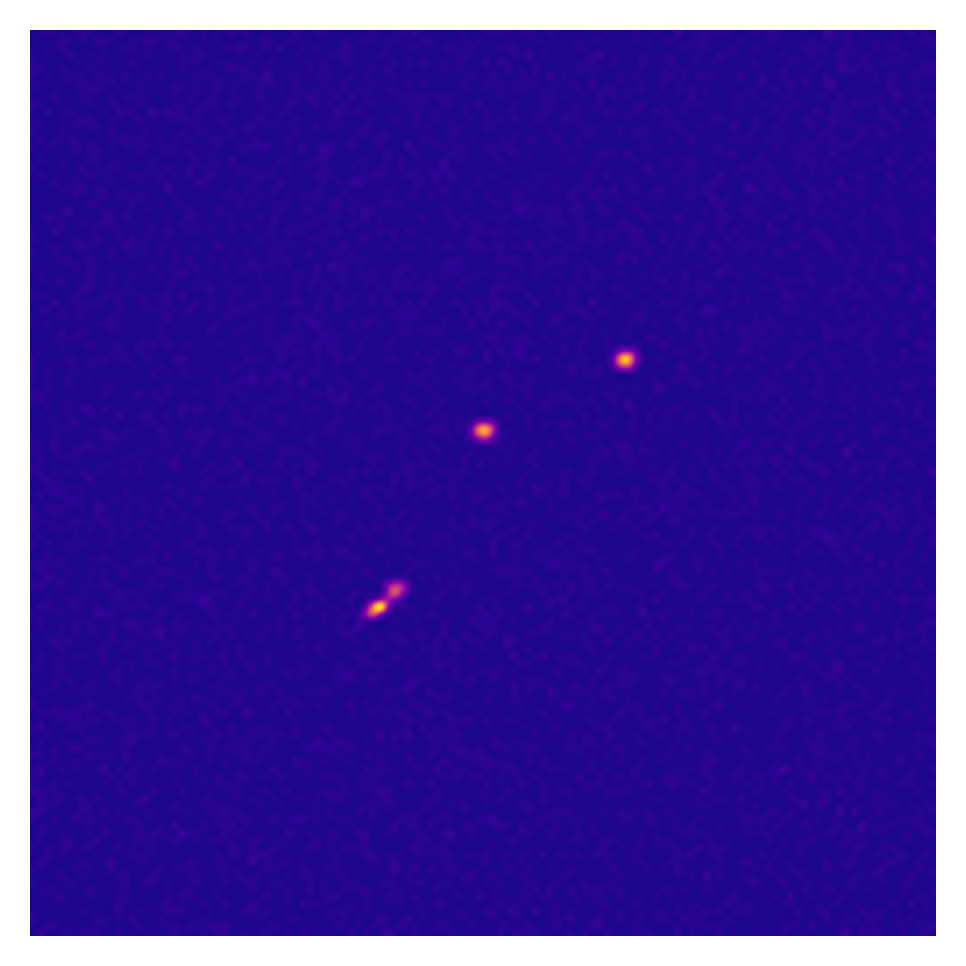}
\end{subfigure}%
\hspace{0.1em}
\begin{subfigure}{\outputswidth}
\centering
\includegraphics[width=\textwidth]{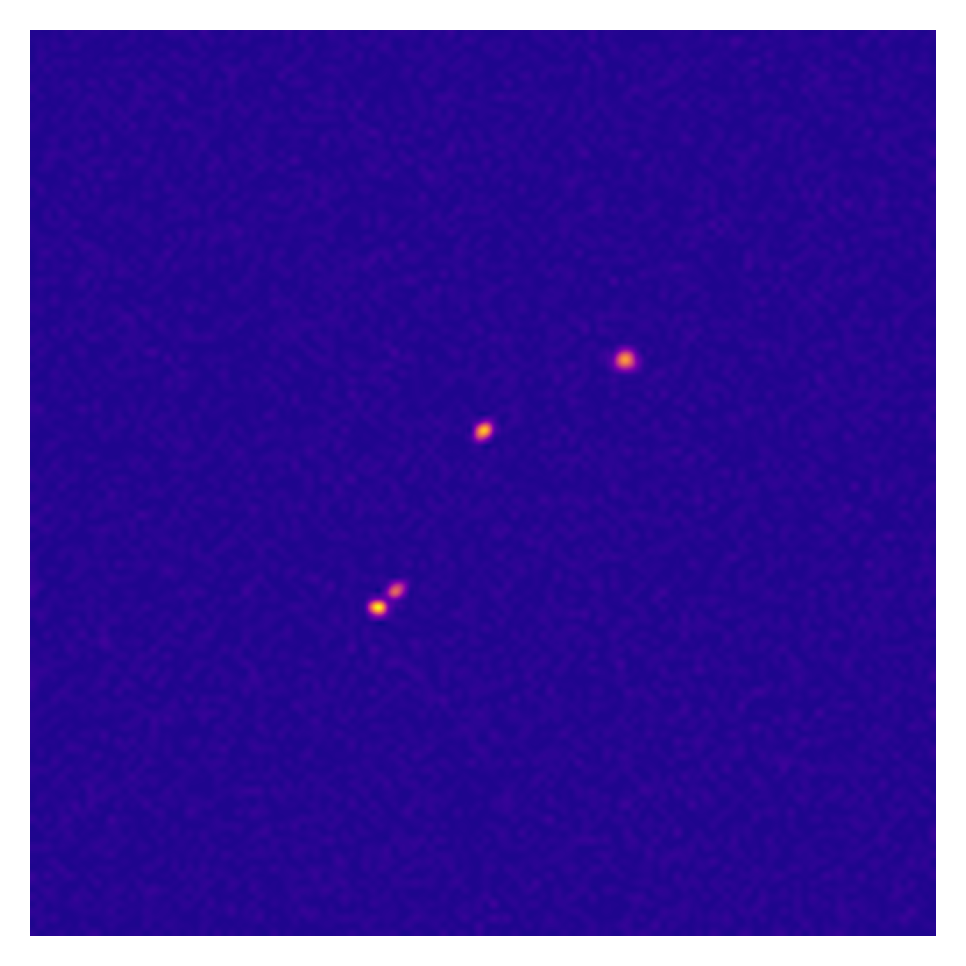}
\end{subfigure}%
\hspace{0.1em}
\begin{subfigure}{\outputswidth}
\centering
\includegraphics[width=\textwidth]{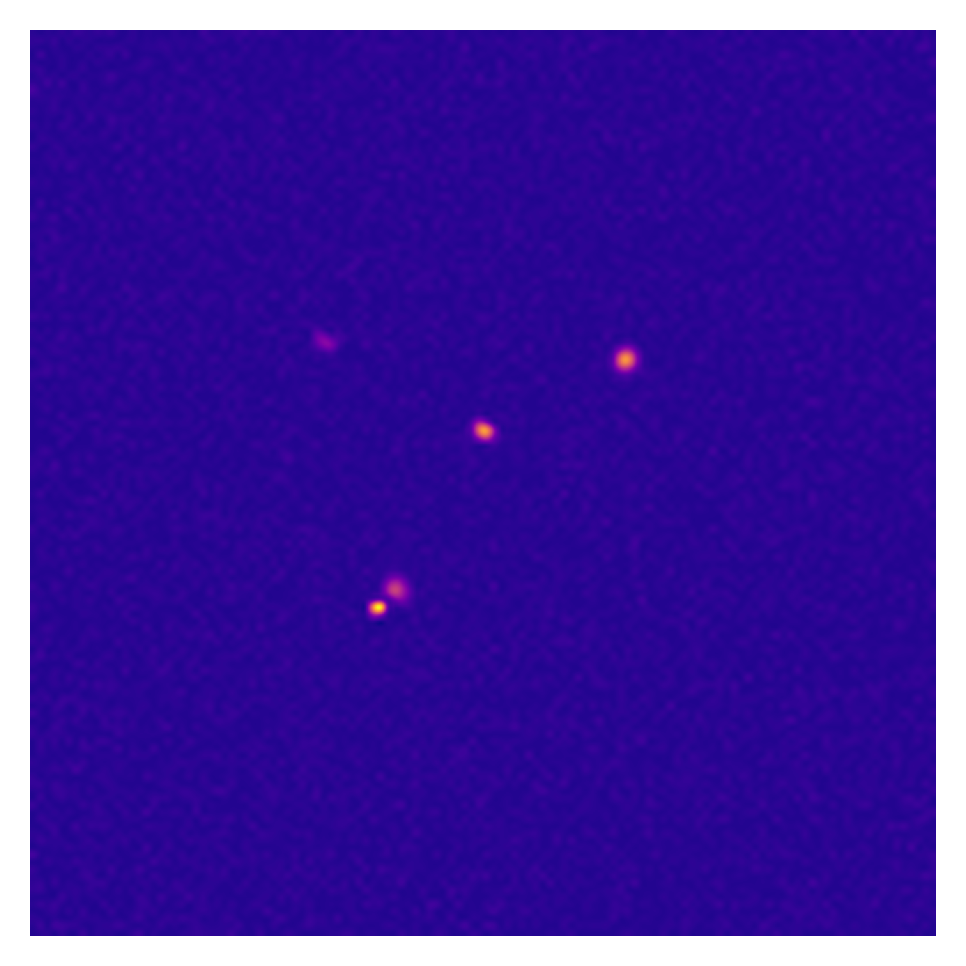}
\end{subfigure}
\hspace{0.1em}
\begin{subfigure}{\outputswidth}
\centering
\includegraphics[width=\textwidth]{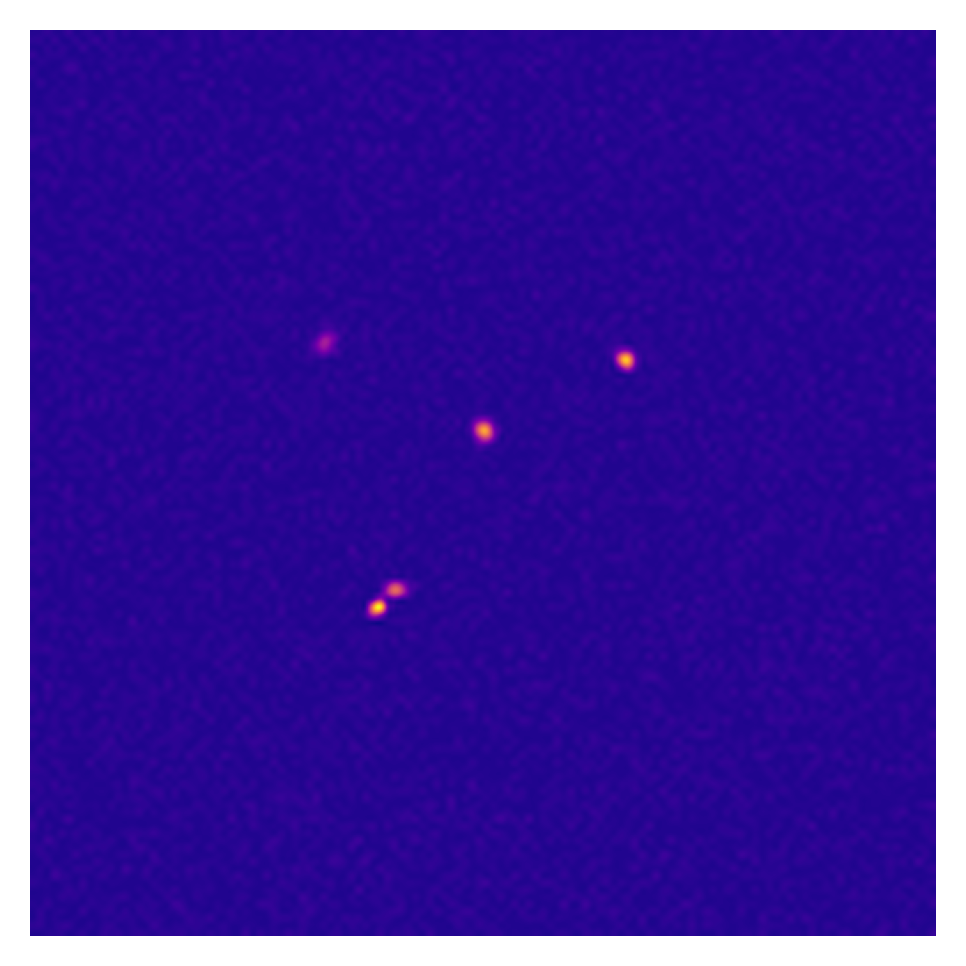}
\end{subfigure}
\caption{Diffusion model workflow: from dirty image to models outputs and uncertainty estimation. Top row: (1) Original dirty image (conditioning), with true sources circled in white. (2) The sky model showing sources detected by the Photutils localization algorithm. (3) The post-processed predicted sky model, aggregated using the median from 20 network outputs, with Photutils-localized sources. (4) Another version of the post-processed predicted sky model, this time aggregated pixel-wise using the mean from network outputs, again with Photutils-localized sources. (5) The standard deviation of multiple predictions before the post-processing step. We refer to this latter as the uncertainty estimation; it is unitless, with brighter colors (like yellow) indicating greater uncertainty. Rows 2-5 present 20 distinct outputs of the diffusion model for the corresponding dirty image, each with a different model noise initialization.}
\label{images:diffusion_output}

\end{figure*}

The model operates in a stochastic manner, producing multiple outputs for the same conditioning image. This stochasticity provides advantages when dealing with sources that exhibit low S/N. By running the model multiple times (20 runs in the experiments), we expect the faint sources to appear in the fraction of the predictions in the same positions. It is important to note that each run takes approximately 80 seconds per image on GPU RTX 3090. The execution time can be further optimized using different generative techniques, which will be the subject of future investigations.

Figure \ref{images:diffusion_output} shows an example of the model output. In the top row, we present samples from the generated data set: the first image is the conditioning (dirty image) and the second one is the corresponding sky model. The next three outputs represent the aggregated mean, the aggregated median, and the standard deviation of the predictions. The raw predictions of the model are in the rows 2-5. We note that there exists a particularly faint source that the proposed model fails to detect in aggregated images. However, in some single runs the source is present. This leads to huge values of standard deviation in this area, which shows that the model is uncertain about its predictions and that further investigation is needed. 

\begin{figure}[ht]
       \centering
        \includegraphics[width=0.7\linewidth]{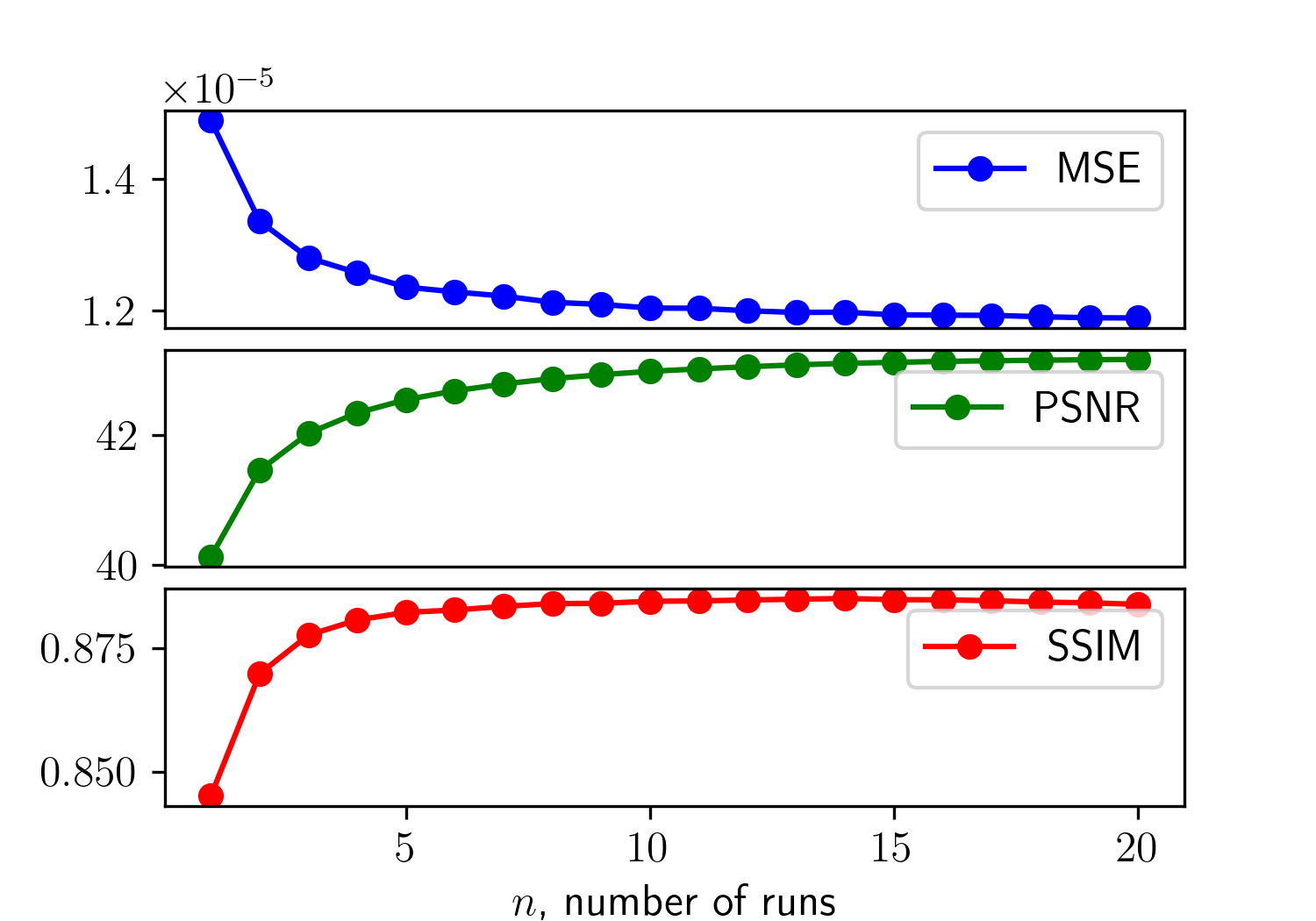}
\caption{Reconstruction metrics with respect to number of runs $\nbruns$ of diffusion model for median aggregation.}
    \label{fig:localization_metrics_scatter}
\end{figure}

This decision to select 20 runs per dirty image is motivated by the observed behavior of reconstruction metrics in Figure \ref{fig:localization_metrics_scatter} for the median aggregation process across multiple images. We notice an improvement in the metrics with an increase in the number of aggregated predicted images. However, this improvement comes at the expense of additional computation time. At the point of $\nbruns=20$, we assume that further increases do not substantially improve the metrics. Consequently, we identify $\nbruns=20$ as an effective compromise between improved performance and computational efficiency.

\begin{table*}[h!]
    \centering

\caption{Reconstruction results: MSE, peak S/N, and SSIM metrics computed on preprocessed sky models for different diffusion settings. }

\begin{tabular}{ccccc}
    \toprule
    \multirow{3}{*}{\textbf{Normalization}} & \multirow{3}{*}{\textbf{Aggregation}} &  \multicolumn{3}{c}{\textbf{Metrics ($\pm$ uncertainty)}} \\
    \cmidrule{3-5}
    && \textbf{MSE} &  \textbf{peak S/N} &  \textbf{SSIM} \\
    && ($\times 10^{-5}$) &  & \\
    \midrule
    \multirow{3}{*}{$\gamma = 1$}
    & single run & $5.276 \pm 0.138$ & $38.031 \pm 1.152$ & $0.977 \pm 0.002$ \\
    & \mystyle{mean} & $4.232 \pm 0.333$ & $\mystyle{42.243 \pm 9.475}$ & $\mystyle{0.998 \pm 0.001}$ \\
    & \mystyle{median} & $\mystyle{3.941 \pm 0.354}$ & $41.587 \pm 9.156$ & $\mystyle{0.998 \pm 0.001}$ \\
    \addlinespace
    \multirow{3}{*}{$\gamma = 2$}
    & single run & $1.492 \pm 0.146$ & $38.999 \pm 0.097$ & $0.830 \pm 0.003$ \\
    & mean & $\mystyle{1.189 \pm 0.604}$ & $41.251 \pm 0.206$ & $0.853 \pm 0.003$ \\
    & \mystyle{median} & $1.190 \pm 0.605$ & $\mystyle{42.148 \pm 0.260}$ & $\mystyle{0.882 \pm 0.003}$ \\
    \addlinespace
    \multirow{3}{*}{$\gamma = 10$}
    & single run & $1.668 \pm 0.171$ & $36.784 \pm 0.464$ & $0.814 \pm 0.004$ \\
    & mean & $\mystyle{1.183 \pm 0.592}$ & $38.008 \pm 2.263$ & $0.835 \pm 0.005$ \\
    & \mystyle{median} & $\mystyle{1.185 \pm 0.586}$ & $\mystyle{38.386 \pm 2.550}$ & $\mystyle{0.862 \pm 0.005}$ \\
    \addlinespace
    \multirow{3}{*}{$\gamma = 20$}
    & single run & $2.264 \pm 0.242$ & $35.105 \pm 0.793$ & $0.892 \pm 0.006$ \\
    & mean & $\mystyle{1.431 \pm 0.645}$ & $37.457 \pm 4.474$ & $0.929 \pm 0.011$ \\
    & \mystyle{median} & $1.442 \pm 0.643$ & $\mystyle{37.651 \pm 4.807}$ & $\mystyle{0.955 \pm 0.008}$ \\
    \addlinespace
    \multirow{3}{*}{$\gamma = 30$}
    & single run & $2.820 \pm 0.313$ & $32.944 \pm 1.086$ & $0.985 \pm 0.002$ \\
    & \mystyle{mean} & $\mystyle{1.553 \pm 0.750}$ & $\mystyle{34.844 \pm 4.585}$ & $0.985 \pm 0.008$ \\
    & median & $1.569 \pm 0.735$ & $34.693 \pm 4.906$ & $\mystyle{0.988 \pm 0.007}$ \\
    \bottomrule
\end{tabular}

    \label{tab:model_comparison_rec}
\footnotesize
\begin{minipage}{\textwidth} Note: We varied the power root used in the normalization during training (Equation \ref{eq:normalization}) and the aggregation method (mean, median over 20 runs, or single run). Single run signifies that the DDPM is run once for a given condition to get the predicted sky model. We evaluate performance on each run individually and then calculate the average score over all 20 runs. The error is calculated by taking the standard deviation over all outputs for the test set for aggregation methods, and over all runs to get a single run score.
\end{minipage}
\end{table*}

The reconstruction metrics are presented in Table \ref{tab:model_comparison_rec}. These metrics are calculated on the direct outputs of the diffusion model; that is, the normalized  sky model. We experiment with different values for the power root $\gamma$ in the normalization function applied to the sky model (Equation \ref{eq:normalization}).

As the original range of values within the images is significantly low, metrics calculated within it would be exceedingly high, rendering them uninformative. We therefore use normalized values. However, as discussed above, higher root powers lead to lower contrast in the images. In this case, the contrast is negatively correlated with the reconstruction metrics –-less contrast in images tends to result in better reconstruction metrics for the same model performance. This relationship is evident in Table \ref{tab:model_comparison_rec}, where lower powers yield superior results compared to higher powers. 

Aggregating multiple outputs from various runs outperforms the output from a single run within the same normalization. This demonstrates that the inherent stochasticity of the process assists in detecting fainter sources, which indicates the value of using statistics across multiple runs to enhance reconstruction. Through experimentation, we find that the mean and median aggregation methods deliver very similar results.

An essential takeaway from this analysis is that reconstruction metrics may not be the most adequate indicators for assessing the quality of astronomical data. In subsection \ref{subsec:sources_localization}, we explore the localization metrics and discuss their performance for the same settings.

\subsubsection{Uncertainty estimation in the image space}

To quantify the uncertainty associated with the sky model predictions, we use the concept of uncertainty estimation in the image space. We define $\x_{\text{uncertainty}}$ using the following formula:

\begin{equation}
\x_{\text{uncertainty}}(h, w) = \sqrt{\frac{1}{\nbruns} \sum_{j=1}^{\nbruns} \left(\xrecon^j(h, w) - \frac{1}{\nbruns} \sum_{k=1}^{\nbruns} \xrecon^k(h, w)\right)^2},
\end{equation}
where $n$ represents the number of runs ($\nbruns=20$), $\xrecon^j$ denotes a prediction of the sky model for the $j$-th run, and 
 $(h, w)$  are the pixel coordinates. This formula denotes the standard deviation   visualized in the top right of Figure \ref{images:diffusion_output}. This latter displays the variability in the predictions across multiple runs. Intuitively, if the model consistently identifies the same sources, the uncertainty is low and we can be certain about their presence. Conversely, if the sources appear and disappear from the predictions across runs, the variance in the image space will be high, indicating a region of the sky that requires further investigation. The variance in the predictions can be juxtaposed with the intensity values estimated in the mean-aggregated image to determine the degree of uncertainty associated with a particular region. This can inform the final decision about whether a given region exhibits significant uncertainty, thereby guiding further analysis.

\subsection{Source localization}
\label{subsec:sources_localization}

After obtaining a predicted sky model, we try to identify the sources present. We evaluate the performance of our source localization using three metrics: purity, completeness, and F1-score, as detailed in Table \ref{tab:localization_metrics}.

The F1-score is a particularly effective measure of model performance because it simultaneously considers both purity (the model's ability to limit FPs) and completeness (the model's ability to identify all real sources in the dataset). We therefore consider F1 as the main indicator for choosing the best model.

We provide the scores for a Photutils method of source localization applied directly to the simulated sky model given by CASA (last line in Table \ref{tab:localization_metrics}). This reference reveals the level of error that comes from the chosen parameters of the algorithm from Photutils.\footnote{The Photutils parameters are:  \texttt{kernel\_x\_size}=3, 
\texttt{kernel\_y\_size}=3,
\texttt{sigma\_clipped\_stats\_sigma}=2.
\texttt{detect\_npixels}=10,
\texttt{deblend\_sources\_npixels}=10,
\texttt{std\_const}=120.
} These parameters stay the same during all experiments, which ensures that any differences in outcomes can be attributed to the models themselves, rather than to variations in the processing steps.

We draw a comparison with the model presented by \cite{Taran_2023}. The distinguishing feature of their model is an additional compression step, which retains only 1450 fixed positions in the UV plane and operates directly in the frequency domain. Contrasting this with the proposed DDPM, which employs 250 passes, the model in \cite{Taran_2023} accomplishes its task in a single pass. This characteristic exemplifies the common trade-off between accuracy and computational efficiency. The model in \cite{Taran_2023} performs better than PyBDSF \citep{mohan2015pybdsf}, which currently thought to be state of the art. Consequently, we chose to compare our results to those of obtained using the model in \cite{Taran_2023}. As can be seen from Table \ref{tab:localization_metrics}, all the models achieve very high F1-scores exceeding 95 \%, compared to 81 \% achieved by the \cite{Taran_2023} model. 

For the images with lower root functions $\gamma \in [1,2] $, the best aggregation methods are \aggregatedetect ~ones. The outputs for the model trained with such normalizations are noisy, which is due to the sparsity of the images. Aggregation on the image level averages the noise in each prediction, producing a higher-quality output. However, as $\gamma$ gets larger and the prediction quality rises, the winning method becomes \detectaggregate.

\begin{figure}[ht]
    \centering
    \includegraphics[width=0.8\linewidth]{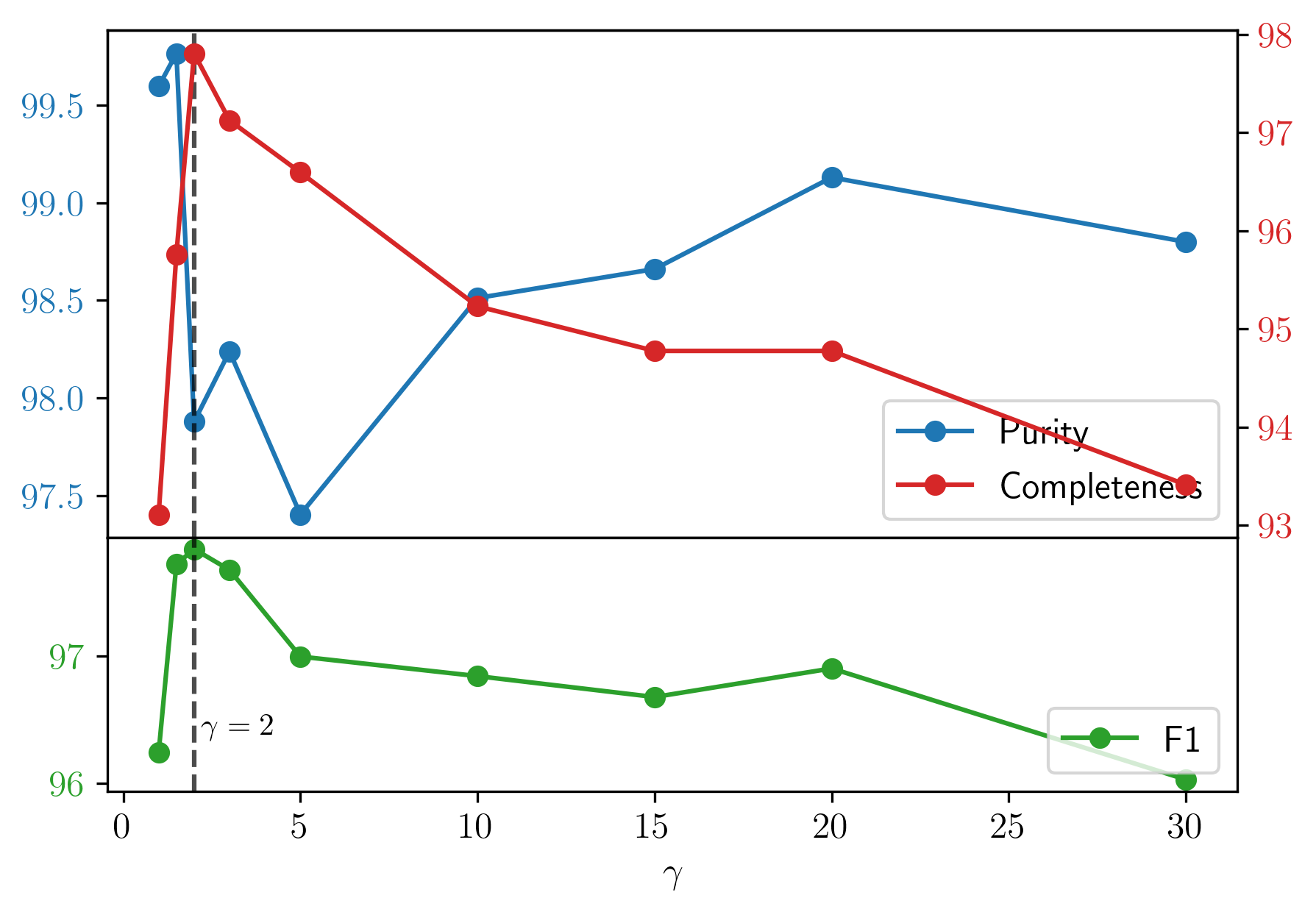}
    \caption{Best scores for purity, completeness, and F1  among all aggregations with respect to different root power $\gamma$ used in normalization during training.}
    \label{fig:optimal_alpha}
\end{figure}

The optimal model employs $\gamma = 2$. We illustrate the highest localization scores relative to various $\gamma$ values in Figure \ref{fig:optimal_alpha}. The $\gamma = 2$ case demonstrates the effectiveness of the proposed normalization technique compared to the $\gamma = 1$ case. However, as $\gamma$ increases, the performance falls because of excessive sharpening of the blob borders, as seen in Figure \ref{fig:images_normalization}.

\begin{figure}[ht]
       \centering
        \includegraphics[width=0.8\linewidth]{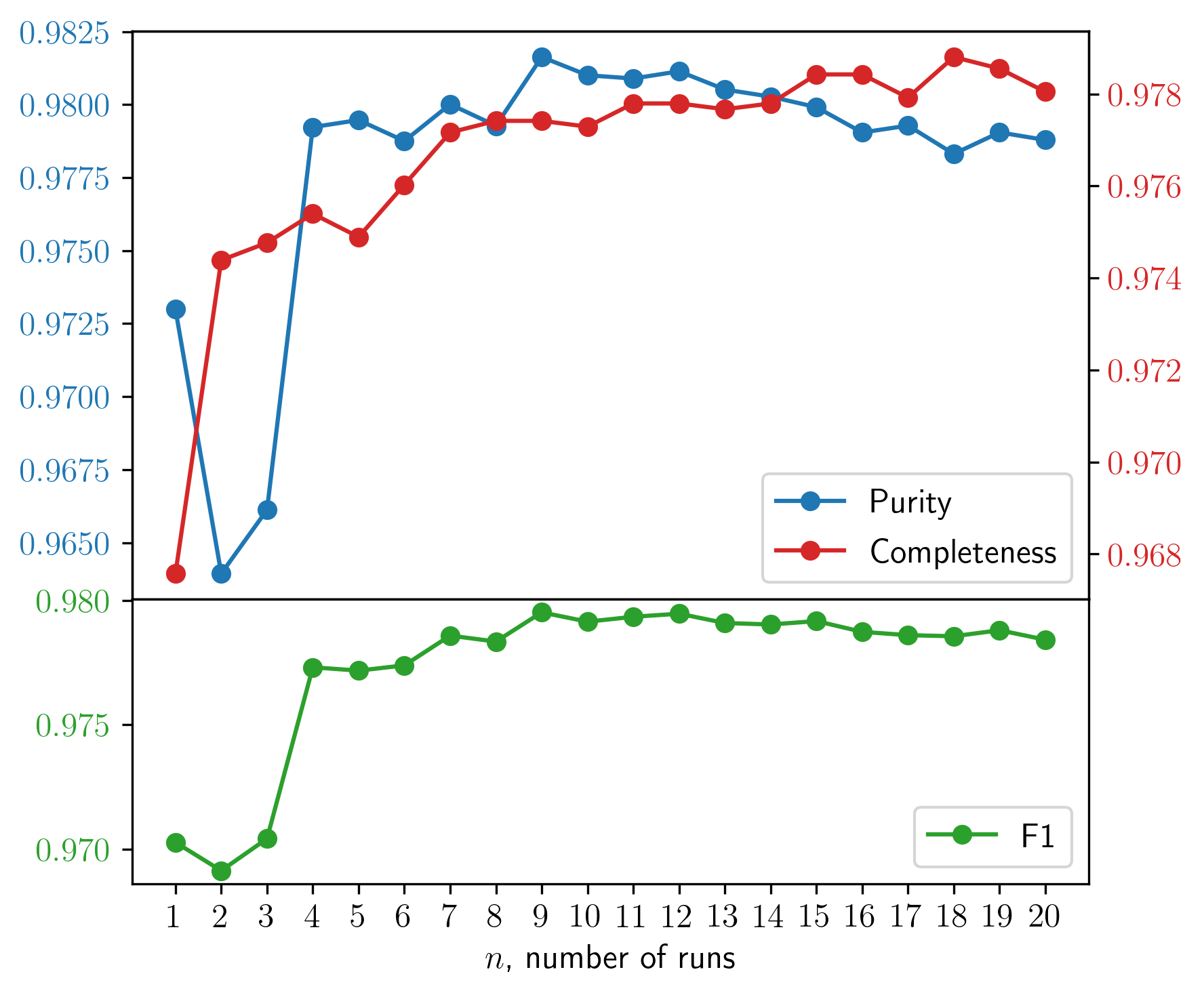}
\caption{Localization metrics with respect to number of runs $\nbruns$ of diffusion model. }
    \label{fig:localization_metrics_n}
\end{figure}

In Figure \ref{fig:localization_metrics_n}, we plot the scores for $\gamma = 2$ against the number of runs. As $\nbruns$ increases, the F1-score saturates, reaching a plateau at $\nbruns = 5$ with no significant further improvement. This behavior contrasts with the reconstruction scenario, where the metrics continue to improve with increasing $\nbruns$. This occurs for reconstruction metrics because the more images we generate, the more errors are corrected when averaging, leading to smooth improvement. However, localization metrics operate differently; their primary focus is on the position of the flux. Consequently, fewer images are required for the flux to appear in the final prediction, and adding more images does not significantly modify the chosen localization metrics. Essentially, these metrics are only concerned with whether the source was detected or not, which is why they exhibit a different behavior.

\subsection{Uncertainty in localization}

The \detectaggregate ~technique confers an advantage, in that it suggests a clear way to find uncertainty on the estimations. We compute the percentage of images where the source is present and then we aggregate those source characteristics through averaging. The fraction of output images where the source is present is interpreted as a measure of source reliability: in this study, the source is not retained if the reliability  is less than 30\%. 

\begin{figure}[ht]
        \centering
        \includegraphics[width=0.8\linewidth]{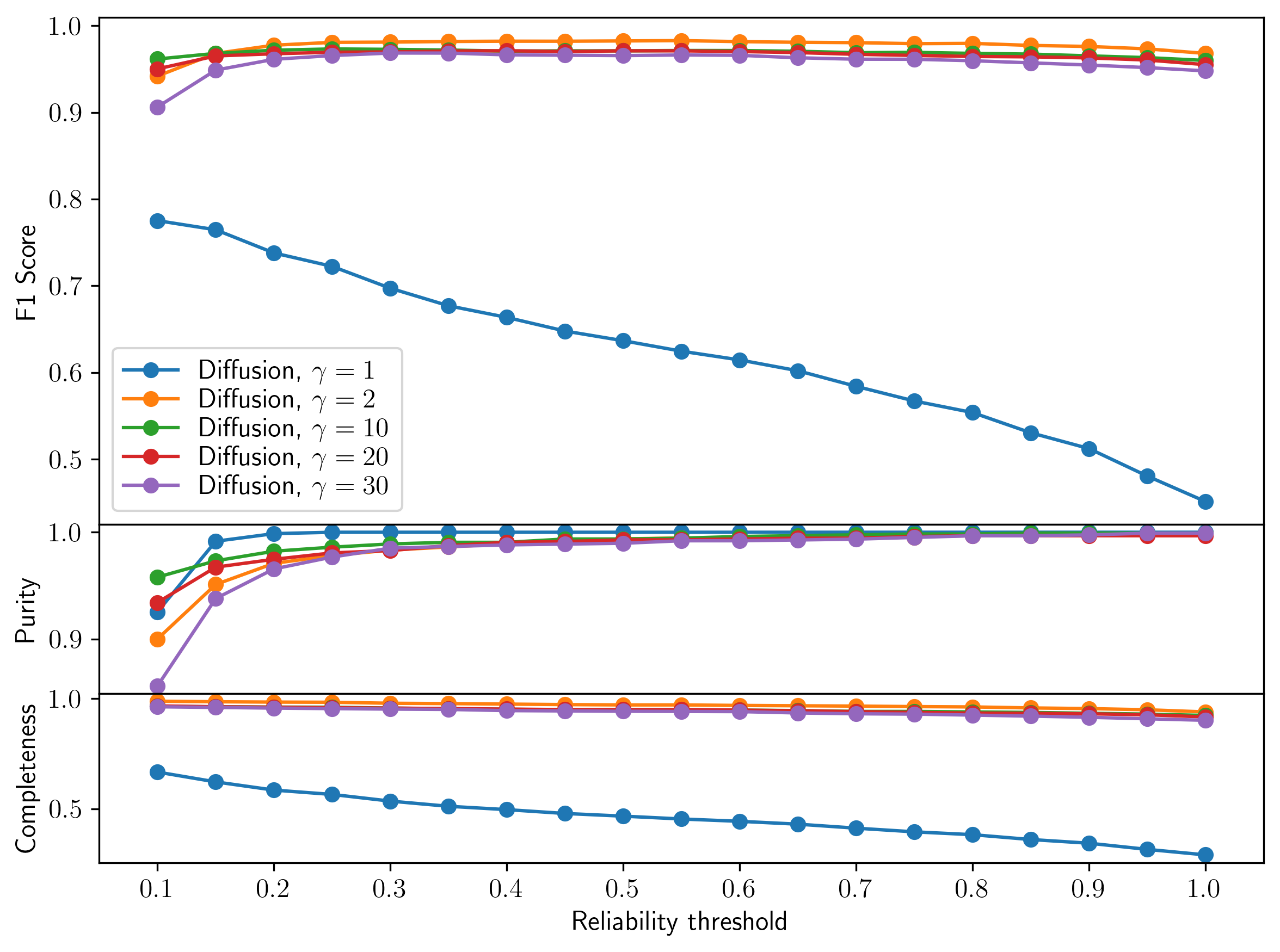}
    \caption{Dependence of F1, purity, and completeness scores on the reliability parameter for all models. Reliability is the fraction of the 20 images in which a source is present. The closest to optimal value we find is 30\%, which means each source should be present at least in 6 images out of 20. }
    \label{fig:F1_score_thresholds}
\end{figure}

To illustrate the rationale behind the choice of the 30\% threshold, we plot the F1-score, purity, and completeness against the reliability threshold for retaining sources in Figure \ref{fig:F1_score_thresholds}. All curves except the one with root power $\gamma=1$ reach saturation around 30\%, have a flat section after, and start to diminish in the end. Purity grows as we demand that the source be more consistent in the images, and completeness falls as we eliminate an increasing number of sources that appear in an insufficient number of images.

We plot the dependence of reliability on the S/N of the sources in Fig. \ref{fig:snr_reliability}. We find that most of the sources with low reliability have low-S/N values. This demonstrates that the stochasticity of the DDPM model mostly affects the low-S/N sources, and they therefore have lower reliability. Consequently, for sources detected with low S/N using the proposed model, reliability can be used as an additional metric to select more robust sources. Choosing a reliability threshold of 30\% is a well-balanced trade-off for the simulated data, allowing the model to have few FPs and FNs even at relatively low S/N. This is discussed in further detail in Section \ref{subsec: snr dependency}.

\begin{table*}[ht]
\centering
\caption{Purity, completeness, and F1-score. }
\label{tab:localization_metrics}
\begin{tabular}{ccccccc}
    \toprule
    \multirow{2}{*}{\textbf{Model}} & \multirow{2}{*}{\textbf{Normalization}} & \multirow{2}{*}{\textbf{Input}} & \multirow{2}{*}{\textbf{Aggregation}} &  \multicolumn{3}{c}{\textbf{Metrics}} \\
    \cmidrule{5-7}
    &&&& \textbf{Purity} & \textbf{Completeness} & \textbf{F1} \\
    \midrule
\multirow{4}{*}{Diffusion} & \multirow{4}{*}{$\gamma=1$}& \multirow{5}{*}{dirty image} & single run & 100.00 & 44.97 & 62.04 \\
 & & & d\detectaggregatetable & 100.00 & 53.52 & 69.72 \\
 & & &  \mystyle{mean} & \mystyle{99.60} & \mystyle{93.11} & \mystyle{96.24} \\
  & & & median & 99.83 & 90.84 & 95.12 \\
\addlinespace
\multirow{4}{*}{Diffusion}  & \multirow{4}{*}{$\gamma=2$}& \multirow{4}{*}{dirty image} & single run & 97.45 & 96.71 & 97.08 \\
  & & & d\detectaggregatetable & 98.33 & 97.88 & 98.10 \\
  & & & mean & 97.88 & 97.80 & 97.84 \\
  & & & \mystyle{median} & \mystyle{99.30} & \mystyle{97.05} &  \mystyle{98.16} \\
\addlinespace
\multirow{4}{*}{Diffusion}  & \multirow{4}{*}{$\gamma=10$} & \multirow{4}{*}{dirty image} & single run & 98.66 & 94.66 & 96.62 \\
  & & & \mystyle{d\detectaggregatetable} & \mystyle{98.90} & \mystyle{95.69} & \mystyle{97.27} \\
  & & & mean & 98.51 & 95.23 & 96.84 \\
  & & & median & 99.29 & 94.78 & 96.98 \\
\addlinespace

\multirow{4}{*}{Diffusion}   & \multirow{4}{*}{$\gamma=20$} & \multirow{4}{*}{dirty image} & single run & 98.26 & 94.43 & 95.81 \\
  & & & \mystyle{d\detectaggregatetable} & \mystyle{99.52} & \mystyle{94.70} & \mystyle{97.05} \\
  & & &  mean & 99.52 & 93.64 & 96.49 \\
  & & &  median & 99.52 & 93.79 & 96.57 \\
\addlinespace

\multirow{4}{*}{Diffusion}   & \multirow{4}{*}{$\gamma=30$} & \multirow{4}{*}{dirty image} & single run & 98.58 & 93.46 & 94.93 \\
  & & & \mystyle{d\detectaggregatetable} & \mystyle{99.20} & \mystyle{94.10} & \mystyle{96.58} \\
  & & & mean & 98.80 & 93.41 & 96.03 \\
  & & & median & 98.81 & 94.25 & 96.47 \\
\addlinespace

 {PyBDSF}
 & - &  clean image & - & 72.18 & 20.82 & 32.31 \\
 \addlinespace
\addlinespace
{Taran et al. (2023)} & - & reduced uv-samples & - & 91.02 & 74.14 & 81.72 \\
\addlinespace
{Photutils localization} & - & {sky model} & - & 99.70 & 99.10 & 99.40 \\
\bottomrule
\end{tabular}

\footnotesize
\begin{minipage}{\textwidth} 
Note: The localization is performed on predictions that have been renormalized to be within the same range as the original sky model. The model configurations vary in terms of normalization power used during training, and also in terms of the aggregation method employed. For the normalization with high root powers $\gamma,$ the optimal aggregation method is \detectaggregate. This approach offers a good trade-off between purity and completeness, resulting in the best F1-score. For the lower $\gamma,$ the best approach turned out to be \aggregatedetect. We compare the results with \cite{mohan2015pybdsf} (PyBDSF) and \cite{Taran_2023}. We run the Photutils algorithm directly on sky models to quantify the error coming from this step, denoted as Photutils localization.
\end{minipage}
\end{table*}

\subsection{Performance vs S/N}  
\label{subsec: snr dependency}

\begin{figure}[ht]
        \centering
        \includegraphics[width=0.8\linewidth]{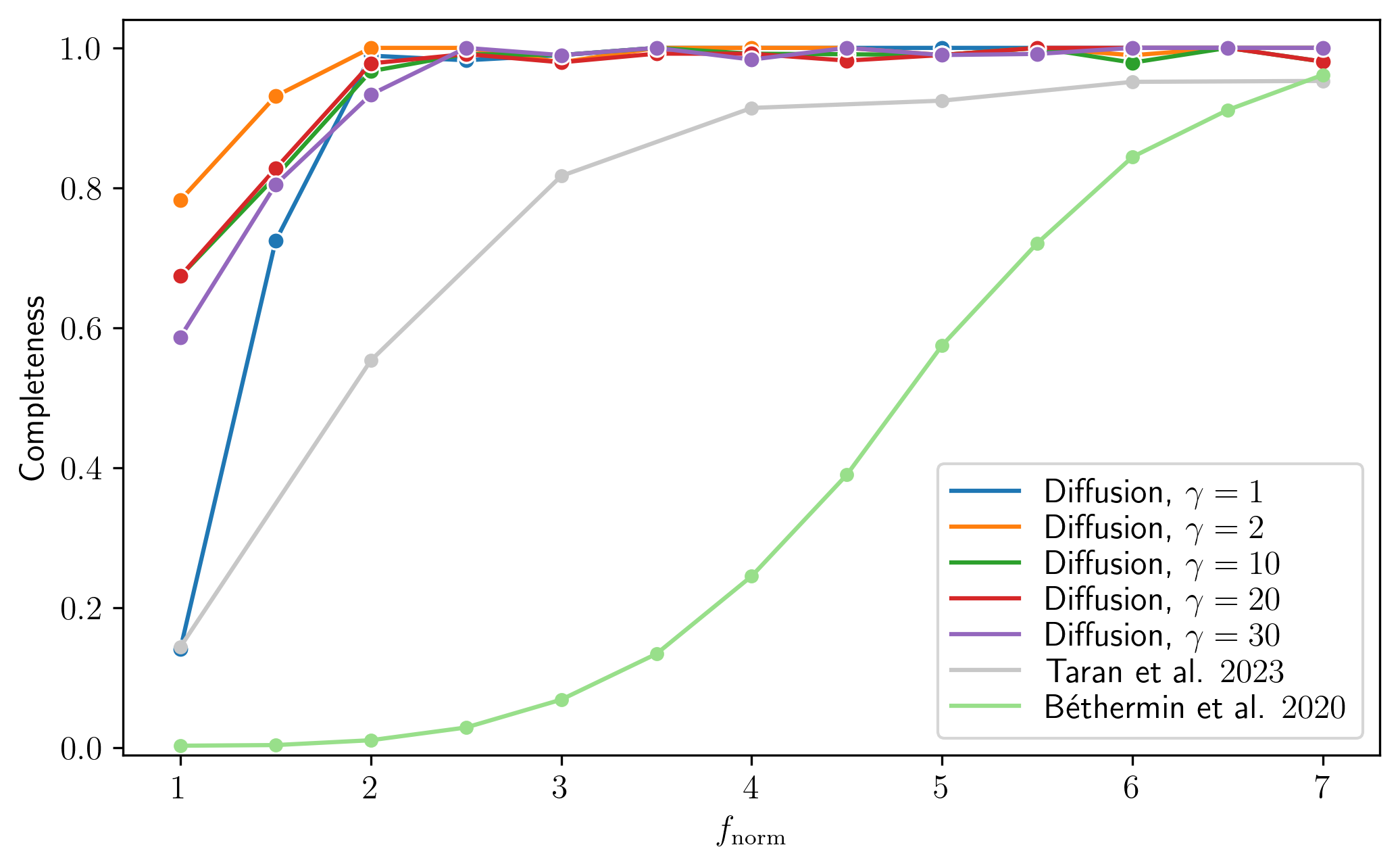}
\caption{Completeness comparison with state-of-the-art methods. These state-of-the-art methods outperform traditional algorithms such as PyBDSF, which is why the comparison is limited to them. All diffusion models demonstrate significant improvement at low S/N. }
 \label{fig:purity_completeness_f1score_snrnorm}
\end{figure}

We study the performance of the model as a function of S/N. We use a definition of normalized injected flux from \cite{B_thermin_2020}:

$$ f_{\text{norm}} = \frac{\text{total flux}}{\sigma_{\text{noise}}} \frac{b_{\text{min}} \cdot b_{\text{maj}}}{{\sqrt{b^2_{\text{min}} + s^2_{\text{min}}}} \cdot \sqrt{b^2_{\text{maj}} + s^2_{\text{maj}}}},
 $$
where $b_\text{min}$ and $b_\text{maj}$ are the minor and major axes of the beam (0.82'' and 0.89'' respectively) and $s_\text{min}$ and $s_\text{maj}$ are the minor and major axes of the source.
The results are shown in Figure \ref{fig:purity_completeness_f1score_snrnorm} for the best aggregation method among the tested diffusion models (see Table \ref{tab:localization_metrics}; the best are highlighted in bold for each model and normalization). We conduct a comprehensive comparison with the model from \cite{Taran_2023}. We also compare the results to those obtained by \cite{B_thermin_2020}. 

All of the models showcase a robust and significant performance boost, particularly for low S/N, in comparison to previous methods. This improvement is highlighted by the fact that all models reach a completeness score exceeding 90\% even at a low S/N of 2, which can be compared to the score of 50\% found by\ \cite{Taran_2023} for the same S/N. For S/N values higher than 3, all models reach close to 100\% completeness, which means that they do not miss any real sources.

\begin{figure}[ht]
        \centering
        \includegraphics[width=0.8\linewidth]{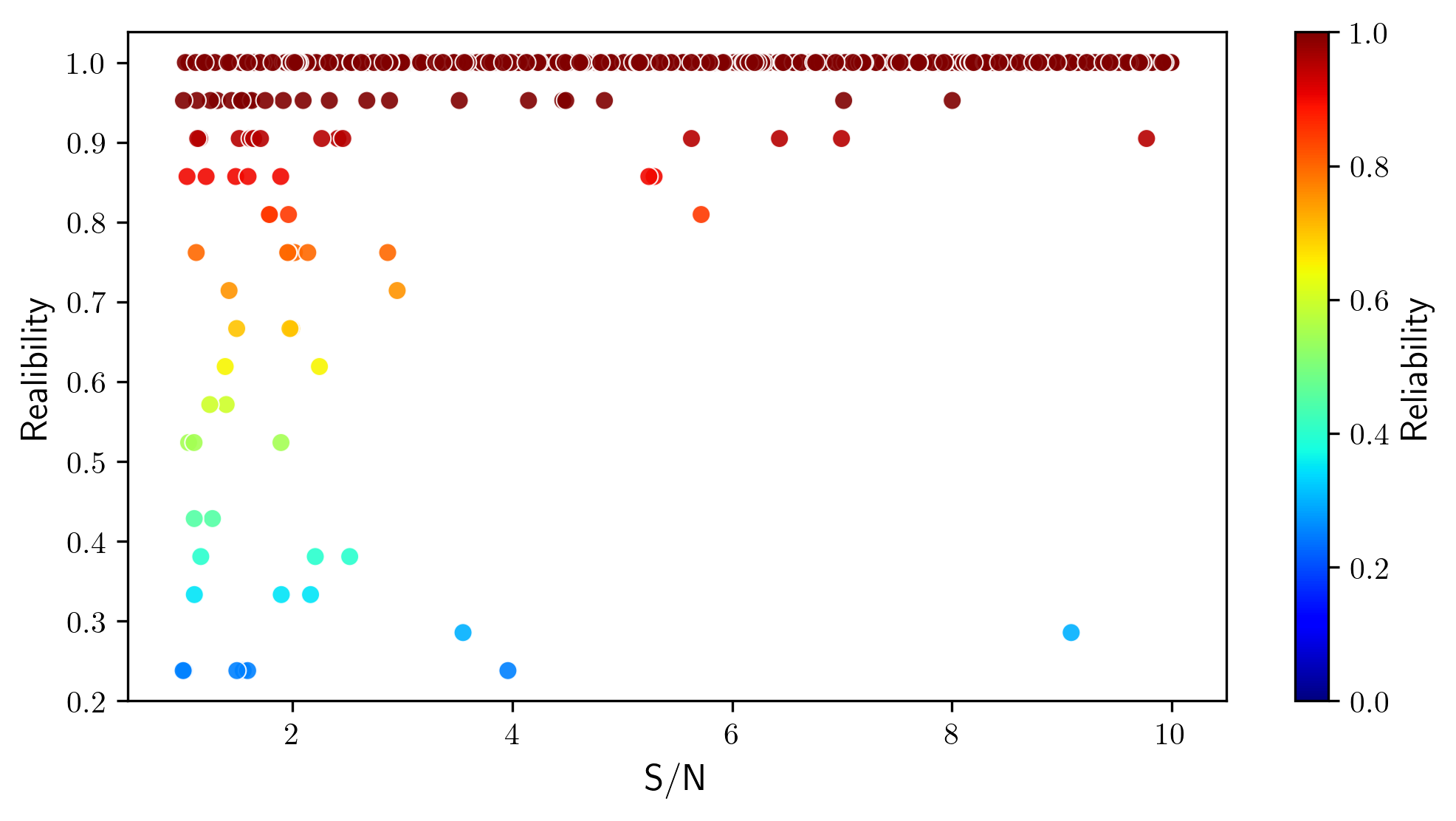}
\caption{Signal-to-noise ratio versus reliability. Lower reliability values are predominantly associated with sources that show lower S/Ns. However, the model often localizes sources with low S/Ns with high confidence.}
    \label{fig:snr_reliability}
\end{figure}

The reliability assigned through the \detectaggregate ~method is anticipated to depend on the S/N of the source, as demonstrated in Figure \ref{fig:snr_reliability}. We observe a trend where sources with lower reliability tend to have lower S/Ns. However, it is notable that the model is able to consistently reconstruct many sources, even those with low S/N.

\subsection{Flux estimation}
\begin{figure}[ht]
        \centering
        \includegraphics[width=0.8\linewidth]{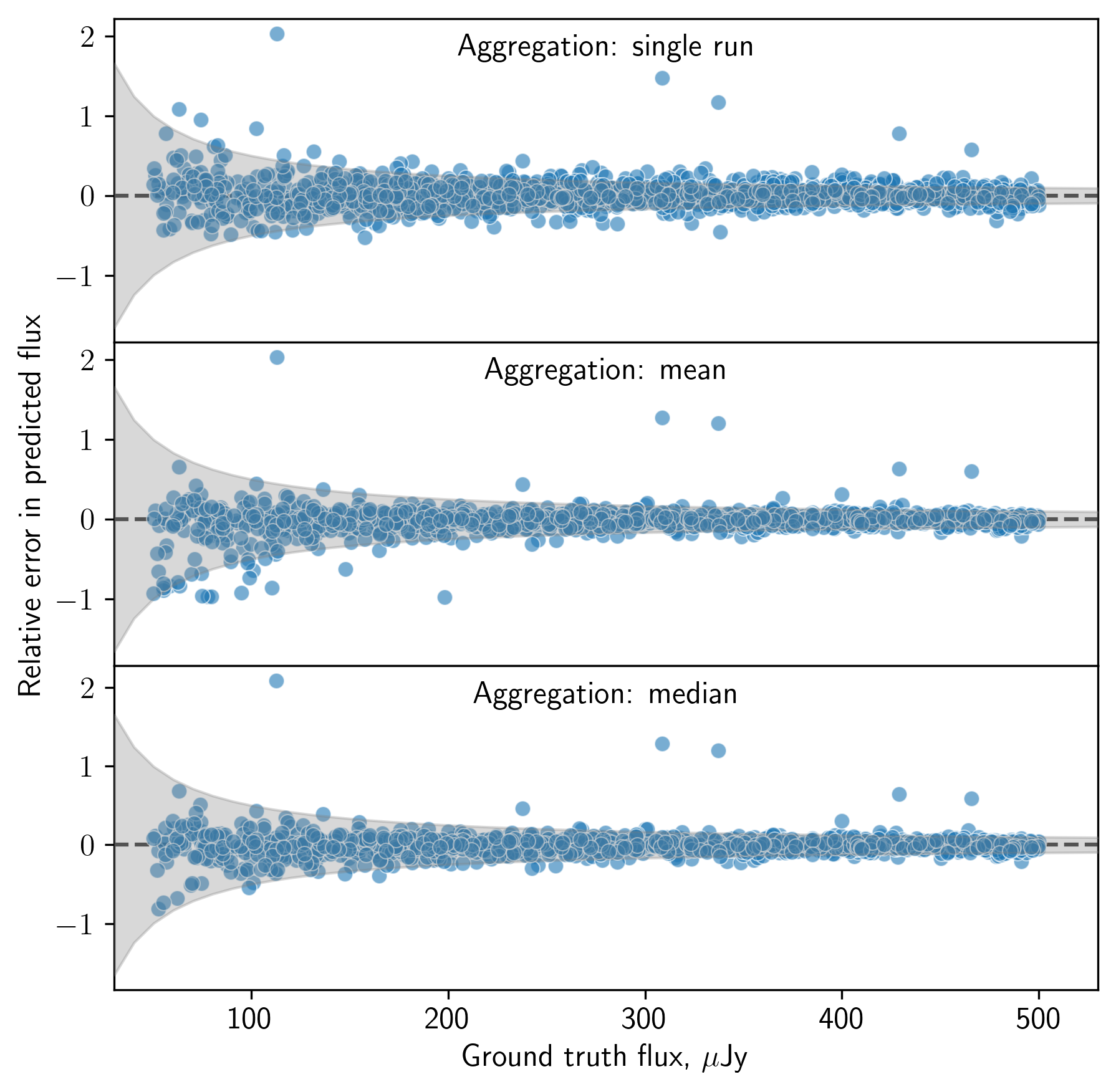}
\caption{Relative error in flux estimation compared to true flux for studied aggregation techniques across images. The gray area indicates the 1$\sigma$ range. Per-pixel aggregation methods, such as the mean and median, yield reliable estimates. Visually, the median provides the most appealing results with the fewest outliers. All results presented here are for the model employing $\gamma = 2$. The outliers are discussed in Appendix \ref{sec: appendix_outliers}. }
    \label{fig:aggregation_for_fluxes}
\end{figure}

\begin{figure}[ht]
        \centering
        \includegraphics[width=0.8\linewidth]{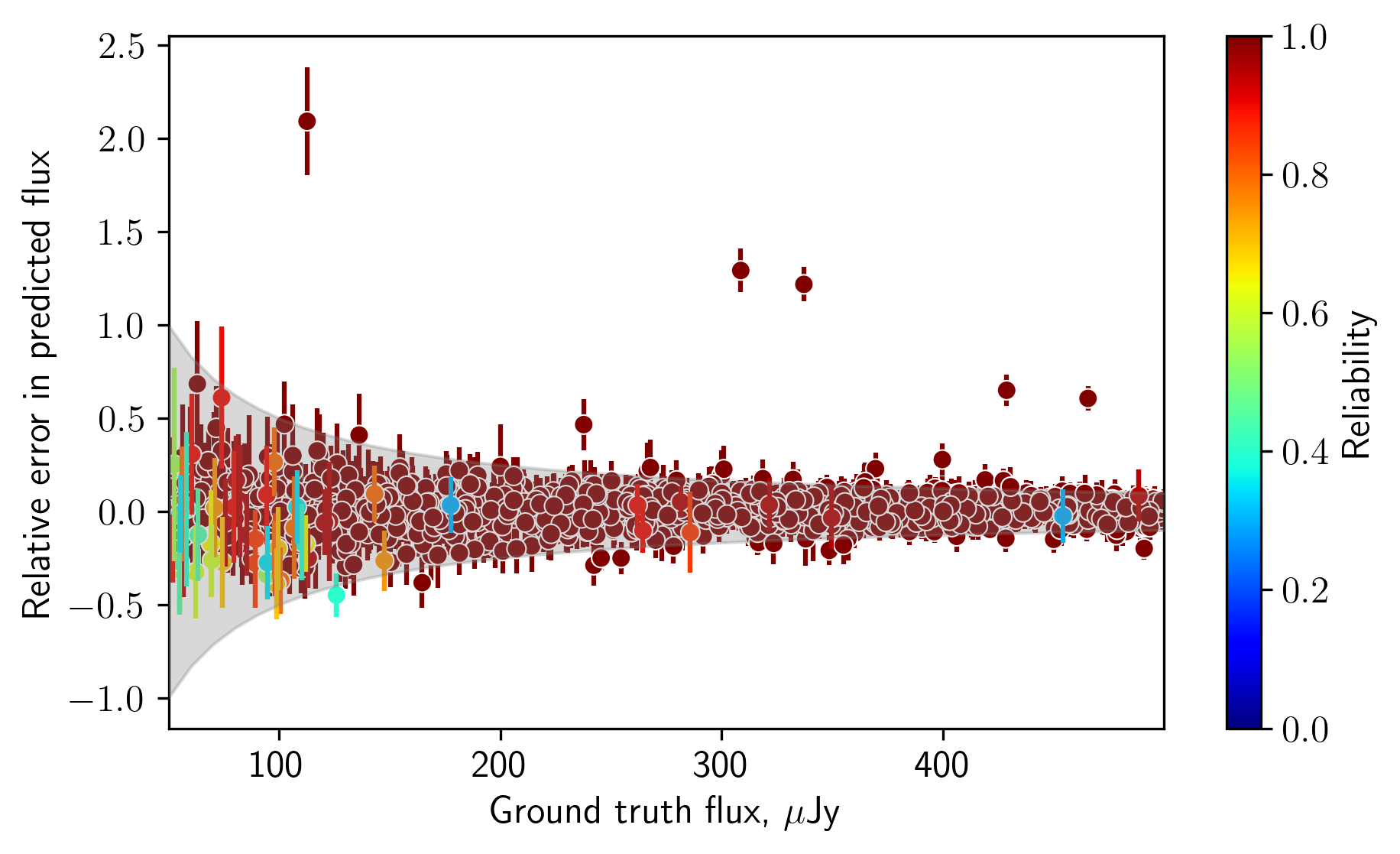}
\caption{Relative error in flux estimation as compared to true flux, aggregated across sources from each predicted image (\detectaggregate ~method). The gray area indicates the 1$\sigma$ range. The error bars represent the standard deviation of the flux estimates obtained from each prediction. The color scale represents the reliability. The results showcased here are based on the model that uses $\gamma = 2$. The outliers are discussed in the Appendix \ref{sec: appendix_outliers}. }

    \label{fig:sources_fluxes}
\end{figure}

As the proposed model is trained to recover the full sky, we also attempted to estimate the flux of the sources in addition to their locations. Here too, we use the Photutils algorithm to estimate the flux of the sources in the predicted sky models. The model predicts a normalized sky model. The normalization is easily invertible, allowing us to estimate the fluxes directly after renormalization of the model outputs. We present the results for all aggregation methods in Figures \ref{fig:aggregation_for_fluxes} and \ref{fig:sources_fluxes}. We plot the relative error  $ \frac{\text{flux}_\text{estimated} - \text{flux}_\text{true}}{\text{flux}_\text{true}} $ on the y-axis and the true flux ($\text{flux}_\text{true}$) on the x-axis, where $\text{flux}_\text{estimated}$ is the estimated flux from the predicted sky model and $\text{flux}_\text{true}$ is known from the simulation. 

The gray area represents the allowable error range: the root-mean-squared (rms) noise of the simulated data is $\sim$50 $\mu$Jy. If the flux estimation error falls within this range, it indicates that the model's estimations are robust. 

In Figure \ref{fig:aggregation_for_fluxes}, we compare different aggregation methods and examine their impact on flux estimation. From this plot, it is clear that aggregation substantially improves the estimates as it averages out errors across all predictions, further highlighting the promise of using stochastic models. Notably, the single run does not yield good results, indicating the importance of per-pixel aggregation. As expected, median aggregation has fewer outliers than mean aggregation. Aggregating through the mean is more efficient than through \detectaggregate, making it a viable alternative. 

The \detectaggregate ~method is shown in Figure \ref{fig:sources_fluxes}. Here for the flux of each source, we can also estimate the uncertainty in the flux by measuring the standard deviation of the flux across several runs. We show this as error bars on the fluxes in Figure \ref{fig:sources_fluxes}. The reliability of each source is color-coded: in the chosen color map, blue corresponds to sources detected in small fractions of predictions, whereas red corresponds to sources that were consistently detected across all of them, giving a reliability of 100 \%. We notice that there are very few outliers in the flux estimation. Interestingly, their reliability scores are high. It is remarkable that the fluxes of the low-S/N sources are estimated relatively accurately.

\begin{table*}[ht]
    \centering
    \caption{Comparative results for the fraction of the detected sources whose estimated flux is within 50 $\mu$Jy of the true flux.}
    \label{tab:comparative_results_fraction}
    \begin{tabular}{ccccccc}
    \toprule
    \multirow{2}{*}{\textbf{Model}} & \multirow{2}{*}{\textbf{Normalization}} & \multicolumn{4}{c}{\textbf{Aggregation}} \\
    \cmidrule{3-6}
    && \textbf{Mean} & \textbf{Median} & \textbf{Single run} & \textbf{D\detectaggregatetable} \\
    \midrule
        CLEAN+PyBDSF & - & \multicolumn{4}{c}{0.57} \\
        
        \multirow{7}{*}{Diffusion} & $\gamma=1$ & 0.85 & 0.66 & 0.048 & 0.015 \\
        & \textbf{$\gamma=2$} &   \textbf{0.96} & \textbf{0.97} &  \textbf{0.88} &  \textbf{0.97}  \\
        & $\gamma=10$ & 0.92 & 0.93 & 0.71 & 0.89  \\
        & $\gamma=20$ &  0.82 & 0.83 & 0.53 & 0.72  \\
        & $\gamma=30$  & 0.74 & 0.75 & 0.47 & 0.66 \\
        \bottomrule
    \end{tabular}
\footnotesize
\begin{minipage}{\textwidth}
Note: The \detectaggregate  ~method results are reported for a reliability threshold of 30\%. We compare our results with PyBDSF.
\end{minipage}
\end{table*}

In Table \ref{tab:comparative_results_fraction}, we compare the proposed method against CLEAN+PyBDSF, measuring the fraction of sources with the correct flux estimation (whithin 50 $\mu$Jy) among all sources detected. The method significantly outperforms PyBDSF scores. We also notice that any aggregation improves the score compared to the single run for $\gamma \geq 2$. 

The model trained with $\gamma = 1$ demonstrates low scores for flux estimation in both single-run and \detectaggregate ~methods. This underperformance can be attributed to the impulsiveness of the predicted images and the inability of the network to effectively reconstruct them in a single run. In this case, the \detectaggregate ~method also fails, as the errors from each individual image are considerable. However, the \aggregatedetect ~approach significantly improves the scores, even for the case of $\gamma = 1$.
For both localization and flux-estimation tasks, the best performance is achieved with the model trained using the sky model normalization approach with $\gamma = 2$ and median aggregation method.

\subsection{Sensitivity to the noise}

The data set used for training the model is characterized by a noise level of 50 $\mu$Jy. During real observations, the noise level does not remain constant. One of the sources of noise is the water vapor present in the atmosphere. To ensure that the model is applicable in real settings, it is vital that it performs well not only under the conditions it was trained on but also when exposed to varying atmospheric conditions. To evaluate how the performance of the  model varies in response to different noise levels, we conducted an analysis in which we alter the amount of water vapor. In order to do this, we used the additional simulations with varying PWV parameter. We tested the model on these data. 

\begin{figure}[ht]
        \centering
        \includegraphics[width=0.8\linewidth]{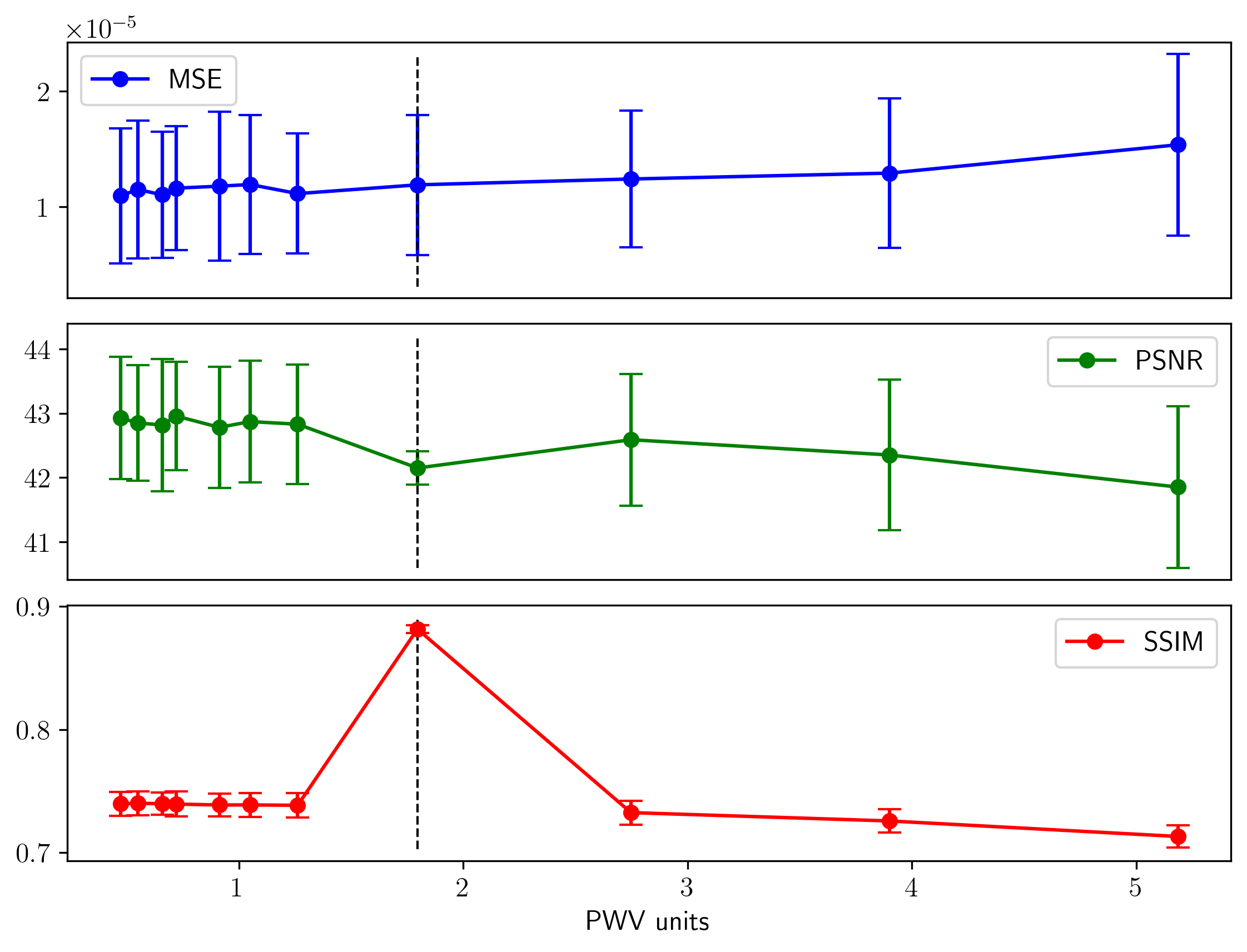}
\caption{Impact of water vapor on reconstruction metrics. The MSE, SSIM, and peak S/N values are normalized to highlight trends. As the amount of water vapor in the atmosphere increases, the quality of the reconstruction diminishes, evidenced by the decreasing SSIM and peak S/N, and the increasing MSE. }
    \label{fig:pwv_reconstruction}
\end{figure}

The reconstruction results are shown in Figure \ref{fig:pwv_reconstruction}. We see that all metrics decrease as the noise increases. The model did experience different S/N during training, which explains the relatively small decrease in the performance. 

\begin{figure}[ht]
    \centering
    \includegraphics[width=0.8\linewidth]{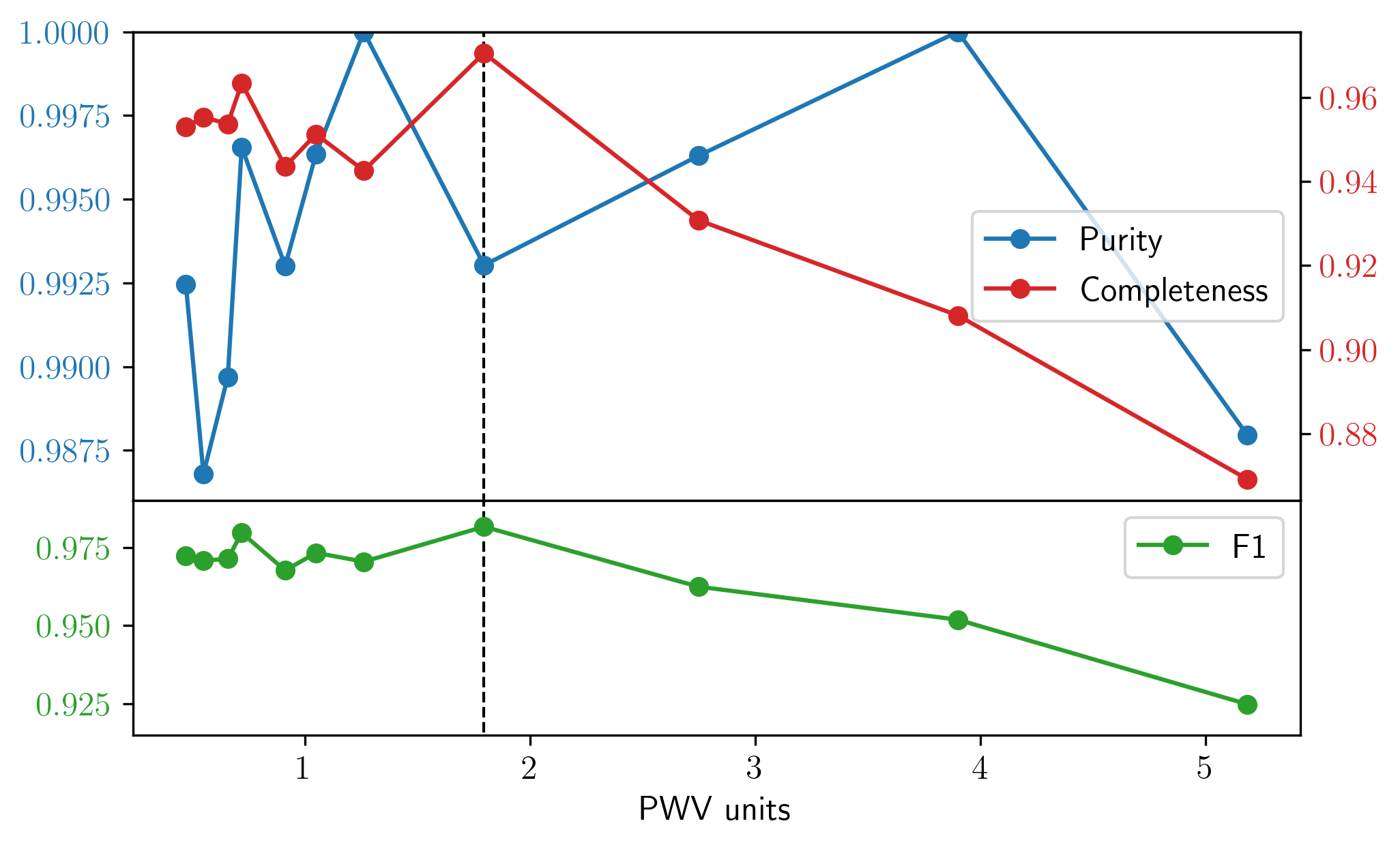}

    \caption{Impact of water vapor on localization metrics. The purity exhibits a relative change of about 1\%, while the completeness decreases significantly by 4\%. The model does not identify many FPs as the noise levels increase, but its capability to accurately detect existing sources diminishes rapidly. Notably, even with the highest noise levels, the completeness score outperforms the previous scores on the test set with the low water vapor noise. The dashed line is at 1.796 PWV units, which corresponds to the training data.}
    \label{fig:pwv_localization}
\end{figure}

\begin{figure}[ht]
    \centering
    \includegraphics[width=0.8\linewidth]{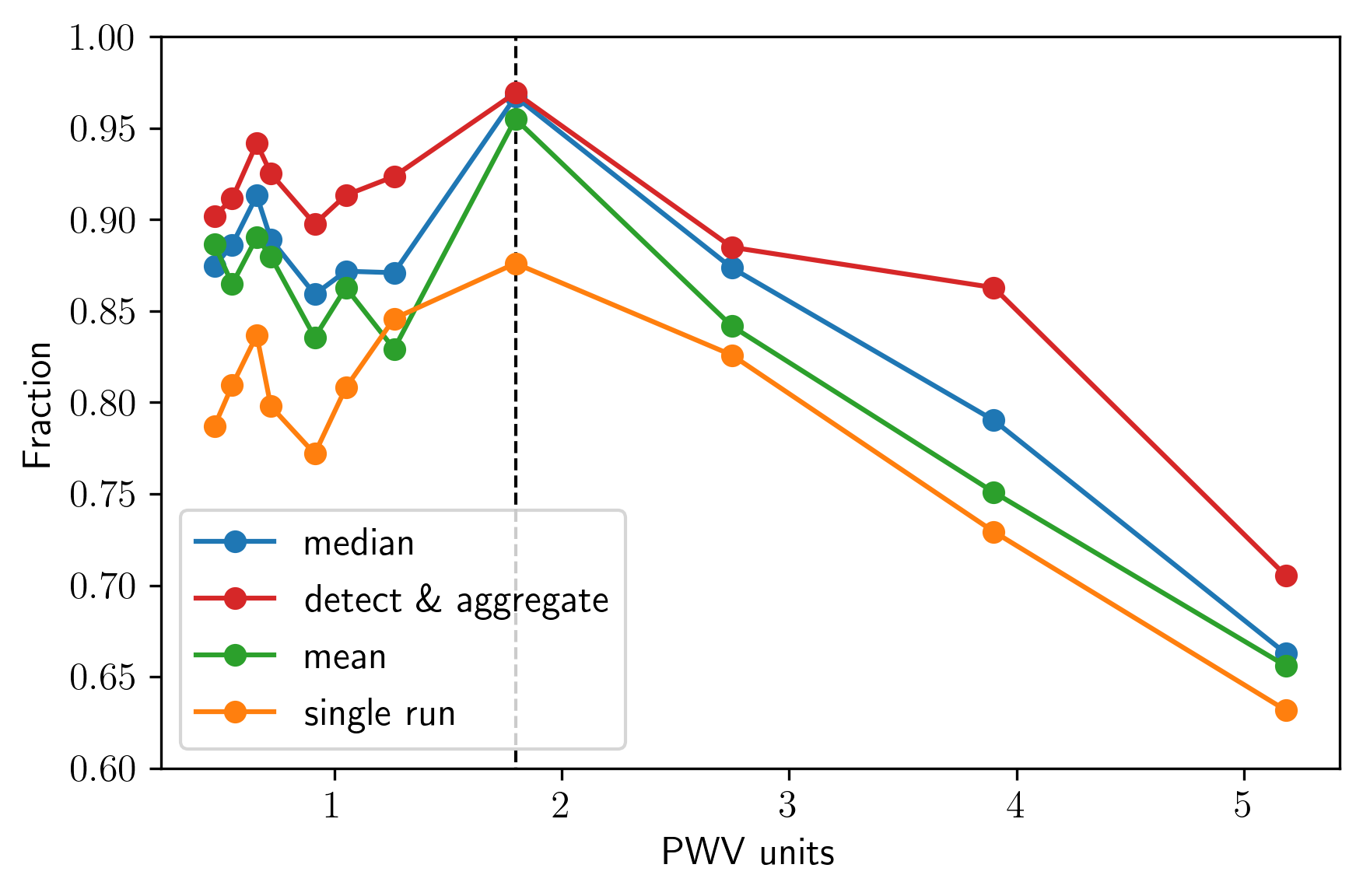}

    \caption{Influence of water vapor on fraction of point sources whose predicted flux falls within noise amplitude from true flux. As water vapor levels rise, this fraction significantly decreases. The dashed line represents a PWV value of 1.796 units, which corresponds to the conditions of the training data. }
    \label{fig:pwv_fractions}
\end{figure}

The localization results are illustrated in Figure \ref{fig:pwv_localization}, where we plot the model’s completeness and purity as a function of the amount of water vapor (and therefore noise level). From the figure, it is evident that as the amount of water vapor increases, the model’s completeness decreases. This decline in completeness indicates that the model is less sensitive to faint sources when they are affected by noise.

In contrast, the purity of the model does not exhibit a significant decrease as the amount of water vapor increases.  The relative stability of purity suggests that, even though the model fails to detect an increasing number of true sources as the noise levels rise (number of FNs rises), it does not produce many FPs by misinterpreting noise as actual sources.

However, changing the water vapor has a significant influence on flux estimation. We present the results of these experiments in Figure \ref{fig:pwv_fractions}. It is evident that the best performance occurs at 1.796 PWV units, which corresponds to the conditions of the training set. Among all the methods, \detectaggregate ~proves to be the most robust to changes in noise. Nevertheless, all methods show a significant decrease in performance as water vapor levels increase. Training the model for specific water vapor values is needed to maintain performance.

\section{Discussion}
\label{sec: discussion}

The method introduced in this paper demonstrates state-of-the-art performance in localizing point sources in  ``dirty'' radio astronomy images, even under low-S/Ns conditions. The results showcase the capability of the presented model to outperform existing methods such as PyBDSF in both localization and characterization of the sources, and particularly in estimating the flux of the sources for the simulated dataset.

The primary limitation of the model is its reconstruction speed. Using an RTX 3090 GPU, the model executes 250 DDPM steps for each of the 20 realizations and takes approximately one minute and eight seconds to produce a single image. This process is conducted with a batch size of 20, allowing for the parallel processing of 20 realizations simultaneously, thereby optimizing the GPU's capabilities. We note that running these 20 parallel realizations requires a GPU with at least 8 GB of memory.This bottleneck presents an area for future development, with a need to explore more efficient generation techniques from diffusion models. Optimizing the speed of a DDPM is one of the main problems for the engineering research community and several solutions have already been suggested by \cite{karras2022elucidating}.

An additional direction of interest is to explore the transferability of the proposed diffusion models. We would like to see whether or not a model trained on data from one telescope can successfully be transferred to another. Similarly, how well a DDPM model trained on simulated data can perform when applied to real data remains unexplored. Also, the portability of the model to SKA is unknown. These queries highlight potential limitations of the current approach and we plan to address them in future work.

\section{Conclusions}
\label{sec: conclusions}

In this work, we introduce a new approach for interferometric image reconstruction in astronomy using diffusion models, in particular the conditional denoising diffusion probabilistic model (DDPM). The DDPM model is trained, validated, and tested on simulated deep ALMA observations of extra-galactic targets previously used by \cite{Taran_2023}. In the proposed approach, we take the dirty images produced from simulated UV visibilities as input with some preprocessing and return a set of different sky models. Given the inherent stochasticity of DDPM models, the described model produces multiple sky models for the same dirty image.  We propose two different approaches to aggregate the predictions: \aggregatedetect ~and \detectaggregate. The properties of the sources are extracted from predicted sky models using the Photutils algorithm. We compare locations and flux values from our DDPM model with the true sky model from the simulations. We devise a simple diagnostic to measure the reliability of a source in predicted sky models. We demonstrate that the DDPM model can be used to perform source localization with high purity and completeness. Finally, we show that it can also be used to accurately estimate source flux. 

Our main results can be summarized as follows:

\begin{itemize}
      \item Aggregating the sky models from multiple runs of the described model using mean or median techniques produces a good representation of the true sky model. Producing a larger number of such runs ($\sim$ 20) produces aggregated sky models with low MSE and high peak S/N\textcolor[rgb]{0.984314,0.00784314,0.027451}{} and SSIM.
      \item The proposed model further performs an even better image reconstruction if we preprocess the true sky models while training with a power $\gamma = 2$ in Equation \ref{eq:normalization}. 
      \item The DDPM-based approach provides a significant improvement in source localization over traditional methods in terms of purity and completeness, particularly at low S/N.
      \item For S/N=2, the described model demonstrates 70\% accuracy without any normalization techniques, outperforming the previous state-of-art algorithm, PyBDSF, which achieves 55\% accuracy.
      \item We find that $n=5$ runs is optimal, allowing the model to obtain an F1-score of above 98 \%.
      \item We introduce a method to estimate the reliability of the predicted sources using the inherent stochastic nature of DDPM.
      \item We also propose a framework for estimating the fluxes of sources as the method focuses on reconstructing the sky model. The performance surpasses that of PyBDSF by 32\% in terms of the fraction of sources for which the flux is estimated within the noise level.
      \item We also examine the influence of the normalization power, $\gamma$, of the root transformation on the efficiency of the method. Our findings suggest that training with $\gamma = 2$ is optimal for both localization and flux estimation. 
      \item We study the impact of noise magnitude by varying water vapor levels. The model's completeness drops to 88\%, yet its purity stays consistent, suggesting that it identifies few false positives despite increased noise. Adapting the model to specific water vapor values is crucial when using it for flux estimation, though the proposed approach consistently outperforms PyBDSF.

\end{itemize}

In summary, we developed a novel DDPM-based method for source detection and flux estimation that operates directly on dirty images. This approach offers significant improvements over existing algorithms in source detection across all S/Ns.

\section*{Acknowledgements}
M. Drozdova and V. Kinakh are supported by the \emph{RODEM: Robust deep density models for high-energy particle physics and solar flare analysis} Sinergia Project funded by the Swiss National Science Foundation, grant number CRSII5-193716. O. Taran and O. Bait are supported by the \emph{AstroSignals} Sinergia Project funded by the Swiss National Science Foundation, grant number CRSII5-193826. We also would like to thank the editor and referee for their useful comments.

\bibliographystyle{plainnat}
\bibliography{paper}
\begin{appendix}

\section{Outliers}
\label{sec: appendix_outliers}

In Figures \ref{fig:aggregation_for_fluxes} and \ref{fig:sources_fluxes}, we observe the outliers that persist irrespective of the aggregation method used. We selected five sources exhibiting the greatest relative error for further analysis. The problematic sources are those that are located in close proximity to each other. The dataset includes two images where two sources share identical coordinates, a scenario that the proposed method cannot discern correctly. Instead of distinguishing the two, the model predicts a single source with a flux equivalent to the sum of the fluxes from the closely situated sources. The fifth error occurs when sources are located near one another, but not with identical coordinates; while the flux of one is accurately determined, the other is predicted to be substantially larger than its actual value. This discrepancy can potentially be influenced by the proximity of the neighboring source, as the predicted flux is of a comparable magnitude. Detailed results are presented in Table \ref{tab:outliers}.

\begin{table}[h]
\centering
\caption{Most extreme outliers: sources with largest estimation errors in flux by DDPM}
\label{tab:outliers}
\begin{tabular}{cccc}
\toprule
Data Type & Ra & Dec & Flux ($\mu$Jy) \\
\midrule
Real & 150.192944 & 3.995500 & 309 \\
Real & 150.192944 & 3.995500 & 429 \\
Predicted & 150.192945 & 3.995501 & 738 \\
\midrule
Real & 149.655167 & 3.643278 & 238 \\
Real & 149.655167 & 3.643278 & 113 \\
Predicted & 149.655165 & 3.643277 & 346 \\
\midrule
Real & 148.163222 & 1.370722 & 126 \\
Real & 148.163222 & 1.370444 & 400 \\
Predicted & 148.163220 & 1.370729 & 399 \\
Predicted & 148.163223 & 1.370444 & 436 \\
\bottomrule
\end{tabular}
\footnotesize
\begin{minipage}{0.46\textwidth} 
Note: These outliers are from the Figures \ref{fig:aggregation_for_fluxes} and \ref{fig:sources_fluxes}.  The proposed algorithm fails when sources are overlapping each other. When there is exact overlap, the prediction is a source with a flux that is approximately equal to the sum of the  fluxes of the overlapping true sources (first two sections).
\end{minipage}
\end{table}

\section{Diffusion without scaling}

\begin{figure}[ht]
        \centering
        \includegraphics[width=0.8\linewidth]{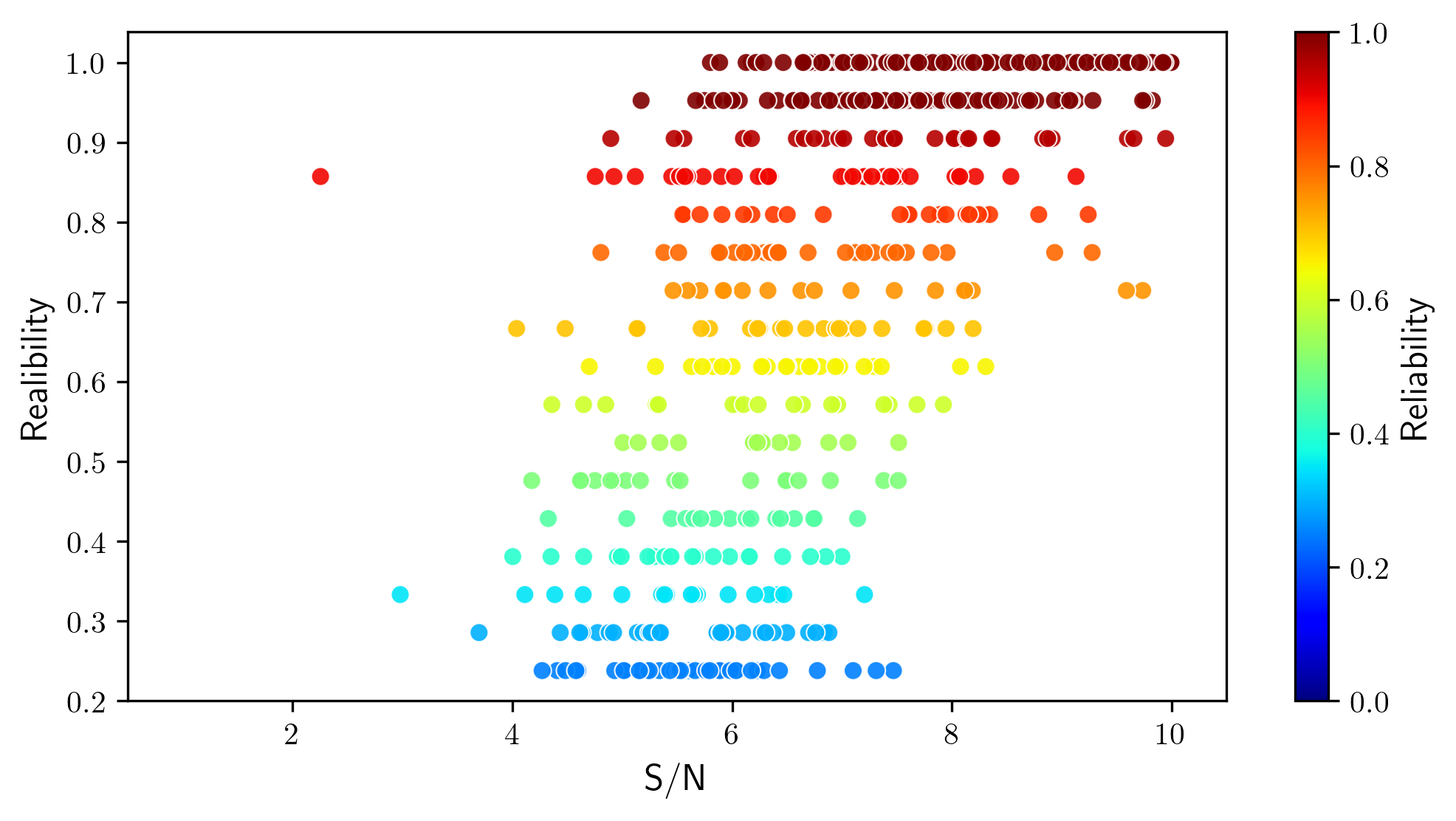}
\caption{S/N versus reliability for $\gamma = 1$.}
    \label{fig:power1_snr_reliability}
\end{figure}

This section presents plots for a DDPM model with $\gamma = 1$ in preprocessing during training in Equation \ref{eq:normalization}. Figure \ref{fig:power1_snr_reliability} illustrates the absence of detected low-S/N sources and a more uniformly spread reliability parameter compared to $\gamma = 2$.

\begin{figure}[ht]
        \centering
        \includegraphics[width=0.8\linewidth]{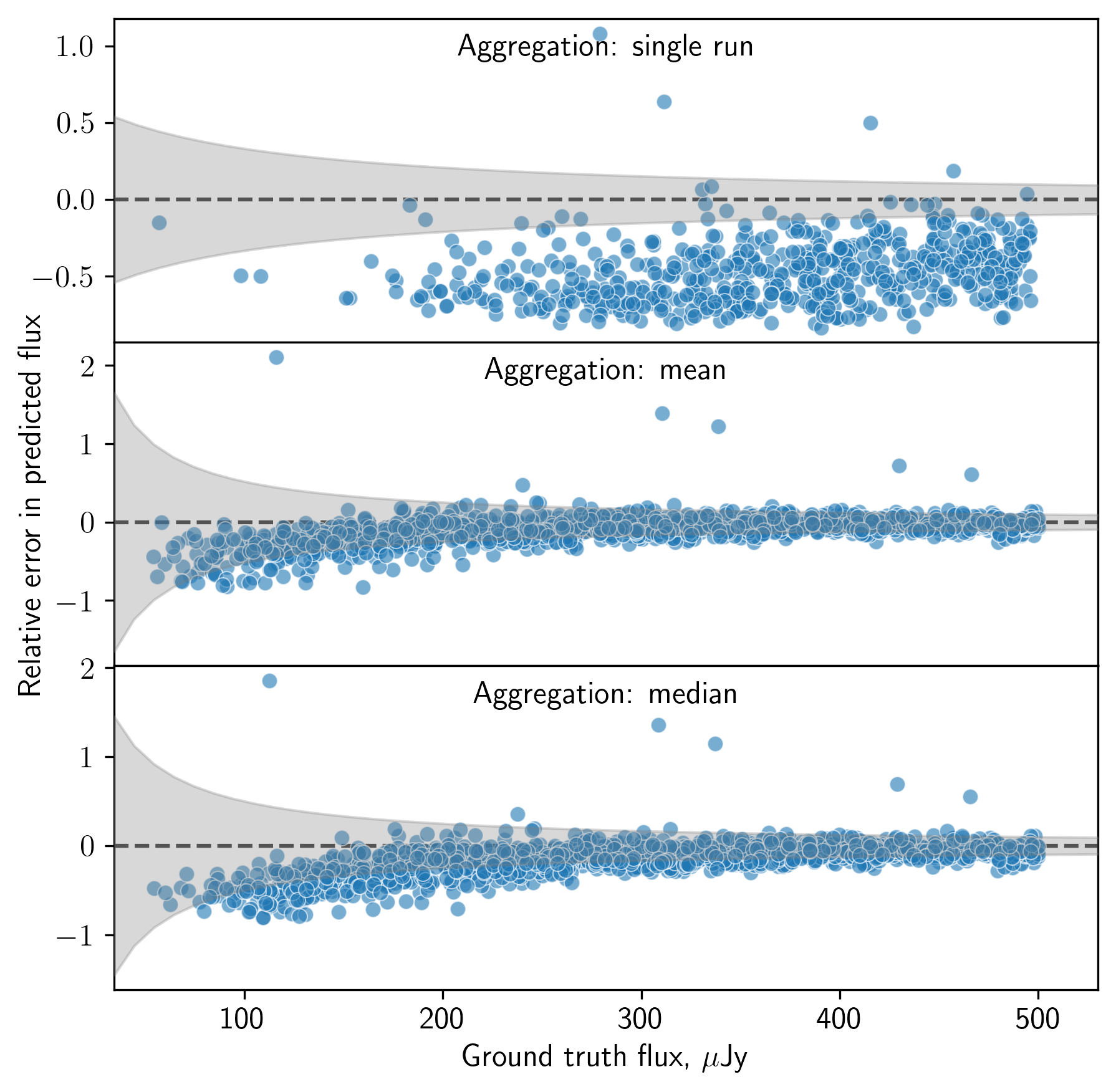}
\caption{Flux estimation for different \aggregatedetect ~methods $\gamma = 1$.}
    \label{fig:power1_flux estimation}
\end{figure}

\begin{figure}[ht]
        \centering
        \includegraphics[width=0.8\linewidth]{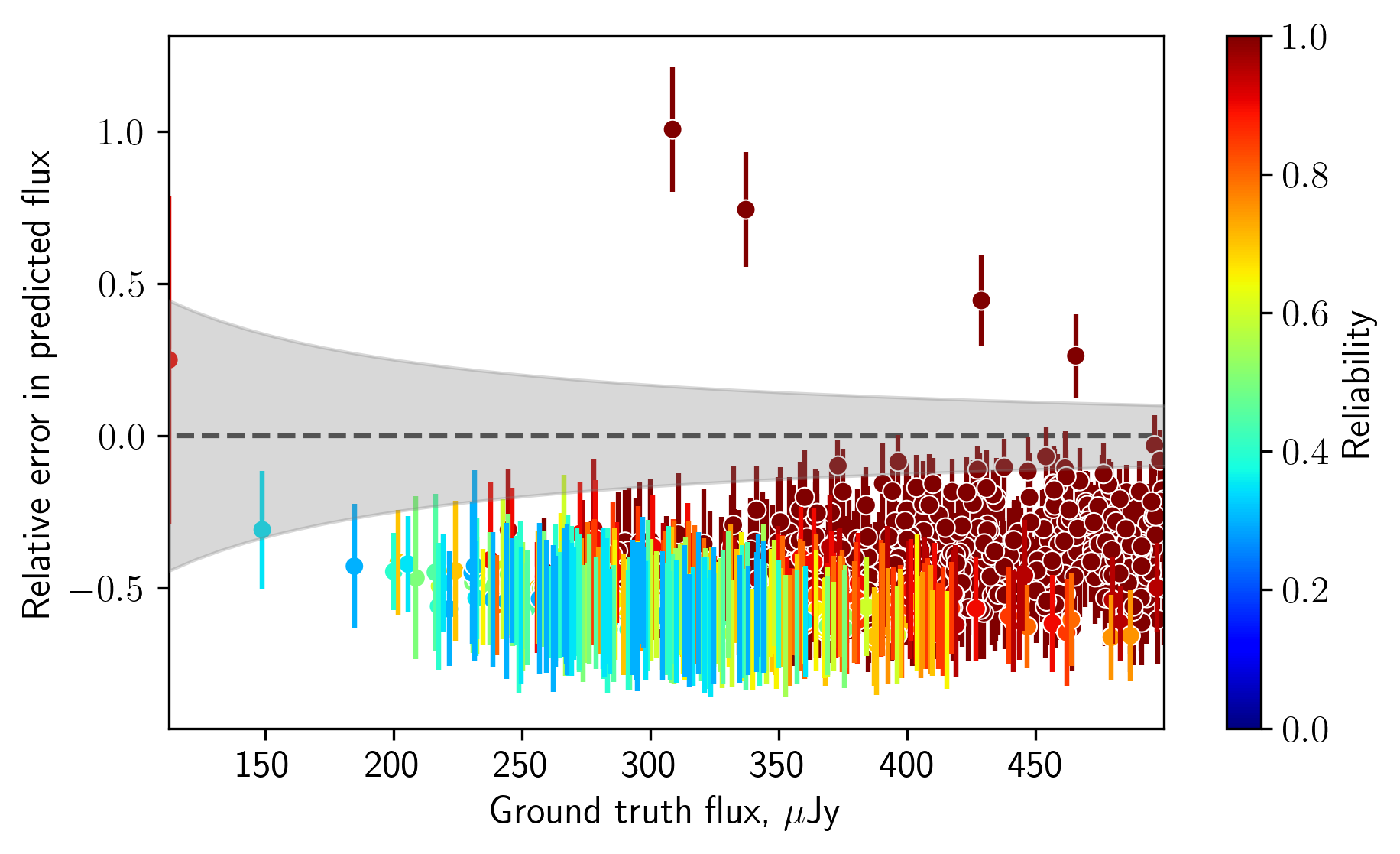}
\caption{Flux estimation for \detectaggregate ~for  $\gamma = 1$.}
    \label{fig:power1_flux estimation2}
\end{figure}

Figures \ref{fig:power1_flux estimation} and \ref{fig:power1_flux estimation2} show that the proposed model consistently underestimates flux values for all ranges in a single run. Nevertheless, aggregation methods, such as mean and median, significantly aid, unlike the \detectaggregate ~method, which performs poorly, which is likely due to the consistent underestimation of sources in a single run. Thus, directly averaging these predictions does not improve the value. However, averaging over image pixels (as \aggregatedetect ~does) increases the source area, correcting overall flux values. 

\begin{figure}[ht]
        \centering
        \includegraphics[width=0.8\linewidth]{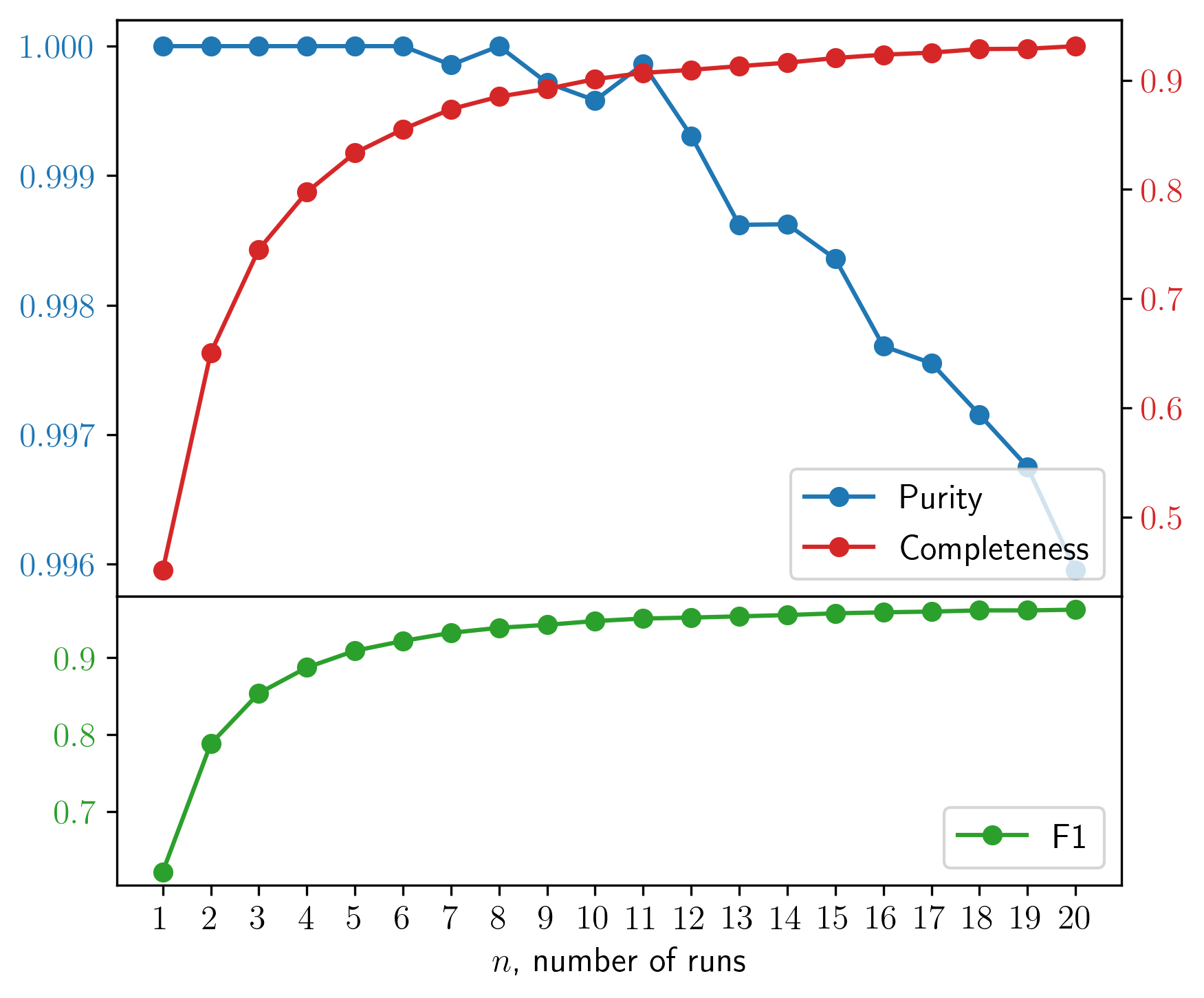}
\caption{Localization metrics for $\gamma = 1$.}
    \label{fig:power1_localisation}
\end{figure}

\begin{figure}[ht]
        \centering
        \includegraphics[width=0.8\linewidth]{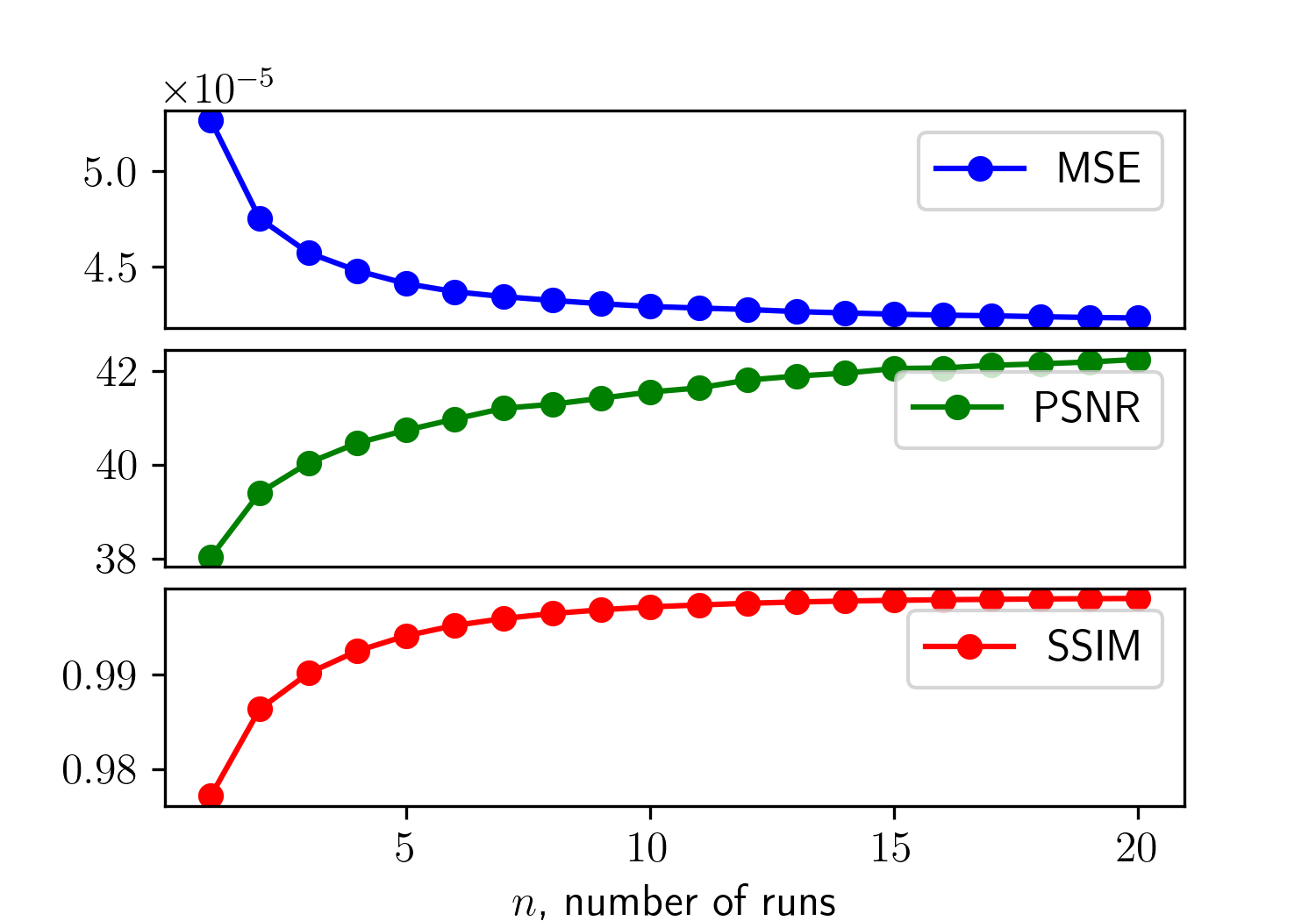}
\caption{Normalized metrics for $\gamma = 1$.}
    \label{fig:power1_reconstruction}
\end{figure}

Figures \ref{fig:power1_localisation} and \ref{fig:power1_reconstruction} demonstrate the dependency of localization and reconstruction metrics on the number of runs for $\gamma = 1$. The reconstruction metrics follow the same pattern as $\gamma = 2$: a greater number of images equates to better metrics. The localization metrics exhibit a different behavior: a single run has very low completeness, less than 50\% compared to 96\% for $\gamma = 2$. For $\gamma = 1$, the completes improves slower than for $\gamma = 2$, not surpassing 95\% even after 20 runs, contrary to 98\% for $\gamma = 2$.

\end{appendix}

\end{document}